\documentclass[%
reprint,
 amsmath,
 amssymb,
 aps,
 prx,
 longbibliography
]{revtex4-1}

\usepackage{graphicx}% Include figure files
\usepackage{dcolumn}% Align table columns on decimal point
\usepackage{bm}% bold math
\usepackage{subfigure}
\usepackage{dsfont} % mathds command

\usepackage{graphicx}
\usepackage[colorlinks=true,citecolor=blue]{hyperref}
\usepackage[margin=1in]{geometry}

\usepackage{multirow}
\usepackage{physics}

\usepackage[capitalise]{cleveref}
\crefname{section}{Sec.}{Secs.}
\crefname{appendix}{App.}{Apps.}

\def\ValueevenOddWinelandPurityNeven{200}
\def\ValueevenOddWinelandPurityNodd{201}
\def\ValueevenOddLocalDissipationNeven{30}
\def\ValueevenOddLocalDissipationNodd{31}
\def\ValueevenOddLocalDissipationgammaphi{0.00005}
\def\ValueevenOddLocalDissipationgammarel{0.0001}
\def\ValueevenOddLocalDissipationgammaphiInseta{0.00005}
\def\ValueevenOddLocalDissipationgammaphiInsetb{0.0005}
\def\ValueevenOddLocalDissipationgammaphiInsetc{0.005}
\def\ValueevenOddLocalDissipationgammaphiInsetd{0.05}
\def\ValueevenOddLocalDissipationgammarelInseta{0.0001}
\def\ValueevenOddLocalDissipationgammarelInsetb{0.001}
\def\ValueevenOddLocalDissipationgammarelInsetc{0.01}
\def\ValueevenOddLocalDissipationgammarelInsetd{0.1}

\def\ValueevenOddTimeEvolutionr{2.5}

\def\ValueevenOddSyProbabilityDistributionNeven{200}
\def\ValueevenOddSyProbabilityDistributionNodd{201}
\def\ValueslowRateDecayRatera{0.2}
\def\ValueslowRateDecayRaterb{0.5}
\def\ValueslowRateDecayRaterc{1.0}
\def\ValueslowRateDecayRaterd{1.4}
\def\ValueslowRateDecayRatere{4.0}
\def\ValueslowRateDecayRateca{0.9}
\def\ValueslowRateDecayRatecb{0.95}
\def\ValueslowRateDecayRatecc{0.98}
\def\ValueslowRateDecayRatecd{0.99}
\def\ValueslowRateTimeEvolutionN{50}
\def\ValueslowRateTimeEvolutionr{1.0}

\def\ValueslowRateTimeEvolutiongammaphiOverGamma{0.005}
\def\ValueslowRateTimeEvolutionkappaOverGamma{0}

\def\ValueslowRateTimeEvolutiongammarelOverGammab{0.001}
\def\ValueslowRateTransitionRatesN{128}

\def\ValueTlsLossr{2.5}

\def\ValuefiniteTemperatureWinelandN{200}

\def\ValuefiniteTemperatureRatioWinelandAtRoptVsHLexp{0.13}

\def\ValuefiniteTemperatureRatioWinelandAtRoptVsHLN{200}
\def\ValuefiniteTemperatureROptVsNthexp{-0.25}

\def\ValuefiniteTemperatureROptVsNexp{0.75}

\def\ValuefiniteTemperatureEvenOddFixedNN{8}
\def\ValuefiniteTemperatureEvenOddFixedNNpo{9}

\def\ValuefiniteTemperatureEvenOddFixedGammaCollgammacoll{0.017}

\renewcommand{\vec}[1]{\boldsymbol{#1}}

\newcommand{\comma}{~,}
\newcommand{\fullstop}{~.}
\newcommand{\ie}{\emph{i.e.}}
\newcommand{\eg}{\emph{e.g.}}
\newcommand{\hc}{\mathrm{h.c.}}

\newcommand{\wineland}{ \xi^{2}_{R} }
\newcommand{\gammaphi}{\gamma_{\phi}}
\newcommand{\gammarel}{\gamma_{\rm rel}}
\newcommand{\mD}[2]{\mathcal{D}\left[{#1}\right] {#2} }
\newcommand{\bigoh}[1]{\mathcal{O}\left( {#1} \right)}
\newcommand{\rhoop}{ {\hat \rho}}
\newcommand{\Crel}{\mathcal{C}_{\rm rel}}
\newcommand{\etarel}{\eta_{\rm rel}}
\newcommand{\etaphi}{\eta_{\phi}}
\newcommand{\psidk}[1]{\ket{\psi_{\rm dk}[#1]}}
\newcommand{\Sp}{{{\hat S}_{+}}}
\newcommand{\Sm}{{{\hat S }_{   -}}}
\newcommand{\Sx}{{\hat S}_{x}}
\newcommand{\Sy}{{\hat S}_{y}}
\newcommand{\Sz}{{\hat S}_{z}}
\newcommand{\Sigmaop}{{\hat \Sigma}}
\newcommand{\am}{{\hat a}}

\newcommand{\sy}{\hat\sigma_{y}}
\newcommand{\sz}{\hat\sigma_{z}}

\newcommand{\smm}{\hat\sigma_{-}}
\newcommand{\dg}{^\dagger}
\newcommand{\A}{{\hat A}}
\newcommand{\ave}[1]{\langle #1 \rangle}

\newcommand{\CSzSz}{\hat{C}_{ZZ}}
\newcommand{\etatilde}{{\tilde{\eta}}}
\newcommand{\Ctilde}{\tilde{\mathcal{C}}}
\newcommand{\kappasqz}{\kappa_{\rm sqz}}
\newcommand{\gammacoll}{\gamma_{\rm coll}}
\newcommand{\kappaint}{\kappa_{\rm int}}
\newcommand{\gammaheat}{\gamma_\mathrm{heat}}

\begin{document}

\title{Reservoir-engineered spin squeezing: macroscopic even-odd effects and hybrid-systems implementations}

%PeterG: there is an issue of getting the asterisk showing up before the comma; have to try to sort that out. 
\author{Peter Groszkowski$^{1}$}
\thanks{These authors contributed to this work equally.}
\author{Martin Koppenh\"ofer$^{1}$}
\thanks{These authors contributed to this work equally.}
\author{Hoi-Kwan Lau$^2$}
\author{A. A. Clerk$^1$}
\affiliation{$^1$Pritzker School of Molecular Engineering, University of Chicago, Chicago, IL, USA \\
$^2$Department of Physics, Simon Fraser University, Burnaby, BC, Canada}

\date{\today}

\begin{abstract}
We revisit the dissipative approach to producing and stabilizing spin-squeezed states of an ensemble of $N$ two-level systems, providing a detailed analysis of two surprising yet generic features of such protocols.  The first is a macroscopic sensitivity of the steady state to whether $N$ is even or odd. We discuss how this effect can be avoided (if the goal is parity-insensitive squeezing), or could be exploited as a new kind of sensing modality 
to detect the addition or removal of a single spin.
The second effect is an anomalous emergent long timescale and a ``prethermalized'' regime that occurs for even weak single-spin dephasing. 
This effect allows one to have strong spin squeezing over a long transient time even though the level of spin squeezing in the steady state is very small.
We also discuss a general hybrid-systems approach for implementing dissipative spin squeezing that does not require squeezed input light or complex multi-level atoms, but instead makes use of bosonic reservoir-engineering ideas. Our protocol is compatible with a variety of platforms, including trapped ions, NV defect spins coupled to diamond optomechanical crystals, and spin ensembles coupled to superconducting microwave circuits.
\end{abstract}

\maketitle


\section{Introduction}
\label{sec:introduction}

Among the most sought-after states in quantum metrology are spin-squeezed states, highly entangled states of spin-$1/2$ ensembles that enable parameter sensing with a sensitivity better than the standard quantum limit, even reaching fundamental Heisenberg-limit scaling \cite{kitagawa1993squeezed,pezze2018quantum}.  
The standard approach for producing these states is to unitarily evolve an initial product state under a collective spin-spin interaction Hamiltonian.  
While many interactions are possible, the most widely studied one is the one-axis twist (OAT) Hamiltonian \cite{kitagawa1993squeezed}, which has been realized in a number of ground-breaking experiments
\cite{LerouxPRL2010,Riedel2010,GrossNature2010,HostenScience2016}.  It unfortunately is not capable of achieving Heisenberg-limited squeezing even in the ideal case \cite{pezze2018quantum}.  
An alternate, more complex interaction Hamiltonian is the two-axis twist (TAT) Hamiltonian \cite{kitagawa1993squeezed,Cappellaro2009,You2011TAT,borregaard2017one,groszkowski2020heisenberg}, which, while more resource intensive, allows achieving Heisenberg-limited scaling.

\begin{figure}[t]
    \centering
    \includegraphics[width=0.46\textwidth]{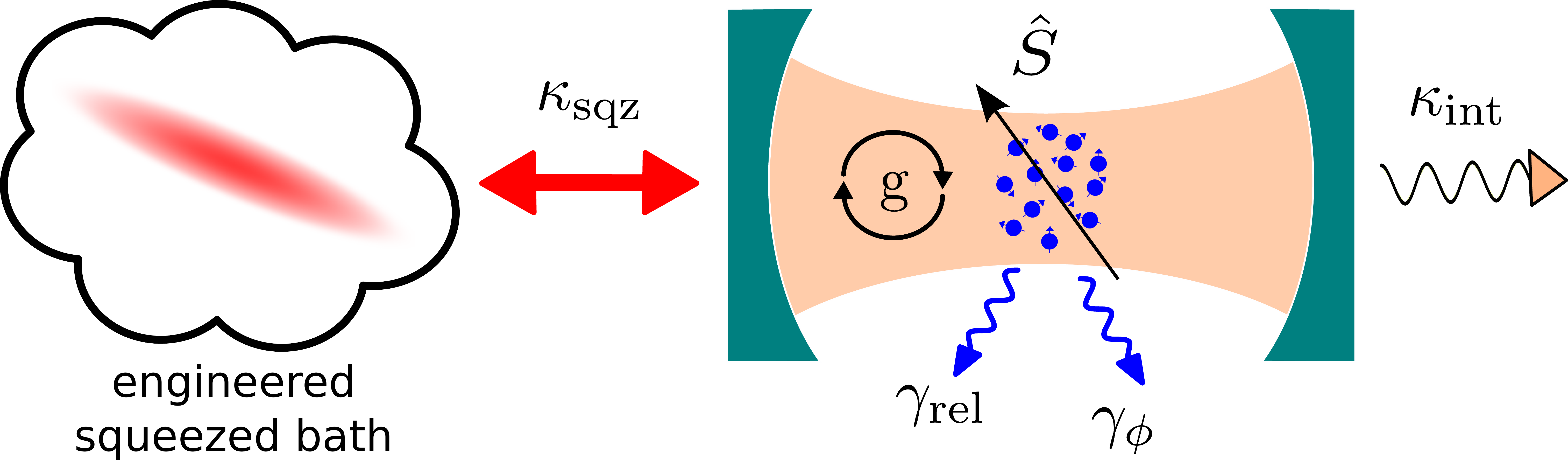}
    \caption{
        Schematic representation of a generic approach to generate dissipative spin squeezing by coupling spins to a bosonic mode that interacts with a squeezed reservoir. 
    	The squeezing rate experienced by the cavity is governed by the parameter $\kappasqz$, while $g$ represents the spin-cavity coupling strength. 
    	Limiting factors to the protocol's performance are the intrinsic photon-loss rate $\kappaint$, the local spin-relaxation rate $\gammarel$, and the local spin-dephasing rate $\gammaphi$. 
    }
    \label{fig:schematic}
\end{figure}

While easy to understand, tailored unitary-evolution is not the only approach to spin squeezing.  
An alternative is to use the general strategy of reservoir engineering \cite{Poyatos1996}, where tailored dissipation is exploited to both produce and \emph{stabilize} a non-trivial state of interest, \ie, a spin-squeezed state (see Fig.~\ref{fig:schematic}).  
The dissipative approach in principle has several advantages:  the spin-squeezed state is stabilized in the steady state (as opposed to just prepared at a specific instant of time), the stabilization is largely insensitive to the initial state of the ensemble, and one can achieve Heisenberg-limited scaling.  
The dissipative stabilization of bosonic squeezed states has been studied extensively both theoretically \cite{1993PhRvL..70..556C,kronwald2013arbitrarily,didier2014perfect} and experimentally \cite{Wollman2015,Kienzler2015us,Teufel2015,Sillanpaa2015,Lei2016,Dassonneville2021}.  
Corresponding schemes for spin squeezing have also been studied theoretically. 
The earliest works analyzed schemes where atoms are directly illuminated with squeezed light.  
Both the cases of two-level atoms \cite{AgarwalPuri1989,AgarwalPuri1990,AgarwalPuri1994} and $V$-type multilevel atoms \cite{Kuzmich1997} were studied. 
More recently, it was shown theoretically that the same effective dissipative dynamics could be realized by using Raman processes in driven multi-level atoms coupled to a lossy cavity \cite{dalla2013dissipative,borregaard2017one}.

%%%%%%%%%%%%%%%%%%%%%%%%%%%%%%%%%%%%%%%%%%%%%%%%
%%%%%%%%%%%%%%%%%%%%%%%%%%%%%%%%%%%%%%%%%%%%%%%%

In this work, we revisit the dissipative approach to spin squeezing.  
Our work complements previous studies both by discussing a powerful, alternative method for implementing these schemes, as well as describing surprising phenomena that had not been fully analyzed in the past.  
In terms of implementation, we analyze a very general hybrid-systems approach that harnesses bosonic dissipative squeezing.  
We consider a spin ensemble which is resonantly coupled to a cavity mode (via a standard Tavis-Cummings \cite{TavisCumming1968} interaction), which is in turn coupled to an effective squeezed reservoir (see \cref{fig:schematic}).  
Previous proposals \cite{AgarwalPuri1990,Kuzmich1997} suggested to implement this squeezed reservoir by driving the cavity with squeezed light, an approach which is limited by losses associated with the transport and injection of an externally prepared optical squeezed state. 
We show that there are also simpler methods to generate the effective squeezed reservoir, which  
can be implemented using only \emph{classical} optical or microwaves drives by harnessing existing dissipative bosonic squeezing schemes. 
Specifically, we consider coupling the cavity to an auxiliary lossy degree of freedom (two level system or bosonic mode) which is driven simultaneously with imbalanced red-detuned and blue-detuned sideband drives. 
Such schemes produce an effective squeezed dissipator for the cavity and have been experimentally implemented in wide variety of platforms, including optomechanics \cite{Wollman2015}, trapped ions \cite{Kienzler2015us} and superconducting circuits \cite{Dassonneville2021}.  
Since only classical radiation is required, this approach is insensitive to the aforementioned transport and insertion losses of squeezed radiation.
We demonstrate that this hybrid-systems approach to dissipative spin squeezing can reach the Heisenberg limit, and also outperform OAT in the presence of single-spin $T_1$ decay.  
Note that unlike the Raman scheme of Ref.~\onlinecite{dalla2013dissipative}, which requires atoms with a specific four-level configuration, the approach here only requires standard two-level atoms, making it compatible with a wide variety of systems (including possibly solid-state systems such as ensembles of NV defect spins in diamond \cite{bennett2013phonon}).

Our work also analyzes surprising phenomena that were not fully discussed previously.  
Perhaps most striking is the extreme sensitivity of dissipative spin squeezing to the parity of the total number of spins $N$:  the steady state is macroscopically different for $N$ spins versus $N+1$ spins.  
While this effect was implicitly contained in the results of Agarwal and Puri~\cite{AgarwalPuri1989,AgarwalPuri1990} (see Sec.~\ref{sec:PreviousWork} for a detailed discussion of the relation to previous works), we provide here a fully qualitative and quantitative analysis. 
We discuss how this effect can be avoided (if one wants strong spin squeezing independent of parity), and how it could also be exploited as a new kind of sensing modality.  
We also make a surprising connection to a non-dissipative many-body system, the antiferromagnetic Lipkin-Meshkov-Glick (LMG) model \cite{LipkinMeshkovGlick1965,Unanyan2003}.

A second surprising and new phenomenon we describe is the interplay between collective dissipative-spin-squeezing dynamics and noncollective single-spin dephasing.  
As we show, this results in an extremely long relaxation timescale in the system (\ie, inverse dissipative gap) which grows with system size $N$.  
At a fundamental level, the effect has parallels to prethermalization behavior observed in weakly nonintegrable systems (see, \eg, \cite{DAlessio2016,Langen2016}).  
At a practical level, we show that even infinitesimally weak single-spin dephasing dramatically impairs the steady-state spin squeezing to at most $-3\,\mathrm{dB}$.  
However, we also show that this need not be a limitation: large amounts of squeezing are possible in the prethermalized regime, \ie, at transient times parametrically shorter than the timescale required to reach the steady state, or by deliberately adding very small levels of single-spin relaxation. 
This effect may allow one to quickly generate strong squeezing even in parameter regimes that had previously been discarded based on the low level of steady-state squeezing.

Finally, we analyze the impact of imperfections in the reservoir-engineering process that lead to 
the engineered squeezed dissipation having a non-zero impurity and effective thermal occupancy.  
We reveal a striking sensitivity of spin squeezing to such imperfections if the squeezing strength is made too large.

The remainder of this paper is organized as follows: In \cref{sec:model}, we outline the key idea behind the standard approach to dissipative spin squeezing as well as summarize our generic protocol. 
In \cref{sec:evenOdd}, we explore the even-odd effect and briefly discuss connections to sensing.
In \cref{sec:scaling}, we carefully analyze the performance of our dissipative spin squeezing protocol in the presence of single-spin dissipation, showing that the steady-state squeezing it generates can outperform the transient squeezing produced by standard OAT.  
In \cref{sec:dephasing}, we discuss the emergence of anomalously slow relaxation times 
and we introduce a dynamical-decoupling protocol to cancel the effect of inhomogeneous broadening.
In \cref{sec:Temperature}, we analyze imperfections in the reservoir-engineering process, while 
in \cref{sec:Implementation}, we discuss in more detail how our protocol could be implemented in a variety of different physical systems.  
In Sec.~\ref{sec:PreviousWork}, we review previous works on dissipative spin squeezing and discuss their relation to our new findings.
Conclusions and a summary are presented in \cref{sec:Conclusions}.

%%%%%%%%%%%%%%%%%%%%%%%%%%%%%%%%%%%%%%%%%%%%%%%%%%%%%%%%%%%%%%%%%%%%%%%%%
%%%%%%%%%%%%%%%%%%%%%%%%%%%%%%%%%%%%%%%%%%%%%%%%%%%%%%%%%%%%%%%%%%%%%%%%%
%%%%%%%%%%%%%%%%%%%%%%%%%%%%%%%%%%%%%%%%%%%%%%%%%%%%%%%%%%%%%%%%%%%%%%%%%

\section{Model and the basic dissipative squeezing protocol}
\label{sec:model}

The reservoir engineering approach to spin squeezing requires one to construct a nontrivial dissipative environment for the spins.  
In this section, we review the idealized spin-only quantum master equation that describes the needed dissipative dynamics \cite{AgarwalPuri1989,AgarwalPuri1990}.  
We then present a more realistic model that corresponds to the generic, experimentally-friendly hybrid-systems setup sketched in Fig.~\ref{fig:schematic}, where a spin ensemble is coupled to a cavity (or other bosonic mode), which is in turn coupled to an engineered squeezed reservoir.

Throughout this paper, we quantify the amount of metrologically-useful spin squeezing (\ie, as relevant to a standard Ramsey measurement) using the Wineland parameter \cite{wineland1992spin,pezze2018quantum}. 
It is defined as 
\begin{align}\label{eq:wp}
    \xi_R^2 \equiv 
        N  \frac{\langle \Delta \hat{S}_{\perp}^2 \rangle}{\langle \vec{\hat{S}} \rangle^2} \comma 
\end{align}
where $ \langle \Delta \hat{S}_{\perp}^{2} \rangle $ is the minimum variance in a direction perpendicular to the direction of the mean of the collective spin and $\vec{\hat{S}} \equiv (\Sx,\Sy,\Sz)$ is the vector of spin operators.

%%%%%%%%%%%%%%%%%%%%%%%%%%%%%%%%%%%%%%%%%%%%%%%%%%%
%%%%%%%%%%%%%%%%%%%%%%%%%%%%%%%%%%%%%%%%%%%%%%%%%%%

\subsection{Idealized spin-only model}
\label{sec:idealCase}

We consider the following quantum master equation acting on the Hilbert space of $N$ spin-$1/2$ particles,
\begin{align}
    \dot\rhoop =&   \Gamma \mD{\Sigmaop[r] }{\rhoop} \comma
    \label{eq:masterEqIdeal1}
\end{align}
where we introduced the operator
\begin{align}
    \Sigmaop[r] &=  \cosh(r) \Sm - \sinh(r) \Sp \fullstop
    \label{eq:sigmaPp}
\end{align}
Here, $\Gamma$ is the coupling rate to the engineered reservoir, $r$ characterizes the squeezing strength, and $\mathcal{D}[\hat{z}]\hat{\rho} = \hat{z} \hat{\rho} \hat{z}^\dagger - \{ \hat{z}^\dagger \hat{z}, \hat{\rho} \}/2$ is the standard Lindblad dissipative superoperator. 
We also introduced the collective spin operators $\hat{S}_{\pm}=\Sx \pm i \Sy$ with $\hat{S}_{k} =\frac{1}{2} \sum_{j=1}^N \hat{\sigma}_k^{(j)}$ for $k \in \{x,y,z\}$, where $\hat{\sigma}_k^{(j)}$ denotes a standard Pauli matrix acting on the $j$th spin.
Here, $\Sigmaop[r]$ is analogous to a standard bosonic Bogoliubov annihilation operator, where bosonic raising and lowering operators have been replaced by $\Sp$ and $\Sm$ respectively.  
Similar to reservoir-engineered bosonic squeezing \cite{kronwald2013arbitrarily}, the desired squeezed state will correspond to the vacuum of this operator  
and the squeezing parameter $r$ characterizes the amount of squeezing, $e^{-2r}$, of the vacuum fluctuations.

To be more explicit, Refs.~\onlinecite{AgarwalPuri1989,AgarwalPuri1990} showed that, for even $N$, \cref{eq:masterEqIdeal1} has pure steady states that correspond to zero-eigenvalue eigenstates (\ie, ``dark states'') of $\Sigmaop[r]$, 
\begin{align}
	\Sigmaop[r] \psidk{j;r} = 0 \fullstop
	\label{eq:DarkStateCondition}
\end{align}
Since \cref{eq:masterEqIdeal1} conserves the total angular momentum $j$, there is a dark state for each allowed value of $j$.
Each $\psidk{j;r}$ has a mean spin polarization in the $z$ direction, and exhibits squeezing (anti-squeezing) of $\Sy$ ($\Sx$). 
The choice of the squeezing axis is determined by the relative phase between the $\Sp$ and $\Sm$ terms in \cref{eq:sigmaPp}, which is chosen here to be $-1$.
If the system is initialized in an arbitrary state with a definite value of $j$, the dissipative dynamics will relax the system to a dark state in this subspace.  
For states in the maximum-angular-momentum subspace $j = j_\mathrm{max} = N/2$, the relaxation timescale (\ie, the inverse dissipative gap of the Liouvillian) is $\propto 1/ N \Gamma$, see Sec.~\ref{sec:evenOdd}.
Note that the dark states with $j < j_\mathrm{max}$ are not unique, since the corresponding angular-momentum subspaces are degenerate \cite{Dicke1954}. 
However, if the initial state and the dynamics are invariant under permutation of spins, the system will only explore permutationally invariant states \cite{ChaseGeremia2008}, and there is a unique dark state for each $j$ subspace, see \cref{sec:App:LiouvillianPerturbationTheorySlowTimescale:StructureHS}.

As detailed in \cref{sec:App:EvenOddEffect}, the dark states can be expressed in the form \cite{AgarwalPuri1990,AgarwalPuri1994}
\begin{align}
	\psidk{j;r} = \mathcal{N}(r) e^{\theta \Sz} \ket{j,0}_y \comma
	\label{eq:darkState0}
\end{align}
where $\ket{j,m}_y$ denotes an eigenstate of $\vec{\hat{S}}^2$ and $\Sy$, $\mathcal{N}(r)$ is a normalization constant, and we defined $\theta = \ln \sqrt{\tanh (r)}$.
In terms of the eigenstates $\ket{j,m}$ of $\vec{\hat{S}}^2$ and $\Sz$, these states read as follows.
\begin{align}
    \psidk{j;r} &=  \sum_{m=-j}^{j} c_{m}^{(j)}(r) \ket{j, m} \comma
    \label{eq:darkState1}
\end{align}
where every second coefficient is nonzero,
\begin{align}
    c^{(j)}_{-j + 2k}(r)  &=   
    \begin{pmatrix}j\\k\end{pmatrix} 
    \sqrt{\begin{pmatrix}2j\\ 2k\end{pmatrix}^{-1}}  
    \tanh^{k}(r) c^{(j)}_{-j}(r) \comma \label{eq:darkStateck} 
\end{align}
for $k\in \{0,...,j \}$, and all other coefficients vanish, $c^{(j)}_{-j+2k+1}(r) = 0$ for $k\in \{0,...,j-1 \}$. The coefficient $c_{-j}^{(j)}(r)$ serves as a normalization constant.

The parameter $r$ controls the amount of squeezing in the steady state. 
If we initialize the system in an arbitrary state with $j = N/2$, the resulting pure steady state is squeezed, with $\wineland \rightarrow 2/(N+2)$ in the large-$r$ limit.  
This corresponds to Heisenberg-limited spin squeezing, and thus outperforms both the standard quantum limit (\ie, $\wineland =1$) as well as the maximum squeezing possible with an ideal OAT interaction ($\wineland \propto 1/N^{2/3}$).
Note that a standard leading-order Holstein-Primakoff approximation could be used to map \cref{eq:masterEqIdeal1} to a bosonic squeezing dissipator; however, this would not let one understand the ultimate saturation of squeezing (with increasing $r$) to the Heisenberg-limited value.

%%%%%%%%%%%%%%%%%%%%%%%%%%%%%%%%%%%%%%%%%%%%%%%%%%%
%%%%%%%%%%%%%%%%%%%%%%%%%%%%%%%%%%%%%%%%%%%%%%%%%%%

\subsection{Hybrid-systems approach to dissipative spin squeezing}
\label{sec:realisticSystem1}

As noted in the introduction, previous studies have analyzed methods for realizing the dissipative dynamics in 
\cref{eq:masterEqIdeal1}. 
These methods either required direct driving of spins with squeezed light \cite{AgarwalPuri1990,Kuzmich1997} (which is experimentally challenging), or the use of Raman processes in structured four-level atoms \cite{dalla2013dissipative,borregaard2017one} (which is not applicable to generic two-level systems). 
We present here an alternate, generic method that takes a hybrid-systems approach:  a cavity (or other bosonic mode) is coupled both to an ensemble of two-level systems, as well as to an engineered, bosonic squeezed reservoir (see Fig.~\ref{fig:schematic}).  
As discussed, such a bosonic squeezed reservoir can be realized using only classical driving fields, and has been implemented in a variety of different experiments \cite{Wollman2015,Kienzler2015us,Dassonneville2021}.
We discuss specific implementation strategies of this general approach in \cref{sec:Implementation}; here, we present the general structure of the overall quantum master equation.

To this end, we consider a spin ensemble that is resonantly coupled to a bosonic mode (with lowering operator $\am$).  
In the rotating frame, the Hamiltonian is 
\begin{align}
    \hat{H} &= g ( \am \dg \Sm + \am \Sp ) \comma
    \label{eq:H}
\end{align}
where $g$ is the spin-cavity coupling strength.
We further assume that this mode is coupled both to an engineered squeezed reservoir (with coupling rate $\kappasqz$ and squeezing parameter $r$) as well as subject to unwanted zero-temperature loss (at rate $\kappaint$).  
The quantum master equation is then  
\begin{align}
    \dot\rhoop 
    &= -i \comm{\hat{H}}{\rhoop}  
    + \kappasqz \mD{\cosh(r) \am + \sinh(r) \am \dg}{\rhoop} \nonumber  \\
    &+ \kappaint \mD{\am}{\rhoop} 
    + \frac{\gammaphi}{2} \sum_{k} \mD{\sz^{(k)}}{\rhoop}       
    + \gammarel \sum_{k} \mD{\smm^{(k)}}{\rhoop} \fullstop 
    \label{eq:masterEqCavitySpins}
\end{align}
We have also included standard single-spin decay and dephasing dissipators (at rates $\gammarel$ and $\gammaphi$, respectively).

At a heuristic level, the cavity serves as a transducer that allows the spins to inherit the squeezed fluctuations produced by the bosonic squeezed reservoir.  
As the squeezed reservoir is engineered, we will treat $r$ and $\kappasqz$ as tuneable parameters that can be optimized. 
In contrast, we will take the coupling $g$ and the unwanted dissipation (\ie, $\kappaint$, $\gammaphi$, and $\gammarel$) to be fixed.  
This then motivates introducing single-spin cooperativities $\eta_k$ and collective cooperativities $\mathcal{C}_{k}$ via:
\begin{align}
    \mathcal{C}_{k} &\equiv  N \frac{4 g^{2}}{\kappaint \gamma_{k}}  \equiv N \eta_{k} \comma
    \label{eq:collCoop}
\end{align}
where $k \in \{\phi, {\rm rel}\}$. 
The goal will be to understand the optimal squeezing possible for a fixed value of $\mathcal{C}_k$.  
As we will show in \cref{sec:scaling}, in the case where single-spin relaxation dominates over dephasing, the optimized dissipative scheme achieves steady-state squeezing scaling as $\wineland \propto 1 / \sqrt{\mathcal{C}_{\rm rel}}$.  
This is significantly better than the optimized transient OAT squeezing in this regime, which only scales as $\wineland \propto 1 / \left( \mathcal{C}_{\rm rel} \right)^{1/3}$ \cite{lewis2018robust}.

To connect our setup to the simpler quantum master equation \eqref{eq:masterEqIdeal1}, we consider the regime where the condition $\sqrt{N} g \ll \kappaint + \kappasqz$ holds, and we adiabatically eliminate the cavity $\am$.  
We obtain (see \cref{sec:App:AdiabaticElimination})
\begin{align}
    \dot\rhoop 
        =&\ \Gamma \mD{\Sigmaop[r]}{\rhoop} 
        + \gammacoll \mD{\Sm}{\rhoop} \nonumber \\
        &+  \frac{\gammaphi}{2} \sum_k \mD{\sz^{(k)}}{\rhoop} 
        + \gammarel \sum_k \mD{\smm^{(k)}}{\rhoop} \comma 
    \label{eq:realisticMasterEq2}
\end{align}
where we have defined 
\begin{align}
    \Gamma&= \frac{4g^{2}}{\left(\kappasqz + \kappaint\right)^{2}}   \kappasqz \comma
    \label{eq:kappasqz}
\end{align}
and
\begin{align}
    \gammacoll&= \frac{4g^{2}}{\left(\kappasqz + \kappaint \right)^{2}}   \kappaint \fullstop
    \label{eq:gammacoll}
\end{align}
We see that the internal loss of the cavity results in a collective relaxation process for the spin ensemble; this is similar to OAT-based protocols that are derived using a strongly detuned cavity-spin ensemble system (in contrast to the resonant regime considered here).

%%%%%%%%%%%%%%%%%%%%%%%%%%%%%%%%%%%%%%%%%%%%%%%%%%
%%%%%%%%%%%%%%%%%%%%%%%%%%%%%%%%%%%%%%%%%%%%%%%%%%
%%%%%%%%%%%%%%%%%%%%%%%%%%%%%%%%%%%%%%%%%%%%%%%%%%

\section{The even-odd effect}
\label{sec:evenOdd}

\subsection{Basic effect}
\label{subsec:EvenOddBasics}

A striking feature of the purely dissipative dynamics described by \cref{eq:masterEqIdeal1} is an extreme sensitivity to the parity of the number $N$ of spins.  
As we will see, the steady state can be macroscopically different for $N$ spins vs.\ $N+1$ spins.
While early work noted that the form of the steady state depends on parity \cite{AgarwalPuri1989,AgarwalPuri1990}, subsequent studies on achievable squeezing focused on the even-$N$ case \cite{AgarwalPuri1994,dalla2013dissipative}. 
Our work reveals important new aspects of this parity effect.  
We show that by appropriate parameter tuning, one can avoid this effect, allowing steady-state squeezing that is near Heisenberg limited regardless of the parity of $N$.  
We also discuss a different regime where the even-odd effect could be used for a new sensing modality based on the macroscopic sensitivity to spin-number parity.  
Crucially, we show that there is no long timescale associated with the emergence of this sensitivity to the addition or removal of a single spin.  
Note that the even-odd effect in dissipative spin squeezing has no counterpart in bosonic dissipative squeezing.

%%%%%%%%%%%%%%%%%%%%%%%%%%%%%
\begin{figure}[t]
	\centering
    \includegraphics[width=0.45\textwidth]{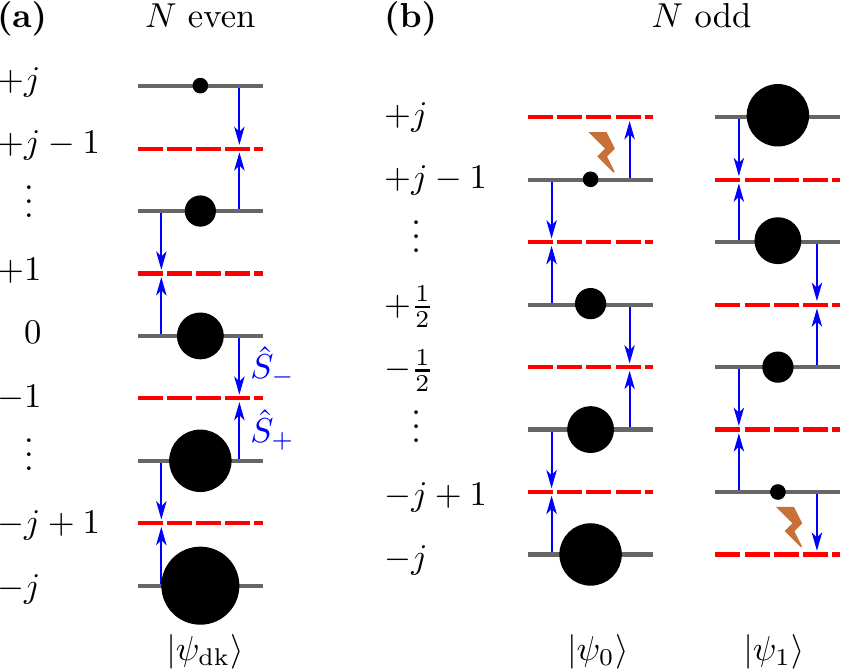}
	\caption{
	    Sketch of the steady state for \textbf{(a)} even $N$ and \textbf{(b)} odd $N$. 
		The size of the black circles represents the population of a level $\ket{j,m}$.
		For even $N$, a pure dark state exists for any squeezing parameter $r$ because the jump operator $\Sigmaop$ leads to destructive interference between adjacent levels (blue arrows) such that every second level is unoccupied (dashed red lines).
		For odd $N$ and large $r$, the interference condition cannot be satisfied for all levels (brown flashes) and the steady state is mixed. 
		The two pure-state contributions with largest statistical weight are sketched here.
	}
	\label{fig:EvenOdd1}
\end{figure}
%%%%%%%%%%%%%%%%%%%%%%%%%%%%%

We start with a simple intuitive picture that explains why the steady state of \cref{eq:masterEqIdeal1} is so sensitive to the parity of $N$.  
Recall that pure bosonic squeezed states are fully paired: they are superpositions of states having even photon numbers only \cite{GerryKnight}. 
A similar structure holds in our spin problem.  
We can think of the fully polarized state $|j,m= -j \rangle$ as being the ``vacuum'', and a state $|j,m=-j + q \rangle$ as having $q$ excitations (\ie, $q$ flipped spins). 
We thus see directly from \cref{eq:darkStateck} that, like bosonic squeezed states, the spin dark states $\psidk{j;r}$ also only involve even numbers of excitations $q$.

Formally, in both the bosonic and spin problem, this paired structure leads to destructive interference that makes the state dark. 
When $\Sigmaop[r]$ acts on a paired state, it creates a state having only {\it odd} number of excitations. 
For a given odd excitation number $q_{\rm odd}$, achieving a dark state requires destructive interference between the two pathways leading to $q_{\rm odd}$: 
$\hat{S}_{-}$ could have acted on the state with $(q_{\rm odd}+1)$ excitations, or $\hat{S}_{+}$ could have acted on the state with $(q_{\rm odd}-1)$ excitations. 
These destructive interference conditions can be directly used to derive the coefficients in \cref{eq:darkStateck} that determine $\psidk{j;r}$.
This structure is shown schematically in Fig.~\ref{fig:EvenOdd1}(a).

With this picture in mind, it is easy to see why we cannot have a pure dark state for odd $N$.  
In this case, the maximum number of excitations $q_{\rm max}$ is \emph{odd}. 
As such, the needed destructive interference is impossible to achieve.  
Starting with a fully paired state, we can create a state with $q_{\rm max}$ excitations by acting with $\hat{S}_+$ on $|j,-j+(q_{\rm max}-1)\rangle$.  
However, there is no complementary $\hat{S}_{-}$ process, as there is no state with $q_{\rm max}+1$ excitations. 
The best one can then do is to construct fully paired states that are only approximately dark due to this incomplete destructive interference [see Fig.~\ref{fig:EvenOdd1}(b)].

\begin{figure}[t]
	\centering
	\includegraphics[width=0.45\textwidth]{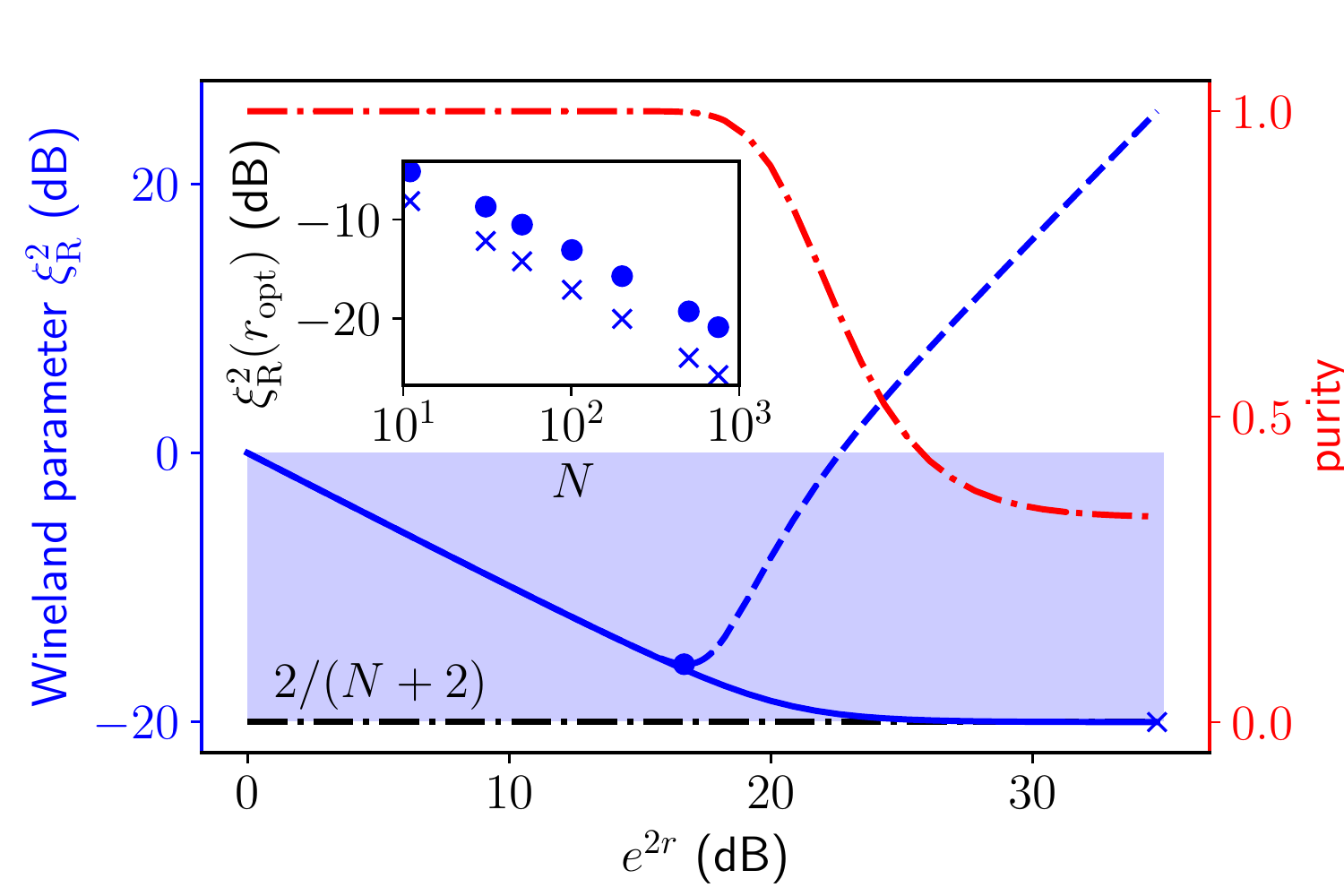}
	\caption{
	    Properties of the steady state of \cref{eq:masterEqIdeal1} for $N=\ValueevenOddWinelandPurityNeven$ vs.\ $N=\ValueevenOddWinelandPurityNodd$ spins. 
		For even $N$, the Wineland parameter (solid blue line) converges to a Heisenberg-limited scaling (dash-dotted black line) in the limit $r \to \infty$.
		For odd $N$, the Wineland parameter (dashed blue line) diverges if $e^{2r} \gg N$ and its purity (dash-dotted red line) approaches $1/3$. 
		\textbf{Inset}: Minimum Wineland parameter for even $N$ (crosses) and odd $N$ (dots) obtained at an optimal squeezing strength $r_\mathrm{opt}$. 
		The squeezed spin component is always $\Sy$.
	}
	\label{fig:EvenOdd2}
\end{figure}

The net result of this ``frustration" is dramatic:  for odd $N$ and large $r$, the dissipative steady state of \cref{eq:masterEqIdeal1} is impure and, moreover, exhibits no spin squeezing for large $r$.  
More specifically, for odd $N$, the steady-state squeezing diverges in the large-$r$ limit, while the purity tends asymptotically to $1/3$.  
This behaviour is shown explicitly in Fig.~\ref{fig:EvenOdd2}.  
One also sees that, for modest $r$, there is no appreciable even-odd effect:  the odd-$N$ steady state is almost pure and has the same squeezing as the even $N$ case. 
This also follows from our heuristic picture:  for small enough $r$, there is very little probability to have a large number of ``excitations'', and hence one is almost insensitive to the frustration resulting from the cut-off on maximum excitation number.

While our discussion has been focused on the ideal quantum master equation~\eqref{eq:masterEqIdeal1}, the even-odd effect persists even in the presence of single-spin relaxation and dephasing [as described by \cref{eq:realisticMasterEq2} in the limit $\kappaint \to 0$]. 
As discussed in \cref{sec:App:EvenOddEffect}, observing the even-odd effect in the steady state requires the single-spin cooperativities $\eta_{\rm rel}$ and $\eta_\phi$ defined in \cref{eq:collCoop} to be order unity or larger.

Finally, we note that the even-odd effect discussed here is distinct from the sensitivity to parity exhibited by unitary evolution under a OAT Hamiltonian $\hat{H}_{\rm OAT} = \chi \Sx^2$ \cite{MolmerSorensen1999,Leibfried2004,Leibfried2005}. 
The unitary evolution generated by $\hat{H}_\mathrm{OAT}$ for a time $\pi / 2 \chi$ maps the initially fully polarized state $\ket{N/2,-N/2}$ to Greenberger Horne Zeilinger (GHZ) states oriented along orthogonal axes in phase space, depending on the parity of $N$.  
This coherent effect results in a strong sensitivity to parity at a particular instant in time; in contrast, in our system, we have a dissipative effect where the sensitivity manifests itself in the \emph{steady-state} of the system. 
Moreover, in our case, the even vs.\ odd states are not equivalent up to a  rotation, but differ both in their purity and the magnitude of their fluctuations.

%%%%%%%%%%%%%%%%%%%
%%%%%%%%%%%%%%%%%%%%%

%%%%%%%%%%%%%%%%%%%%%%%%%%%%%%%%%%%%%%%%%%%%%%%%%%
%%%%%%%%%%%%%%%%%%%%%%%%%%%%%%%%%%%%%%%%%%%%%%%%%%

\subsection{Parity-independent Heisenberg-limited squeezing}
\label{subsec:EvenOddHL}

In most experimental situations, the even-odd effect will be a nuisance:  one aims for strong steady-state squeezing without needing to control $N$ at the single particle level.  
We therefore derive a quantitative estimate on the maximum squeezing parameter $r$ that can be used without any parity sensitivity.  
For small $r$, the system and its steady state are well described by a Holstein-Primakoff approximation \cite{HolsteinPrimakoff1940}; one recovers bosonic squeezing physics \cite{Kronwald2014}, which is independent of the parity of $N$.  
However, the correspondence between bosonic squeezing and spin squeezing will break down if the populations of the states $\ket{j,m \approx j}$ become nonzero. 
Using the steady-state occupation number of the Holstein-Primakoff bosons $\ave{\hat{b}^\dagger \hat{b}}_\mathrm{ss} = \sinh^2(r)$, one can estimate that this breakdown happens if the condition $\ave{\hat{b}^\dagger \hat{b}}_\mathrm{ss} \approx N/2$ holds.  
This yields the breakdown criterion
\begin{align}
	e^{2 r} \gtrsim N \comma
\end{align}
which provides an estimate
for the maximum squeezing parameter $r$ possible with no even-odd effect.

While working with $e^{2r} \ll N$ avoids parity effects, one might worry that this constraint precludes ever reaching Heisenberg-limited scaling of the steady-state spin squeezing. 
This is not the case. 
As shown in the inset of Fig.~\ref{fig:EvenOdd2}, the minimum Wineland parameter for odd $N$ (obtained at a squeezing parameter $r_\mathrm{opt}$, which depends on $N$) exhibits Heisenberg-like scaling, and the spin squeezing differs only by a constant prefactor $\approx 2.6$ from the maximum achievable spin squeezing of $\xi_\mathrm{R,HL}^2 = 2/(N+2)$ for even $N$, which is obtained in the limit $e^{2r_\mathrm{opt}} \gg N$.

%%%%%%%%%%%%%%%%%%%%%%%%%%%%%%%%%%%%%%%%%%%%%%%%%%
%%%%%%%%%%%%%%%%%%%%%%%%%%%%%%%%%%%%%%%%%%%%%%%%%%

\subsection{Connections to the LMG Model}
\label{subsec:EvenOddLMG}

Despite first appearances, the extreme even-odd effect of our system is more than a nuisance.  
At a fundamental level, the effect has a surprising connection to a seemingly unrelated closed-system many-body model, the LMG model \cite{LipkinMeshkovGlick1965}.  
To see this, recall that in a quantum trajectories formulation of the master equation in \cref{eq:masterEqIdeal1}, the evolution of a state vector in the absence of quantum jumps is governed by the non-Hermitian Hamiltonian $(-i/2) \hat{H}_\mathrm{LMG}$, where
\begin{align}
	\hat{H}_{\rm LMG} \equiv \Sigmaop^\dagger(r) \Sigmaop(r) = e^{-2 r} \hat{S}_x^2 + e^{2r} \hat{S}_y^2 + \hat{S}_z \fullstop
	\label{eq:SigmaDagSigma}
\end{align}
$\hat{H}_\mathrm{LMG}$ is precisely the Hamiltonian of the anisotropic antiferromagnetic LMG model \cite{LipkinMeshkovGlick1965}, a generalized transverse field Ising model with all-to-all Ising couplings.
For even $N$, we are thus dissipatively stabilizing the many-body ground states $\psidk{j;r}$ of the antiferromagnetic LMG model \cite{Unanyan2003,Vidal2004afmLMG} and converge to one of them depending on the total angular momentum $j$ of the initial condition. 
Note that the physics here has crucial differences from the more studied \emph{ferromagnetic} LMG model, which is also known to exhibit spin squeezing in its ground state \cite{Vidal2004fmLMG,ma2011quantum} 
(but has no simple connection to a dissipative protocol).

Focusing on the case where $N$ is odd and $r > 0$, $\hat{H}_{\rm LMG}$ is positive and the steady state of \cref{eq:masterEqIdeal1} in a given total-angular-momentum subspace $j$ can be written as (see \cref{sec:App:EvenOddEffect})
\begin{align}
	\hat{\rho}_\mathrm{ss}^{(j)} = \frac{1}{\sum_{k=0}^{2j} \frac{1}{\lambda_k}} \sum_{k=0}^{2j} \frac{1}{\lambda_k} \ket{\psi_k}\bra{\psi_k} \comma
	\label{eq:SteadyStateOddN}
\end{align}
where $\lambda_k$ and $\ket{\psi_k}$ are the ordered eigenvalues and eigenvectors of $\hat{H}_\mathrm{LMG}$. 
We can thus directly connect the properties of the odd-$N$ steady state to the spectrum of the LMG Hamiltonian.  
Consider first the limit $r \to 0$, where $\hat{H}_\mathrm{LMG} \to \hat{\vec{S}}^2 - \Sz^2 + \Sz$.  
Then, the Hamiltonian has a \emph{unique} ground state $\ket{\psi_0} \to \psidk{j;0} = \ket{j,-j}$.
Moreover, the ground-state energy is zero for any $N$ and the gap to the double-degenerate first excited states is finite, \ie, $\lim_{r \to 0} \lambda_0 = 0$ and $\lim_{r \to 0} \lambda_{1,2} = 2 j$. 
As a result, the steady state is approximately pure even when $N$ is odd, as $\ket{\psi_0}$ dominates the sum in \cref{eq:SteadyStateOddN}.

In the opposite limit $r \to \infty$, the LMG Hamiltonian is dominated by the $\Sy^2$ term, $\hat{H}_\mathrm{LMG} \approx e^{2 r} \Sy^2$, and its eigenvalues are the eigenstates $\ket{j,m}_y$ of $\Sy$ with energy $\lambda_m \approx m^2 e^{2 r}$. 
Now, there is no zero energy ground state for odd $N$ (because $m$ takes half-integer values), the ground state is double degenerate, and the steady state converges to an incoherent mixture of $\Sy$ eigenstates,
\begin{align}
	\lim_{r \to \infty} \hat{\rho}_\mathrm{ss}^{(j)} \propto \sum_{m=-j}^j \frac{1}{m^2} \ket{j,m}_{y\,y}\!\!\bra{j,m} \fullstop 
\end{align}
A direct computation shows that the purity converges to $\lim_{N \to \infty} \Tr[(\hat{\rho}_\mathrm{ss}^{(N/2)})^2] = 1/3$. 
In the limit $r \to \infty$, there is no mean spin polarization, but the variance of $\hat{S}_y$ remains finite, $\ave{\hat{S}_y^2} \geq 1/4$. 
As a result, the Wineland parameter will diverge as shown in Fig.~\ref{fig:EvenOdd2}.

The connection to the LMG model thus provides useful intuition into the odd-$N$ steady state. 
For even $N$, the dark state $\psidk{j;r}$ remains an exact zero mode of $\Sigmaop^\dagger(r) \Sigmaop(r)$ for any value of $r$ and interpolates smoothly between the limits $\psidk{j;0} = \ket{j,-j}$ and $\lim_{r\to\infty} \psidk{j;r} = \ket{j,0}_y$. 
In terms of the LMG model, this implies that for even $N$, the ground-state gap does not close as a function of $r$ \cite{Unanyan2018}. 
This feature of the antiferromagnetic LMG model has been discussed previously in the context of a \emph{closed system} quantum phase transition \cite{Unanyan2003,Vidal2004afmLMG,Unanyan2005}.

%%%%%%%%%%%%%%%%%%%%%%%%%%%%%%%%%%
\begin{figure*}[t]
	\centering
    \includegraphics[width=0.45\textwidth]{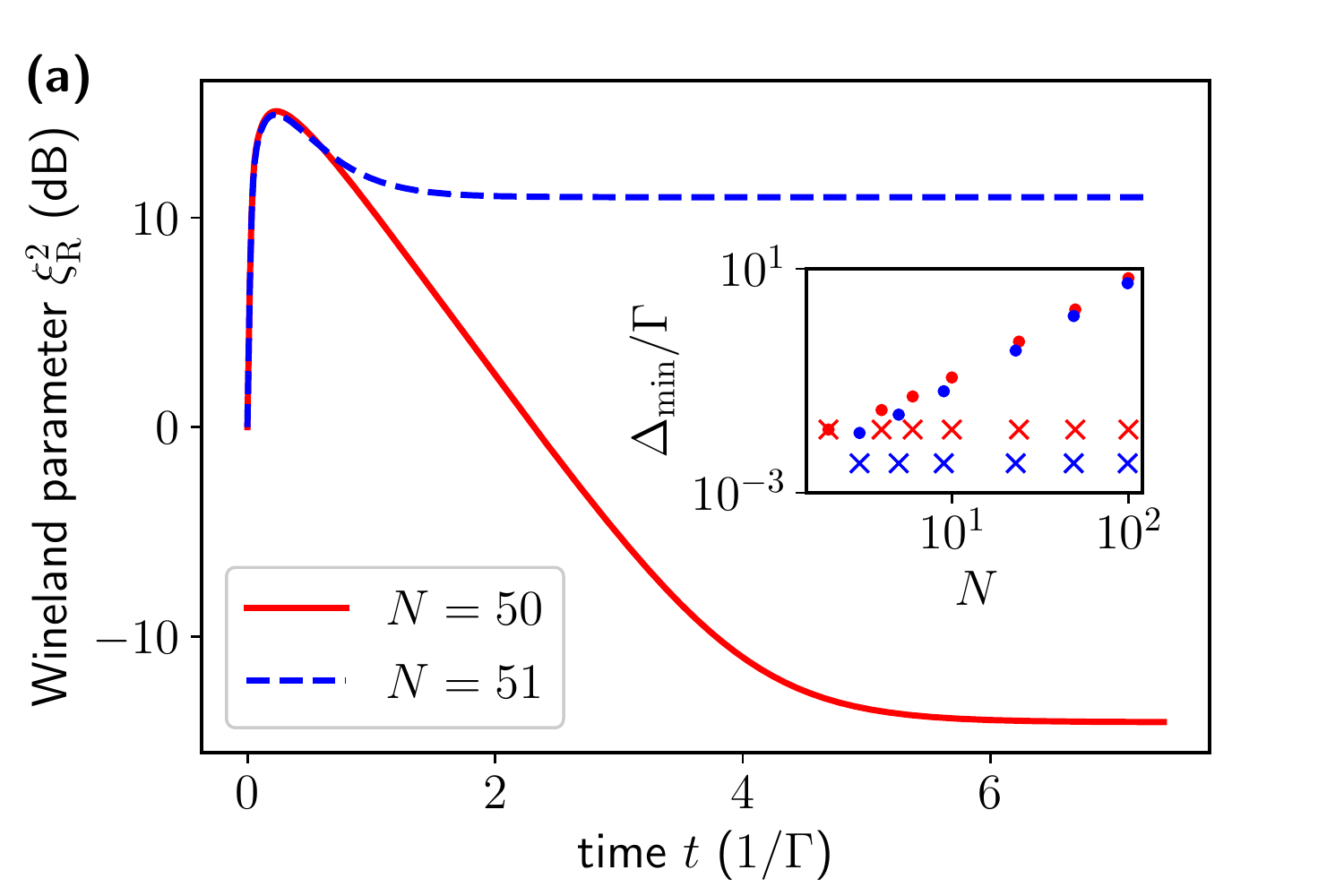}
	\includegraphics[width=0.45\textwidth]{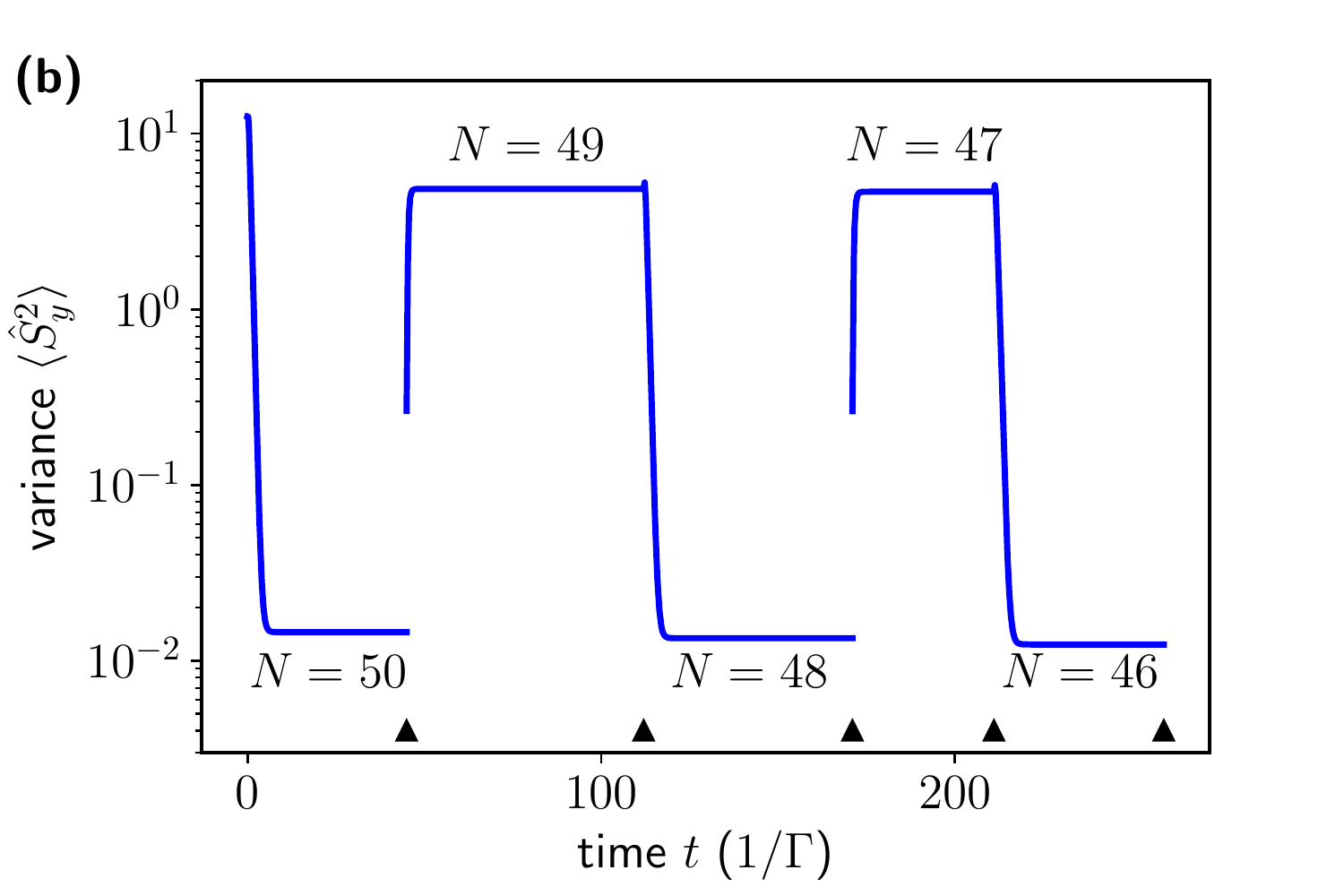}
	\caption{
	    \textbf{(a)} Time evolution of the coherent spin state $\ket{N/2,-N/2}$ under the quantum master equation~\eqref{eq:masterEqIdeal1} for even vs.\ odd $N$ with a squeezing parameter $r=\ValueevenOddTimeEvolutionr$.
		\textbf{Inset}: 
		Minimum spectral gap $\Delta_\mathrm{min}$ of the Liouvillian associated with \cref{eq:masterEqIdeal1} in the maximum-angular-momentum subspace (points) and global minimum evaluated over all angular momentum subspaces (crosses). 
		\textbf{(b)}
		Example of how the extreme even-odd sensitivity of the dissipative steady state could be used for sensing.  
		The variance $\ave{\Sy^2}$ is plotted for a system described by the ideal quantum master equation~\eqref{eq:masterEqIdeal1} with $r=\ValueTlsLossr$, starting from the state $\ket{N/2,-N/2}$.  The evolution is interrupted at randomly chosen times (black triangles), where a single (randomly chosen) spin is removed from the system.
		These spin-loss events cause the system to relax to a new steady state, leading to dramatic swings in the value of $\ave{\Sy^2}$ after each loss event.  
		Note the logarithmic scale used for the $y$ axis.  
    }
	\label{fig:EvenOdd3}
	\label{fig:EvenOdd4}
\end{figure*}
%%%%%%%%%%%%%%%%%%%%%%%%%%%%%%%%%%

%%%%%%%%%%%%%%%%%%%%%%%%%%%%%%%%%%
%%%%%%%%%%%%%%%%%%%%%%%%%%%%%%%%%%

\subsection{Enhanced sensing}
\label{subsec:EvenOddSensing}

The dramatic even-odd sensitivity of the steady state, which has no counterpart in bosonic spin squeezing, could enable
a new kind of sensing modality:  it provides a means for detecting changes in $N$ at the single-spin level.  
This kind of sensing has long been of interest for both fundamental studies and applications \cite{McKeever2004,Teper2006,Fortier2007,Brahms2011,Zhang2012,Chen2014}; a recent experiment has even used dispersive sensing to measure real-time changes in atom number in an atomic ensemble dispersively coupled to a cavity \cite{zeiher2021}. 
Our dissipative setup could provide an alternative route for an analogous kind of sensing.

As discussed above, for a large squeezing parameter $e^{2r} \gtrsim N$, the squeezing of the collective steady state depends exponentially on the parity of $N$ (see Fig.~\ref{fig:EvenOdd2}). 
A simpler quantity, the variance of $\hat{S}_y$, also exhibits this strong sensitivity in the large $r$ limit:  for even $N$, it vanishes like $N^2 e^{-4 r}/8$ whereas, for odd $N$, it converges to the constant value $N/\pi^2$ if $N \gg 1$.
We thus see that measuring $\hat{S}_y^2$ provides a direct means for estimating the parity of $N$.  
Such collective spin fluctuation measurements have been implemented in variety of systems \cite{Appel2009,Takano2009,SchleierSmith2010,Bohnet2016,Beguin2018,Guarrera2021,Luecke2014}. 

While the parity sensitivity is in principle a steady-state effect, the relatively fast relaxation timescale here means that it can be harnessed for real-time sensing.  
We stress that the strong sensitivity to parity does not come at the expense of a vanishingly small bandwidth:  if a spin is suddenly lost, the relaxation time to the new opposite-parity steady state is (at worst) set by the inverse coupling rate $1/\Gamma$.  
This timescale does not grow with system size [see inset of Fig.~\ref{fig:EvenOdd3}(a)].  
The relaxation is even faster if one is in the maximum-$j$ subspace; here, the relaxation rate is  collectively enhanced by a factor of $N$. 

We thus have a means for detecting spins leaving or decoupling from the cavity one by one, as each such event causes a large change in  $\hat{S}_y^2$ [see Fig.~\ref{fig:EvenOdd3}(b)].  
Note that the variance detection requires multiple repetitions of the measuement to distinguish even $N$ from odd $N$:
although the probability to obtain a measurement result with $\abs{m_y} > 1/2$ is negligible for even $N$ in the large-$r$ limit, it is only between $15\,\%$ and $19\,\%$ for odd $N$ [cf.\ Eq.~\eqref{eq:SteadyStateOddN} and App.~\ref{sec:App:EvenOddEffect}]. 
One thus has to wait for a probabilistic measurement outcome with sufficiently large $\abs{m_y}$ (the value depends on the detector resolution) to determine the spin-number parity unambiguously. 

Imperfections of the squeezing process, \eg, an impure engineered reservoir, and local dissipation will reduce the visibility of the even-odd effect (see Sec.~\ref{sec:Temperature} and App.~\ref{sec:App:EvenOddEffect}, respectively).
In Sec.~\ref{sec:implementation:evenodd}, we discuss that even-odd effect can be observed in a state-of-the-art trapped-ion platform for $N \lesssim 10$. 
This opens the exciting possiblity to experimentally verify the even-odd effect in spin squeezing, which has no counterpart in bosonic squeezing.

%%%%%%%%%%%%%%%%%%%%%%%%%%%%%%%%%%%%%%%
%%%%%%%%%%%%%%%%%%%%%%%%%%%%%%%%%%%%%%%
%%%%%%%%%%%%%%%%%%%%%%%%%%%%%%%%%%%%%%%

\section{Enhanced protection against single-spin relaxation}
\label{sec:scaling}

The dissipative approach to spin squeezing also provides strong advantages when unwanted single-spin dissipation is included.  
In this section, we focus on the case where local relaxation is dominant, \ie, we study \cref{eq:realisticMasterEq2} in the limit $\gammarel \neq 0, \gammaphi \rightarrow 0$.  
For atomic systems, this can be viewed as a fundamental limit arising from spontaneous emission, whereas single-spin dephasing is a technical imperfection.  
As noted in Ref.~\onlinecite{lewis2018robust}, in this limit, standard OAT achieves an optimized squeezing that yields the scaling $\wineland \sim  \Crel^{-1/3}$ for large $N$. 
This work also introduced an alternate Hamiltonian protocol involving two mutually-interacting spin ensembles, which could achieve a more favourable $\wineland \sim  \Crel^{-1/2}$ scaling at a specific optimized time.  
As we show below, our dissipative approach can achieve an identical scaling, but now for the \emph{steady state}, and only using a single ensemble of standard two-level systems.  
We also show that this enhanced performance over OAT holds even for small-$N$ ensembles.  Note that single-spin dissipation was also studied in Ref.~\onlinecite{dalla2013dissipative}, but only for spontaneous emission in an ensemble of four-level atoms with a specific structure.  This is distinct from the more generic model \cref{eq:realisticMasterEq2} we study.

Focusing on the limit of large $N$ and a small single-spin cooperativity, we can approximate our system well using a standard mean-field theory based on linearizing the equations of motion for the system's covariance matrix.  
Solving these in the steady state and considering the limit of a sufficiently large $r$ (see \cref{sec:App:coopScalingDerivation}), one finds that the steady-state squeezing is
\begin{align} 
    \xi_{R}^{2} & \approx \frac{N \gammacoll +\Gamma +\gammarel}{N  \gammacoll + \Gamma N +\gammarel} \fullstop
    \label{eq:WinelandMFT}
\end{align}
The numerator here describes unwanted heating by both  single-spin relaxation and the collective decay $\gammacoll$ associated with internal cavity loss. 
The only parameter left to optimize over is $\kappasqz$, the coupling between the cavity and the squeezed reservoir, which enters \cref{eq:WinelandMFT} via \cref{eq:kappasqz,eq:gammacoll}. 
There is a non-trivial minimum here.  
Suppression of unwanted collective heating requires a large $\kappasqz$, as this reduces the ratio $\gammacoll / \Gamma$.  
In contrast, suppressing the effects of $\gammarel$ requires a large $\Gamma$ and hence small $\kappasqz$.

Minimizing with respect to $\kappasqz$, we find
\begin{align}
    \xi_{R}^{2} &\approx 
    \frac{2}{\sqrt{\Crel}} + \bigoh{\frac{1}{\Crel}} \comma
    \label{eq:scaling1}
\end{align}
where the optimal value of $\kappasqz$ satisfies 
\begin{align}
    \kappasqz^{\text{opt}} =&   
    \kappaint \sqrt{\Crel}
    + \bigoh{\Crel^0} \fullstop
    \label{eq:optkappasqz}
\end{align}
We thus obtain an optimized squeezing that scales significantly better with collective cooperativity in this relaxation-dominated regime than the OAT result of $\wineland \sim  \Crel^{-1/3}$.  
In \cref{sec:App:coopScalingSimsMFT} we show numerical simulations of a more accurate non-linear mean-field theory that confirm these results.  
As we have stressed, the squeezing here is also achieved in the steady state (and not just at one optimal time).  
While we assumed a large value of $r$ to derive these results, in practice one only needs $\exp(-2r) \ll  1/\sqrt{\Crel}$ for this scaling to hold.

\begin{figure}[t]
	\centering
    \includegraphics[width=0.45\textwidth]{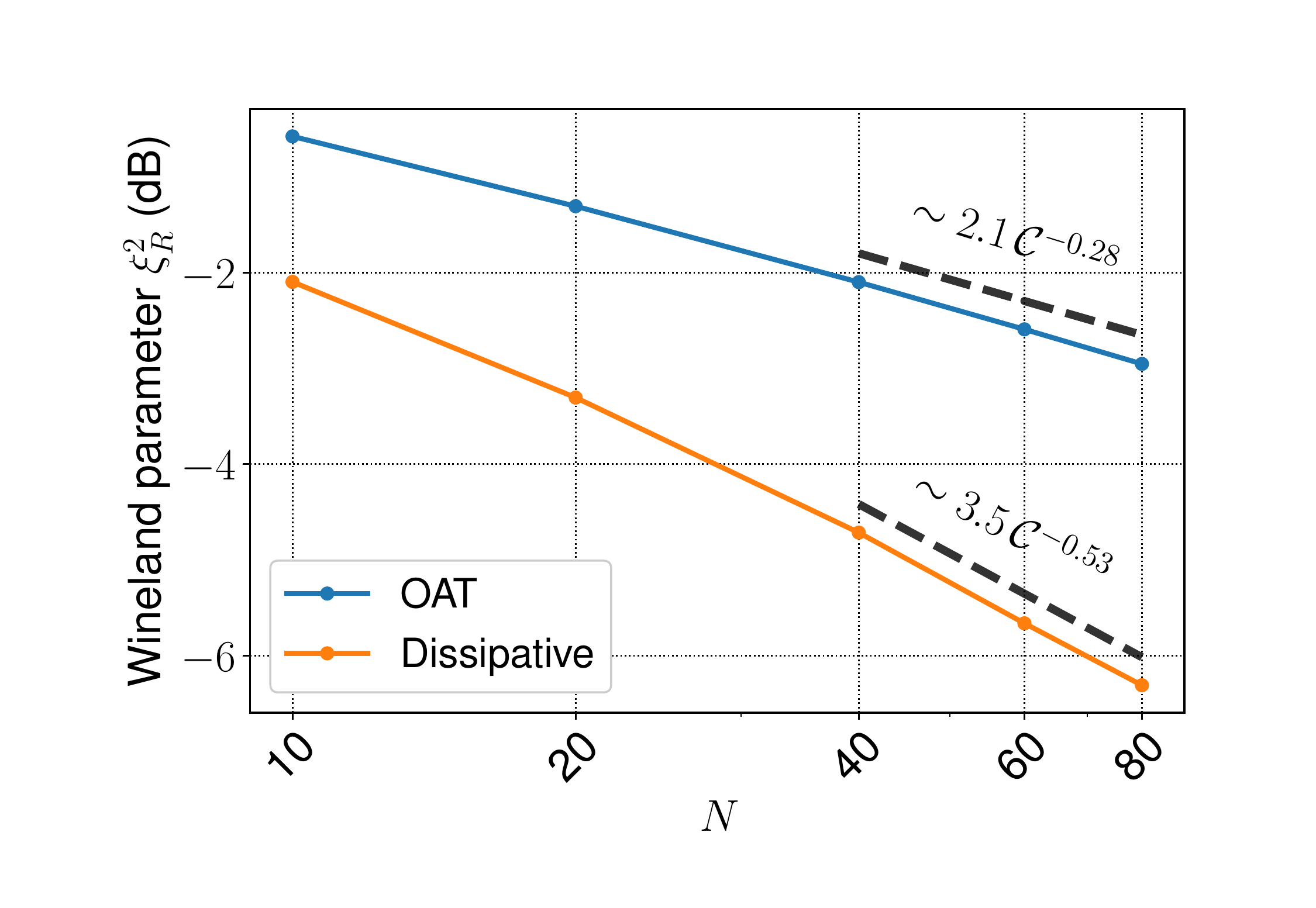}
    \caption{
        A comparison of the Wineland parameter $\wineland$ between the dissipative (blue) and OAT (orange) protocols as a function of $N$ for a small number of spins. 
        The simulations have been performed by evolving the spin-only quantum master equation [see \cref{eq:realisticMasterEq2} for the case of dissipative protocol, and \cref{sec:App:oat} for the details on OAT]. 
        In the dissipative protocol, at each value of $N$, both $r$ as well as $\kappasqz$ have been optimized, while in the case of OAT, the optimization has been performed over the cavity-spin detuning. 
        The parameters used in both cases are $\gammarel=0.02g$ and $\kappaint=100g$, resulting in single spin cooperativity of $\etarel=2$. 
        The two dashed lines show fits of the last few data points. 
    }
	\label{fig:meOATvsDissip}
\end{figure}

The advantage over OAT in this relaxation-dominated regime also persists for smaller-sized spin ensembles. 
To study this regime, we numerically solve \cref{eq:realisticMasterEq2} for the steady state.  
Figure~\ref{fig:meOATvsDissip} shows the obtained results for the steady-state squeezing (orange curve) as a function of $N$, where we have fixed $g$, $\kappaint$, and $\gammarel$ so that the single-spin cooperativity is  $\etarel = 2$. 
For each value of $N$, we optimize the parameters of the squeezed reservoir ($\kappasqz,r$) to minimize the steady state $\wineland$; the optimized values are presented in \cref{sec:App:optParamsME}. 
For comparison, we also plot the optimized \emph{transient} squeezing achievable using OAT (blue curve) in an identical cavity-spin system \cite{lewis2018robust,bennett2013phonon} (see \cref{sec:App:oat} for details). 
For the OAT setup, there is no squeezed reservoir, $\kappasqz = 0$, and there is a large detuning $\Delta$ between the spins and cavity, which is optimized for each value of $N$.

Figure~\ref{fig:meOATvsDissip} shows that, even for small $N$, the dissipative protocol yields an advantage over OAT.
While for these small values of $N$ and large $\etarel$, the linearized mean-field theory scaling predictions are not expected to hold exactly, there is a qualitative agreement with the predicted power laws (as indicated by black dashed lines).

In \cref{sec:App:directSqueezedDriving}, we provide a brief performance analysis of a special case where $\kappaint=0$. 
Mathematically, such a scenario is equivalent to a setup where one directly shines squeezed light onto the spin ensemble. 
We show that in the limit of large spin number, one can achieve the scaling of $\wineland \propto (N\Gamma/\gammarel)^{-1}$, although naturally, having either $\kappaint=0$ or irradiating a spin ensemble directly, would likely be difficult to realize experimentally.

%%%%%%%%%%%%%%%%%%%%%%%%
%%%%%%%%%%%%%%%%%%%%%%%%

%%%%%%%%%%%%%%%%%%%%%%%%%%%%%%%%%%%%%%%%%%
%%%%%%%%%%%%%%%%%%%%%%%%%%%%%%%%%%%%%%%%%%
%%%%%%%%%%%%%%%%%%%%%%%%%%%%%%%%%%%%%%%%%%

\section{Dephasing-dominated regime}
\label{sec:dephasing}
\subsection{Pre-thermalization and emergent slow timescales}
\label{sec:slowTimescale}

%%%%%%%%%%%%%%%%%%%%%%%%
\begin{figure*}[t]
	\centering
    \includegraphics[width=0.45\textwidth]{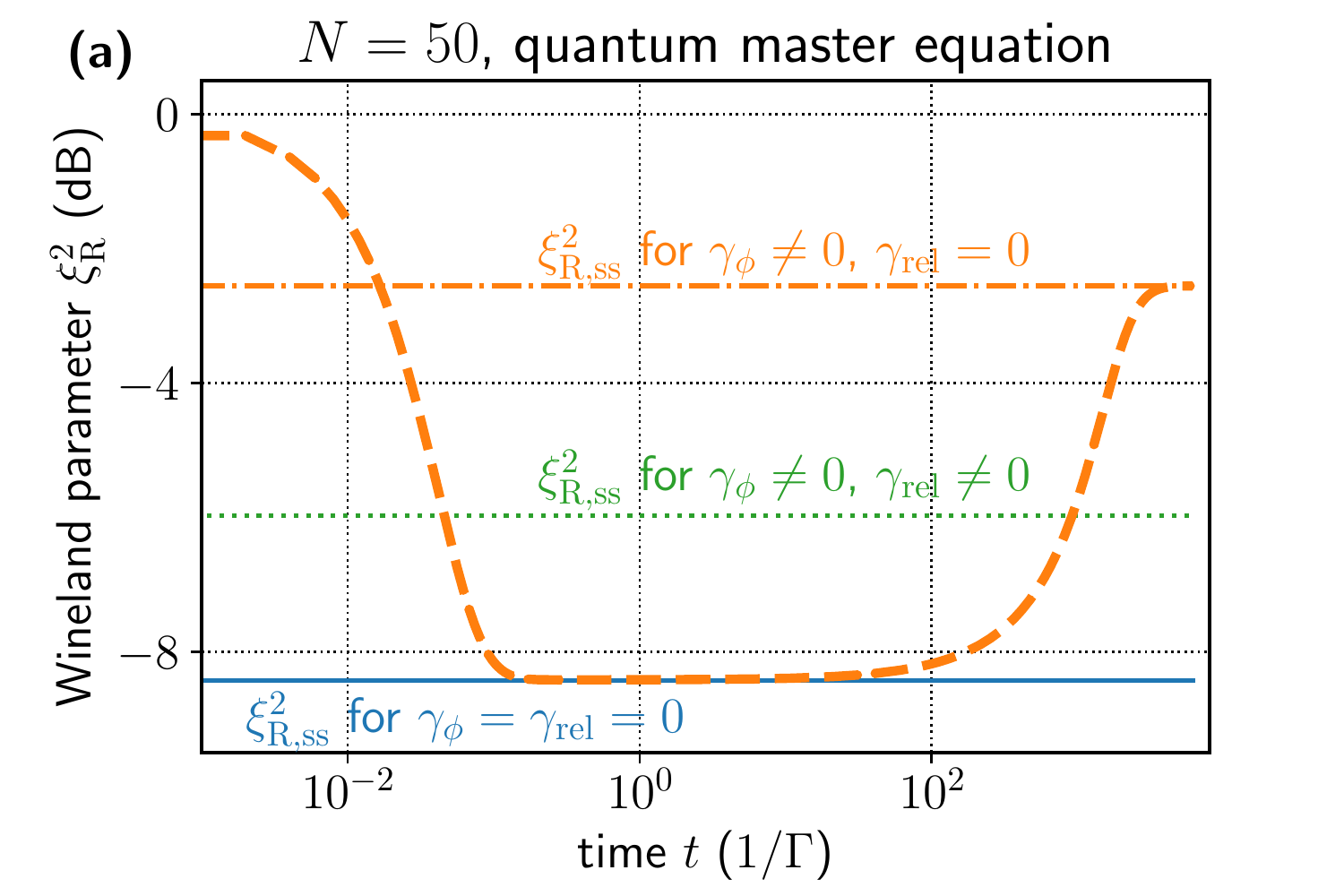}
    \includegraphics[width=0.45\textwidth]{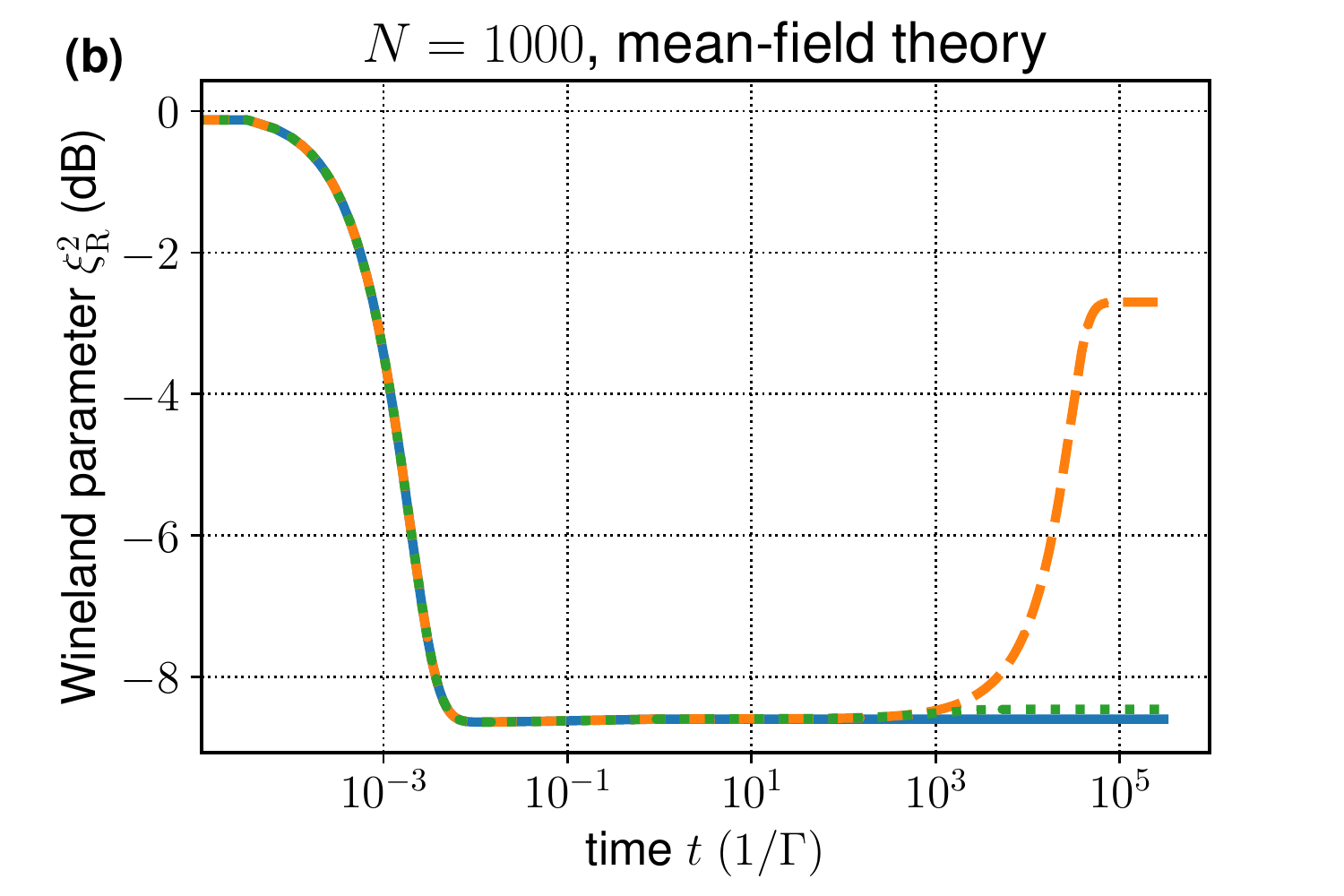}
    \caption{Time evolution of the Wineland spin-squeezing parameter $\xi_\mathrm{R}^2$ in the presence of local dissipation (note the logarithmic scale of the time axis).
		\textbf{(a)} Wineland parameter calculated using the quantum master equation~\eqref{eq:realisticMasterEq2} for weak local dephasing, $r=\ValueslowRateTimeEvolutionr$, and $N=\ValueslowRateTimeEvolutionN$ spins (thick dashed orange line, $\gammaphi/\Gamma = \ValueslowRateTimeEvolutiongammaphiOverGamma$, $\gammarel/\Gamma = \ValueslowRateTimeEvolutionkappaOverGamma$, and $\gammacoll/\Gamma = 0$).
		The final amount of steady-state spin squeezing is indicated by the thin dash-dotted orange line.
		Local dephasing deteriorates the amount of steady-state spin squeezing compared to an ideal system without local dissipation (solid blue line, $\gammaphi/\Gamma = \gammarel/\Gamma = \gammacoll/\Gamma = 0$).
		Local relaxation counteracts this effect and partially restores the steady-state spin squeezing (dotted green line, $\gammaphi/\Gamma = \ValueslowRateTimeEvolutiongammaphiOverGamma$, $\gammarel/\Gamma = \ValueslowRateTimeEvolutiongammarelOverGammab$, and $\gammacoll/\Gamma=0$). Note that the \emph{transient} state is strongly spin squeezed even in the presence of local dissipation since the collective dissipator $\Sigmaop$ induces spin squeezing on a short timescale $\propto 1/N \Gamma$ whereas the system approaches its steady state on a longer system-size-dependent timescale $\propto N/\gammaphi$.
		\textbf{(b)} Wineland parameter calculated using the mean-field equations of motion detailed in \cref{sec:App:MeanFieldTheory} for $N=1000$ spins and the same sets of dissipation rates as in (a).
	}
	\label{fig:slowTimeScale1}
\end{figure*}
%%%%%%%%%%%%%%%%%%%%%%%%

We now consider the effects of weak single-spin dephasing [\ie, the $\gammaphi$ term in \cref{eq:realisticMasterEq2}] on our dissipative spin squeezing protocol. 
For very weak dephasing, such that $\gammaphi < \gammarel / N$ holds, the mean-field theory results of the previous section still provide a good description; one simply substitutes $\gammarel \rightarrow \gammarel + 2 \gammaphi$ in \cref{eq:scaling1,eq:optkappasqz}.  
The more interesting case is when dephasing is the dominant form of single-spin dissipation, but is still weak compared to the rate $\Gamma$ associated with the collective spin squeezing dissipator (\ie, $\Gamma \gg \gammaphi \gg \gammarel$).  
In this case, the dynamics is surprisingly rich, exhibiting features reminiscent of prethermalization behavior observed in weakly nonintegrable systems \cite{DAlessio2016,Langen2016}.  
Prethermalization is associated with approximately conserved quantities that can only be dynamically randomized on extremely long timescales; this results in an intermediate-time quasi-steady state whose form is contingent on the initial value of the conserved quantities. 
Here, a similar phenomenon arises, with total angular momentum playing the role of the approximately conserved quantity.  
We discuss this more in what follows.

Starting from an initial product state, we find that a seemingly tiny amount of single-spin dephasing is enough to completely destroy spin squeezing in the eventual steady state.  
Using a mean-field analysis, one can show that in the presence of arbitrarily weak but non-zero single-spin dephasing (and $\gammarel=\gammacoll=0$), the steady-state squeezing is bounded by $-3\,\mathrm{dB}$ in the large $N$ limit:  
\begin{align}
	\lim_{\gammaphi \to 0} \xi_\mathrm{R}^2 \geq \frac{1}{2} + \frac{\sqrt{N}}{N + 1} \comma
\end{align}
where the optimal value is achieved with $r = \frac{1}{8} \ln N$.

Despite this, there exists an extremely long-lived intermediate-time regime (a quasi-steady state) where strong spin squeezing is observed.  The system's dissipative dynamics is thus characterized by two vastly different timescales,    
as shown in Fig.~\ref{fig:slowTimeScale1}.
The system first evolves into a transient spin-squeezed state on a fast timescale $\propto 1/N \Gamma$.  
In contrast, the eventual relaxation to the true steady state (which has minimal squeezing) occurs on a much slower timescale $\propto N/\gammaphi$.
For a large system size $N$, the ratio of these timescales can be dramatic.  
We also note that the slow relaxation time is parametrically slower than the single-spin dephasing time $1/\gammaphi$.

The emergence of this surprisingly long timescale, and the corresponding fragility of the steady state to weak dephasing, are both surprising; we stress that single-spin relaxation (as discussed in the previous section) does not give rise to an analogous behavior.  
In \cref{sec:App:LiouvillianPerturbationTheorySlowTimescale}, we analyze this effect using Liouvillian perturbation theory \cite{Li2014} and develop an intuitive physical picture:
Single-spin dephasing enables transitions between subspaces of different total angular momentum \cite{ChaseGeremia2008} such that an initial state in the $j=j_\mathrm{max}$ subspace evolves into a steady state populating subspaces with $j < j_\mathrm{max}$.
The degeneracy of the $j < j_\mathrm{max}$ subspaces gives rise to anomalously small matrix elements between the subspaces, which represent bottlenecks for the relaxation to the steady state.

We stress that the surprising impact of dephasing need not be problematic for experiments.  The spin squeezing exhibited by the Wineland parameter $\xi_\mathrm{R}^2$ in the ``prethermalized'' intermediate-time regime is comparable to $\xi_\mathrm{R}^2$ of the steady state obtained in an \emph{ideal} system without single-spin dissipation, as long as the conditions $\gammaphi \ll \Gamma, N \gammaphi$ are satisfied. 
Moreover, there is a simple but effective way to improve the spin squeezing of the steady state by deliberately adding a competing single-spin relaxation process $\gammarel$. 
If this relaxation rate satisfies the condition $\gammarel \gtrsim \gammaphi/N$, population will be pushed back to the large-angular-momentum subspaces, which decreases the steady-state Wineland parameter significantly and increases spin squeezing beyond the $-3 \,\mathrm{dB}$ limit, as shown in Fig.~\ref{fig:slowTimeScale1}.

%%%%%%%%%%%%%%%%%%%%%%%%%%%%%%%%%%%%%%%%%%
%%%%%%%%%%%%%%%%%%%%%%%%%%%%%%%%%%%%%%%%%%
%%%%%%%%%%%%%%%%%%%%%%%%%%%%%%%%%%%%%%%%%%

\subsection{Inhomogeneous broadening and dynamical decoupling}
\label{sec:App:DynDec}

In addition to the Markovian mechanism described in the previous section, in some platforms dephasing due to inhomogeneous broadening of the spin ensemble can also play a role.  
A major advantage of spin squeezing generated by OAT dynamics is that it is compatible with dynamical decoupling and thus allows for an effective cancellation of the impact of inhomogeneous broadening by a simple sequence of $\pi$ pulses about the $x$ axis \cite{bennett2013phonon}. 
At first glance, this does not seem to be the case for our dissipative scheme. 
However, we will show here that our dissipative scheme is in fact compatible with a slightly modified dynamical decoupling sequence.

\begin{figure}
	\includegraphics[width=0.48\textwidth]{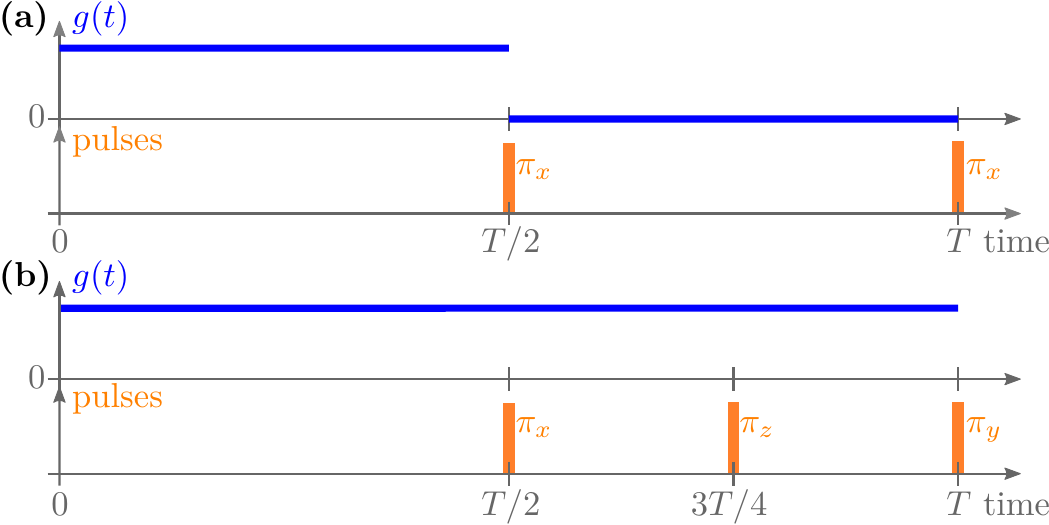}
	\caption{
		Dynamical decouping sequences to cancel inhomogeneous broadening in Eq.~\eqref{eqn:app:dd:H0}.
		Each sequence has a total duration time $T$, where $\kappasqz T \ll 1$ and $\kappaint T \ll 1$, and is repeated multiple times.
		\textbf{(a)} 
		    A simple sequence of $\pi$ pulses about the $x$ axis will generate unwanted $\Sigmaop^\dagger(r)$ antidamping terms [cf.\ Eq.~\eqref{eqn:app:dd:H1}].
		    Therefore, the coupling to the spins, $g(t)$, must be switched off after every second $\pi_x$ pulse. 
		\textbf{(b)}
		    By adding two additional $\pi$ pulses about the $y$ and $z$ axes, the detrimental impact of the first $\pi_x$ pulse can be cancelled and the coupling $g(t)$ can be kept constant.
	}
	\label{fig:dyndec}
\end{figure}

Our starting point is a generalization of the Hamiltonian of Eq.~\eqref{eq:masterEqCavitySpins},
\begin{align}
	\hat{H} = \sum_{j=1}^N \Delta_j \frac{\hat{\sigma}_z^{(j)}}{2} + g \left( \hat{a}^\dagger \hat{S}_- + \hat{a} \hat{S}_+ \right) \comma
	\label{eqn:app:dd:H0}
\end{align}
where $\Delta_j = \omega_j - \omega_0$ denotes the detuning of spin $j$ from the resonance frequency of the bosonic mode $\hat{a}$ due to inhomogeneous broadening.
Using average Hamiltonian theory \cite{Haeberlen1968}, we will now derive an effective Hamiltonian for the two different decoupling sequences shown in Fig.~\ref{fig:dyndec}, which are designed such that the effects of the $\Delta_j$ terms cancel out on average. 
Instead of transforming the state of the system at each decoupling pulse, it is more convenient to consider a Heisenberg picture where the Hamiltonian changes at each pulse, the so-called toggling frame \cite{Haeberlen1968}.

A single $\pi$ pulse about the $x$ axis will flip the sign of $\hat{\sigma}_z^{(j)}$ terms and will exchange the $\hat{S}_+$ and $\hat{S}_-$ operators in the interaction term of Eq.~\eqref{eqn:app:dd:H0}, 
\begin{align}
	\hat{H}_1 = - \sum_{j=1}^N \Delta_j \frac{\hat{\sigma}_z^{(j)}}{2} + g \left( \hat{a}^\dagger \hat{S}_+ + \hat{a} \hat{S}_- \right) \fullstop
	\label{eqn:app:dd:H1}
\end{align}
The first effect is desired and will cancel inhomogeneous broadening, but the second effect is unwanted because it will turn damping of the Bogoliubov mode $\hat{\Sigma}(r)$ into anti-damping. 
If one can control the coupling constant $g$ as a function of time, one can switch off the undesired interaction after every second $\pi$ pulse, as shown in Fig.~\ref{fig:dyndec}(a), and obtains the average Hamiltonian 
\begin{align}
	\overline{\hat{H}} 
	= \frac{\hat{H}}{2} + \frac{\hat{H}_1 \vert_{g=0}}{2} 
	= \frac{g}{2} \left( \hat{a}^\dagger \hat{S}_- + \hat{a} \hat{S}_+ \right) \comma
\end{align}
where the inhomogeneous broadening has been canceled at the cost of a reduction of $g$ by a factor of $2$. 
Experimentally, the coupling $g$ could be switched off by detuning the spins rapidly from the cavity.

Instead of switching off the interaction between the spins and the bosonic mode for half of the period $T$, one could also use the dynamical decoupling sequence shown in Fig.~\ref{fig:dyndec}(b). 
By applying a $\pi$ rotation about the $z$ axis, one can flip the sign of the second term without disturbing the first one, 
\begin{align}
	\hat{H}_2 = - \sum_{j=1}^N \Delta_j \frac{\hat{\sigma}_z^{(j)}}{2} - g \left( \hat{a}^\dagger \hat{S}_+ + \hat{a} \hat{S}_- \right) \fullstop
	\label{eqn:app:dd:H2}
\end{align}
The sequence is terminated by a $\pi$ rotation about the $y$ axis which reverts all signs and restores the original Hamiltonian~\eqref{eqn:app:dd:H0}.
If the waiting times between the pulses have a ratio of $2:1:1$, both inhomogeneous broadening and the unwanted interaction terms will be canceled in the average Hamiltonian,
\begin{align}
	\overline{\hat{H}} 
	= \frac{\hat{H}}{2} + \frac{\hat{H}_1}{4} + \frac{\hat{H}_2}{4}
	= \frac{g}{2} \left( \hat{a}^\dagger \hat{S}_- + \hat{a} \hat{S}_+ \right) \fullstop
\end{align}
Generating the additional $\pi$ pulse about the $z$ axis may seem challenging because it requires a controlled detuning of the spins from the cavity such that the accumulated phase is exactly $\pi$.
Experimentally, this would likely be even more difficult than turning the coupling off for half a period, and one may conclude that this scheme is harder to implement than the first one. 
However, it is well-known from NMR that pulses about the $z$ axis can also be realized using a so-called composite pulse which is a suitable combination of $x$ and $y$ rotations \cite{Freeman1981}. 
Specifically, a $\pi$ pulse about the $z$ axis can be decomposed into pulses about the $x$ and $y$ axes as follows:
\begin{align}
	e^{-i \pi \hat{\sigma}_z/2} 
	&= e^{-i \pi \sigma_x/2} e^{-i \pi \sigma_y/2} \nonumber \\
	&= e^{+i (\pi/2) \sigma_x/2} e^{i \pi \sigma_y/2} e^{-i (\pi/2) \sigma_x/2} \comma
\end{align}
which does not require any detuning between the spins and the cavity.
Higher order contributions to the average Hamiltonian $\overline{\hat{H}}$ will become negligible if the period $T$ of the decoupling sequence satisfies the conditions $\kappa_\mathrm{sqz} T \ll 1$ and $\kappa_\mathrm{int} T \ll 1$.

%%%%%%%%%%%%%%%%%%%%%%%%
%%%%%%%%%%%%%%%%%%%%%%%%

\subsection{Non-uniform single-spin couplings}
Another experimentally very relevant source of imperfections are inhomogeneities in the coupling strength $g$ between the spins and the common bosonic mode, \ie, one has to replace the Hamiltonian in Eq.~\eqref{eq:masterEqCavitySpins} with
\begin{align}
	\hat{H} = \sum_{j=1}^N g_j \left( \hat{a}^\dagger \hat{\sigma}_-^{(j)} + \hat{a} \hat{\sigma}_+^{(j)} \right) \fullstop
\end{align} 
This breaks the permutational symmetry of the spins reflected in the collective spin operators $\hat{S}_\pm$. 
The standard strategy to analyze this effect is based on an expansion of the mean-field equations of motion around the average coupling $\bar{g} = \sum_{j=1}^N g_j/N$ \cite{Agarwal1970}.
Variants of this approach have been applied to study superradiance \cite{Agarwal1970,Julsgaard2012}, microwave quantum memories \cite{Julsgaard2013}, and spin squeezing \cite{SchleierSmith2010squeezing,Leroux2010,Rudner2011,Leroux2012,Bennett2013}.
Non-uniform couplings then lead to an approximate collective model with a renormalized coupling parameter and a reduced effective length of the collective spin vector.
Similar results will hold in our case if the squeezing parameter is not too large, $e^{2r} \lesssim N$. 

To analyze the case of strong squeezing, $e^{2r} \gtrsim N$, more powerful theoretical methods need to be developed, which is an interesting and relevant subject for further study.
We expect non-uniform single-spin couplings to reduce the visibility of the even-odd effect and to influence the prethermalization physics, since the suppression of transition rates crucially relies on the indistinguishability of the spins in the ensemble.

%%%%%%%%%%%%%%%%%%%%%%%%%%%%%%%%%%%%%%%%%%
%%%%%%%%%%%%%%%%%%%%%%%%%%%%%%%%%%%%%%%%%%
%%%%%%%%%%%%%%%%%%%%%%%%%%%%%%%%%%%%%%%%%%

\section{Impure engineered reservoir}
\label{sec:Temperature}

In this section, we explore a different kind of imperfection that has not been studied in previous discussions of dissipative spin squeezing: the engineered reservoir may have an imperfect purity (or equivalently, mimic thermal squeezed light rather than vacuum squeezed light).  
At the most basic level, this corresponds to modifying our ideal quantum master equation~\eqref{eq:masterEqIdeal1} by
\begin{align}
	\dot{\rhoop} = \Gamma (n_\mathrm{th} + 1) \mD{\Sigmaop}{\rhoop} + \Gamma n_\mathrm{th} \mD{\Sigmaop^\dagger}{\rhoop} \comma
	\label{eq:masterEqFiniteTempForSigma}
\end{align}
where $n_\mathrm{th} \geq 0 $ parameterizes the effective temperature of the squeezed reservoir.
We will discuss below in Sec.~\ref{sec:Implementation} and in App.~\ref{sec:Appendix:FiniteTemperatureModel} how this generic model can be related to more microscopic mechanisms (including collective decay, $\gammacoll > 0$). 
Even though the hybrid-systems reservoir-engineering approach we focus on is not limited by losses associated with the transport and injection of an externally prepared optical squeezed state, many reservoir engineering techniques will inevitably result in a $n_\mathrm{th} \neq 0$, hence it is important to understand the impact of this unwanted heating.  
Note that for $n_\mathrm{th} \neq 0$, the steady-state of the spin ensemble will necessarily be impure.  
This can have a deleterious impact on Ramsey-type sensing experiments if one is interested in signal phases that are not infinitesimally small (as has been discussed in the context of OAT \cite{Braverman2018,Braverman2019}).

%%%%%%%%%%%%%
\begin{figure*}
	\includegraphics[width=0.3\textwidth]{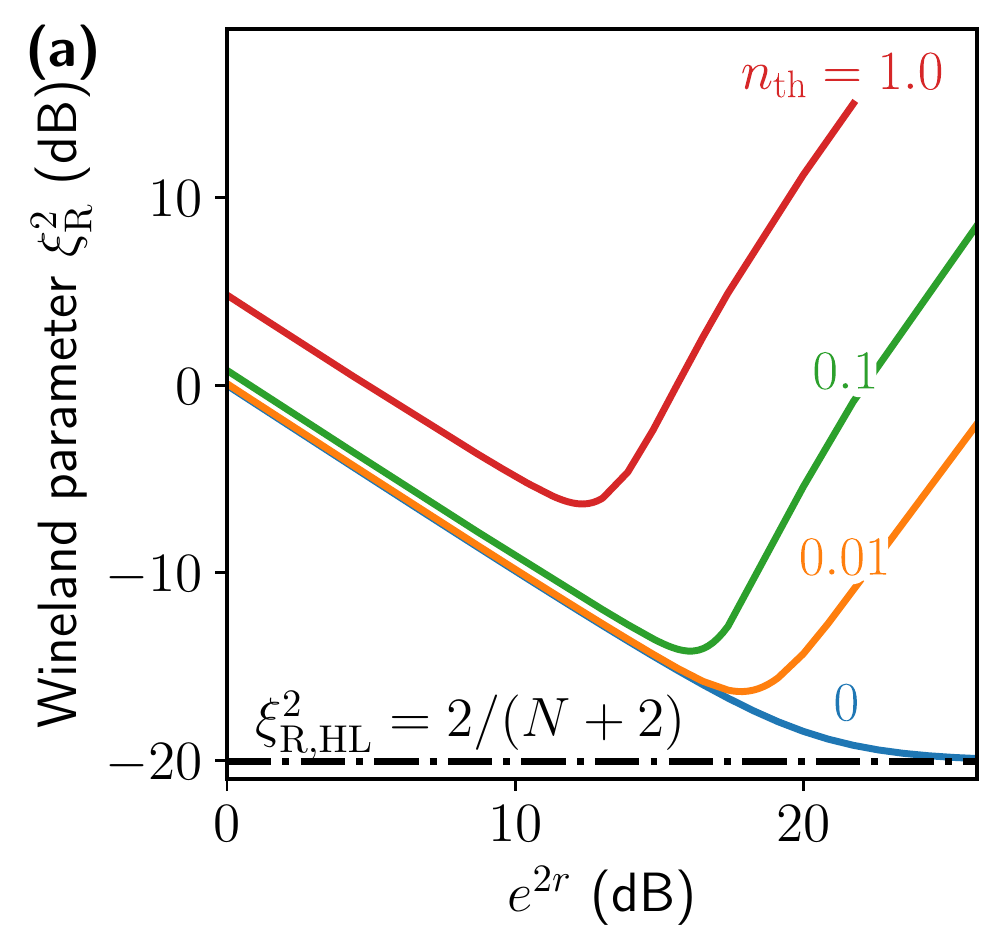}
	\includegraphics[width=0.3\textwidth]{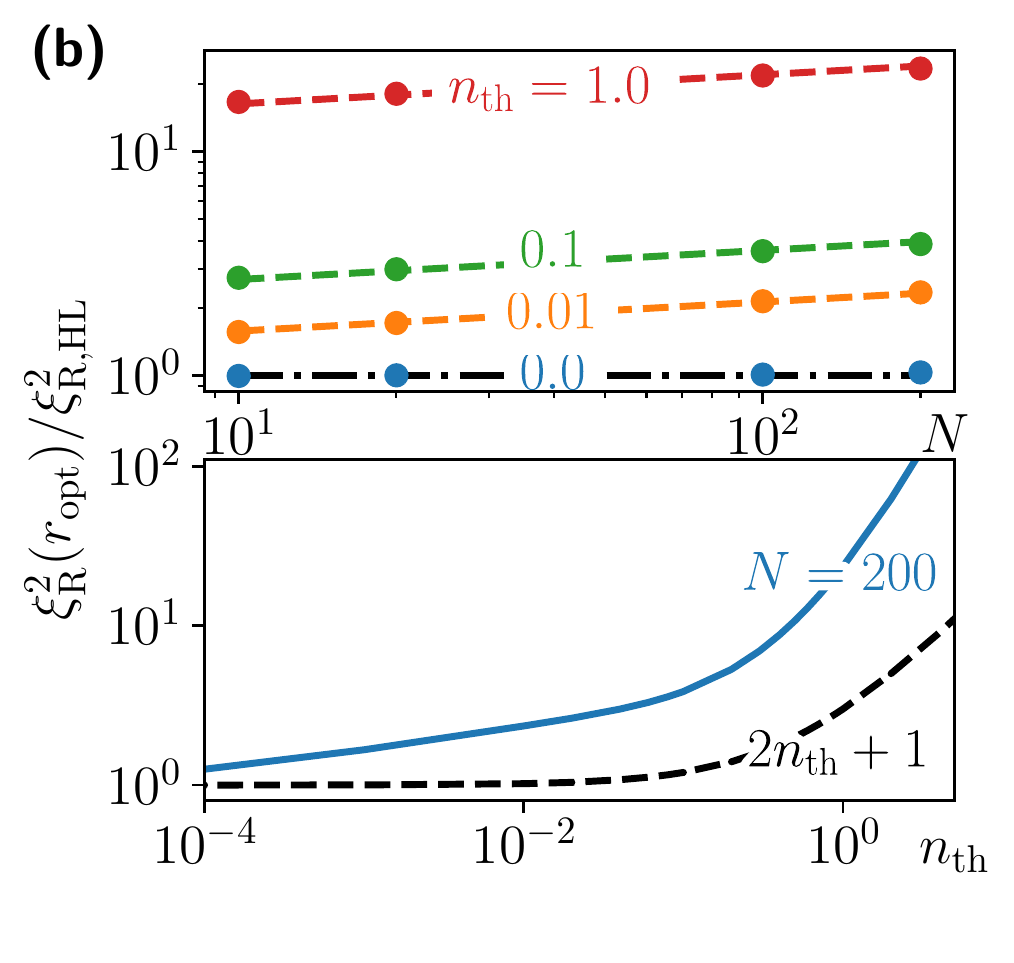}
	\includegraphics[width=0.3\textwidth]{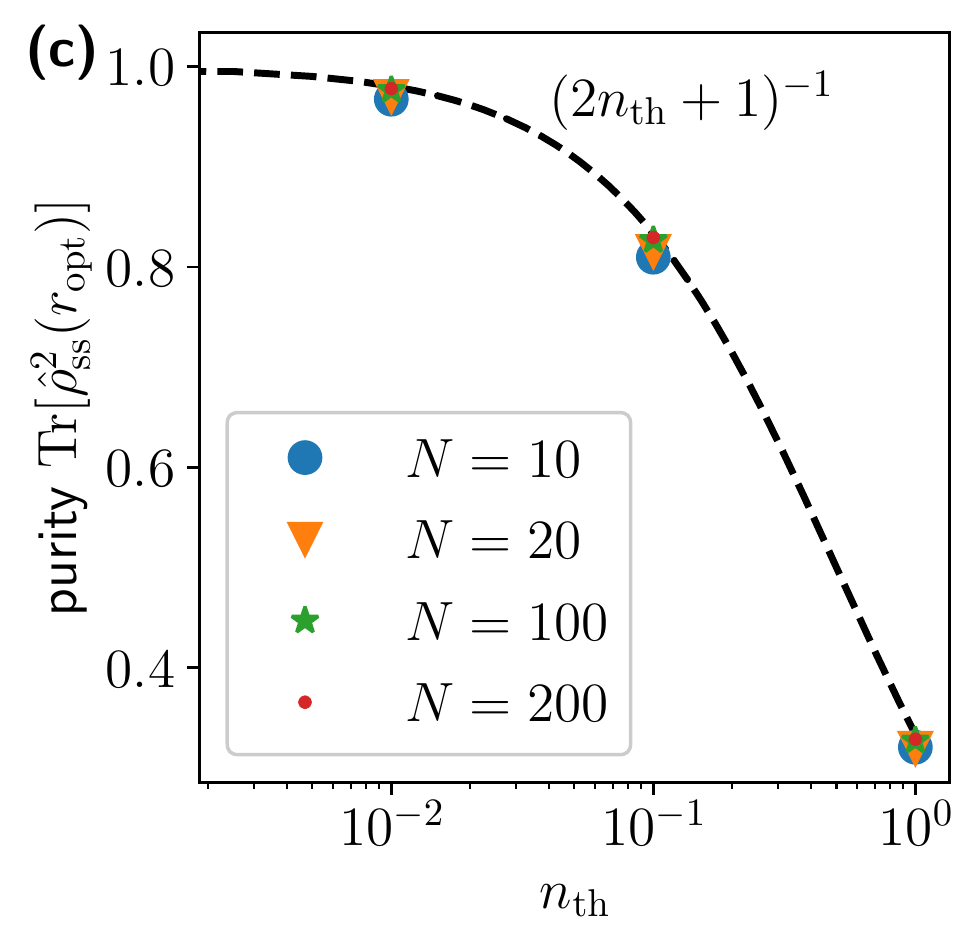}	
	\caption{
		\textbf{(a)} 		
			Steady-state Wineland parameter [obtained by numerically solving Eq.~\eqref{eq:masterEqFiniteTempForSigma}] if the engineered reservoir stabilizes an impure squeezed state for $N = \ValuefiniteTemperatureWinelandN$ and different effective thermal steady-state occupation numbers $n_\mathrm{th}$.
			The dash-dotted black line indicates the Heisenberg limit $\xi_\mathrm{R,HL}^2 = 2/(N+2)$. 
			For $e^{2r} \ll N$, the Wineland parameter follows the bosonic theory~\eqref{eqn:finiteTemp:BosonicWineland}, \ie, spin squeezing is quite robust to $n_\mathrm{th}$ and the Wineland parameter \emph{decreases} with $r$. 
		However, beyond an optimal squeezing strength $r_\mathrm{opt}$, even a small effective thermal occupation $n_\mathrm{th} \ll 1$ deteriorates squeezing quite strongly and the Wineland parameter \emph{increases} with $r$.
		\textbf{(b)}
		    Ratio between the minimum Wineland parameter for $n_\mathrm{th} > 0$ and its ideal Heisenberg-limited value $\xi_\mathrm{R,HL}^2 = 2/(N+2)$ for $n_\mathrm{th}=0$.
		Top panel: 
			Ratio as a function of $N$. 
			The dashed fitted lines indicate a $N^{\ValuefiniteTemperatureRatioWinelandAtRoptVsHLexp}$ scaling. 
		Bottom panel:
			Ratio as a function of $n_\mathrm{th}$ for $N=\ValuefiniteTemperatureRatioWinelandAtRoptVsHLN$, which clearly deviates from bosonic relation~\eqref{eqn:finiteTemp:BosonicWineland} indicated by the dashed black line.
		\textbf{(c)}
			Purity of the steady state of Eq.~\eqref{eq:masterEqFiniteTempForSigma} at $r_\mathrm{opt}$ as a function of $n_\mathrm{th}$, which closely follows the purity of the corresponding bosonic squeezed state (dashed black line) if Eq.~\eqref{eq:masterEqFiniteTempForSigma} is mapped onto bosonic dissipative squeezing using a Holstein-Primakoff approximation.
	}
	\label{fig:finiteTemp:Squeezing}
\end{figure*}
%%%%%%%%%%%%%

For $e^{2r} \ll N$, the spin squeezing described by Eq.~\eqref{eq:masterEqFiniteTempForSigma} is well-approximated by a linearized bosonic master equation (via use of the Holstein-Primakoff approximation \cite{HolsteinPrimakoff1940}). 
In this limit, one thus expects that a small  $n_\mathrm{th}$ will have only a small impact on the steady-state squeezing \cite{kronwald2013arbitrarily}, \ie, 
\begin{align}
	\frac{\xi_\mathrm{R}^2(n_\mathrm{th})\vert_{\mathrm{HP}}}{\xi_\mathrm{R}^2(0)\vert_{\mathrm{HP}}} = 1 + 2 n_\mathrm{th} \fullstop
	\label{eqn:finiteTemp:BosonicWineland}
\end{align}

The interesting question is whether $n_\mathrm{th}$ also has innocuous effects for the larger values of $r$ needed to approach the Heisenberg limit.  
Fig.~\ref{fig:finiteTemp:Squeezing}(a) 
shows that this is not the case.   
The linearized bosonic theory breaks down if the squeezing parameter $r$ is too large, with dramatic consequences: 
even small imperfections, $n_\mathrm{th} \ll 1$, cause the steady-state Wineland parameter to strongly deviate from its ideal limit $\xi_\mathrm{R,HL}^2 = 2/(N+2)$.  
Further, one finds that the steady state Wineland parameter \emph{increases} with increasing $r$ (irrespective of the parity of $N$), \ie, the steady-state squeezing is a non-monotonic function of $r$ and exhibits a minimum at an optimal value $r_\mathrm{opt}$ (which depends on $N$ and $n_\mathrm{th}$).

We thus have an important caveat:  if the engineered squeezed reservoir has a non-zero effective temperature, increasing the reservoir squeezing parameter $r$ \emph{does not} result in ever-increasing steady-state spin squeezing.  We stress that this is true even when $N$ is even.  
Numerical simulations indicate that the minimum Wineland parameter at $r_\mathrm{opt}$ almost follows a Heisenberg-like scaling with $N$, as shown in Fig.~\ref{fig:finiteTemp:Squeezing}(b), but with a significantly larger prefactor than the ideal result $\xi_\mathrm{R,HL}^2 = 2/(N+2)$. 
Further numerical results exploring the parameter dependence of the optimal squeezing parameter $r_\mathrm{opt}$ are given in App.~\ref{sec:Appendix:r_opt}, while \cref{sec:App:ImpureResMFT} briefly discusses scaling obtained using a mean-field theory approach. 

Interestingly, Fig.~\ref{fig:finiteTemp:Squeezing}(c) shows that, although the ratio $\wineland(r_\mathrm{opt})/\xi_\mathrm{R,HL}^2$ deviates from the bosonic expectation, the purity of the steady state at the optimal squeezing parameter $r_\mathrm{opt}$ does closely follow the corresponding relation $(2 n_\mathrm{th} + 1)^{-1}$ valid for bosonic dissipative squeezing.

At a heuristic level, the much stronger sensitivity of dissipative spin squeezing to $n_\mathrm{th} > 0$ and the similar purity (in comparison to dissipative bosonic squeezing) can be understood by the following simplified picture (see also App.~\ref{sec:Appendix:r_opt}). 
Similar to dissipative bosonic squeezing, the dominant contributions to the mixed steady state of Eq.~\eqref{eq:masterEqFiniteTempForSigma} are the dark state of $\Sigmaop$, $\ket{\chi_0} = \psidk{r}$, and the ``first excited'' state $\ket{\chi_1} \propto \Sigmaop^\dagger \psidk{r}$.
Their statistical weights are in a thermal ratio $p_1/p_0 = n_\mathrm{th}/(n_\mathrm{th}+1)$, which explains the similar $n_\mathrm{th}$-dependence of the purity for $r \leq r_\mathrm{opt}$. 
However, the $\ave{\Sy^2}$ variance of $\ket{\chi_1}$ differs strongly from its counterpart in a bosonic squeezed state for $r \gtrsim r_\mathrm{opt}$: 
in this limit, the dark state converges to the $m_y=0$ eigenstate of $\Sy$, $\psidk{r} \to \ket{j,0}_y$ [cf.\ Eq.~\eqref{eq:darkState0}], with a vanishing $\ave{\Sy^2}$ variance.
The operator $\Sigmaop^\dagger$ expressed in the $\Sy$ basis contains spin raising and lowering operators, \ie, the state $\ket{\chi_1}$ is an equal superposition of the $\ket{j,\pm 1}$ states and has a \emph{finite} $\ave{\Sy^2}$ variance.
This leads to an \emph{increase} of the Wineland parameter with increasing $r$ as soon as the $\ave{\Sy^2}$ variance approaches to its nonzero minimum value. 

Thus, the exponential increase of the Wineland parameter for $r \geq r_\mathrm{opt}$ has a very similar origin as the corresponding effect in the the odd-$N$ zero-$n_\mathrm{th}$ case discussed in Sec.~\ref{sec:evenOdd}.
These results are yet another demonstration of the fact that dissipative spin squeezing is more complicated than dissipative bosonic squeezing, due to the finite-dimensional Hilbert space and the intrinsic nonlinearity of spin systems. 

A consequence of these findings is that, for large squeezing parameters $r \gg r_\mathrm{opt}$, an impurity of the engineered reservoir reduces the even-odd effect. 
We will discuss the consequences for the even-odd effect below in Sec.~\ref{sec:implementation:evenodd}, where we will show that the even-odd effect can nevertheless be observed on state-of-the-art experimental platforms.

%%%%%%%%%%%%%%%%%%%%%%%%%%%%%%%%%%%%%%%%%%
%%%%%%%%%%%%%%%%%%%%%%%%%%%%%%%%%%%%%%%%%%
%%%%%%%%%%%%%%%%%%%%%%%%%%%%%%%%%%%%%%%%%%

\section{Hybrid-systems implementation using dissipative bosonic squeezing}
\label{sec:Implementation}

As discussed in Sec.~\ref{sec:realisticSystem1}, the dissipative spin-squeezing setup described by the general quantum master equation~\eqref{eq:realisticMasterEq2} can be realized using standard two-level systems (unlike the more structured four-level atoms in Refs.~\cite{dalla2013dissipative,borregaard2017one}), and \emph{without} requiring the use of non-classical squeezed input light. 
Instead, one harnesses a standard (resonant) Tavis--Cummings coupling between a spin ensemble and a bosonic mode, along with the dissipative squeezing of this bosonic mode which is engineered by coupling the bosonic mode to a lossy auxiliary mode that is driven by imbalanced red-detuned and blue-detuned sideband drives. 
The second element here has been experimentally realized in a variety of systems. 
In this section, we provide more details on the physical implementation of our hybrid-systems approach to dissipative spin squeezing in three promising platforms: 
trapped ions, solid-state optomechanical devices, and superconducting circuits.

%%%%%%%%%%%%%%%%%%%%%%%%%%%%%%%%%%%%%%%%%%
%%%%%%%%%%%%%%%%%%%%%%%%%%%%%%%%%%%%%%%%%%
\subsection{Trapped ions}
\label{sec:implementation:trappedions}

In trapped ions, the relevant spin degree of freedom usually corresponds to two metastable internal states (spin or orbital) of each individual ion.  
In contrast, the bosonic ``cavity'' mode corresponds to a collective motional mode of the ions \cite{Haffner:2008tg} and the coupling parameter $g$ now characterizes the spin-\emph{phonon} coupling.
Recent experiments have already utilized the spin-motion coupling for over 50 ions in a 2D Penning trap \cite{Bohnet2016} and 1D linear Paul trap \cite{2017Natur.551..601Z}. 
The desired Tavis-Cummings coupling is commonly realized by applying a laser field that is resonant with the red motional sideband of the spin-level transition (see, \eg, Ref.~\onlinecite{Porras:2004di}). 
Motional dissipation is, in turn, mediated by coupling the motional mode to a dipole-allowed transition of an ion. 

\begin{figure}
	\centering
	\includegraphics[width=0.48\textwidth]{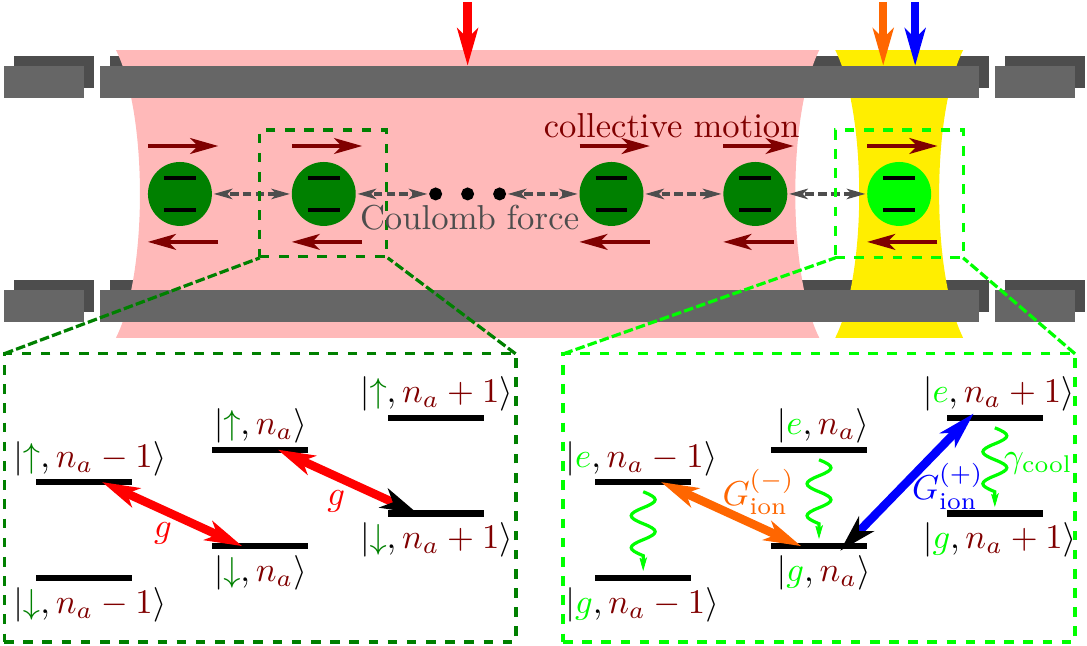}
	\caption{
			Sketch of the trapped-ion setup discussed in Sec.~\ref{sec:implementation:trappedions} using $N$ system ions (dark green circles) and one cooler ion (light green circle) oscillating in a collective motional mode $\hat{a}$ (thin brown arrows). 
			The system ions are coupled to the motional mode at a coupling rate $g$ by a red sideband drive of a metastable transition $\ket{\uparrow} \leftrightarrow \ket{\downarrow}$.
			The squeezed bath for the motional mode $\hat{a}$ is engineered by driving the cooler ion simultaneously on the red and blue motional sidebands (at rates $G_\mathrm{ion}^{(-)}$ and $G_\mathrm{ion}^{(+)}$, respectively) of a short-lived transition $\ket{e} \leftrightarrow \ket{g}$ with a decay rate $\gamma_\mathrm{cool}$ (squiggly light green arrows).
	}
	\label{fig:Implementation:Sketch}
\end{figure}

To realize dissipative spin squeezing with $N$ spins, we imagine a setup that consists of $N+1$ ions.  
$N$ of these ions make up the spin ensemble that we wish to squeeze; the remaining additional ion serves as a ``cooler'' ion that is used to dissipatively squeeze the collective motional mode, as shown in Fig.~\ref{fig:Implementation:Sketch}.
A squeezed bath can be engineered by applying two laser fields that are resonant with the red and blue sideband transitions of the cooler ion \cite{1993PhRvL..70..556C}, leading to an effective Hamiltonian
\begin{equation}
    \hat{H}_\textrm{ion} = G_\textrm{ion}^{(-)} \am \dg \smm + G_\textrm{ion}^{(+)} \am \smm + \hc \comma
\end{equation}
where $G_\textrm{ion}^{(-)}$ ($G_\textrm{ion}^{(+)}$) is the red (blue) sideband coupling, and $\smm$ is the lowering operator of the cooler-ion transition. 
The squeezing strength can be controlled by the ratio of the couplings, \ie, $\tanh(r) = |G_\textrm{ion}^{(+)}/G_\textrm{ion}^{(-)}|$, and the squeezed axis is determined by their relative phase.  
We stress that the sideband transitions are implemented by classical drives, \ie, no squeezed radiation is required for the bath engineering. 
Reading out the spin-squeezed state can be performed by individually measuring each ion in the Pauli $\sy$ basis. 
The collective spin variance can then be obtained from the statistics of the measurement outcomes collected in multiple runs.

To assess the practical requirement of our proposal, we adopt the state-of-the-art system parameters in the dissipative motional squeezing experiment by Kienzler \emph{et al.} \cite{Kienzler2015us}. 
This experiment employs the bath engineering techniques described above to prepare the motion of a single trapped ion in a $-12.6\,\mathrm{dB}$ squeezed ground state with up to $12\%$ infidelity. 
Utilizing this scheme for spin squeezing requires two additions. 

First, the reservoir engineering technique has to be extended to the collective motion of an ion chain. 
Since the center-of-mass mode frequency is not altered by the number of ions in the trap, and the frequencies of other motional modes are well resolved \cite{James:1998tr}, the same sideband drives used in the single-ion case can be applied to engineer the same squeezed bath for the collective center-of-mass mode. 
Note that, for incoherent electric field noise, the center-of-mass heating rate does not depend on the number of ions in the trap \cite{James1998}.

Second, spin-motion coupling has to be applied between the center-of-mass mode and the $N$ system ions with a spin-phonon coupling strength $g$.
To implement the collective spin dissipator in Eq.~\eqref{eq:realisticMasterEq2}, a sufficient but not necessary condition is that the coupling strength $g$ is sufficiently weak to guarantee the adiabatic elimination of the motional mode, while sufficiently strong such that qubit decoherence does not deteriorate the spin squeezing, \ie, $\kappasqz/\sqrt{N} \gg g \gg \sqrt{\kappasqz \gammaphi}$. 
From the time evolution of squeezed-state pumping in Ref.~\onlinecite{Kienzler2015us}, we estimate the experimentally realized squeezed-reservoir coupling rate to be $\kappasqz \approx 0.5\,\mathrm{kHz}$.
Moreover, $\gammaphi \approx 0.1\,\mathrm{Hz}$ has been realized in many experiments (\eg, Ref.~\onlinecite{2005PhRvL..95f0502L}). With these realistic parameters, a coupling $g \approx 30\,\mathrm{Hz}$ should fit in this regime for a modest chain with $N \lesssim 10$ ions. 
Since spin-motion coupling as strong as kHz has been routinely implemented in trapped-ion experiments (\eg, Ref.~\onlinecite{Burd:2021bo}), our desired range is thus well achievable by simply using a weaker drive.

The remaining issue is whether the thermal excitation due to motional heating will mask the desired even-odd effect. 
Motional heating leads to collective spin excitation and relaxation processes at equal rates $\gammaheat$. 
As discussed in App.~\ref{sec:Appendix:FiniteTemperatureModel}, this can be mapped onto a quantum master equation of the form~\eqref{eq:masterEqFiniteTempForSigma} with an effective squeezing parameter $\tilde{r}$, \ie, $\Sigmaop(r) \to \Sigmaop(\tilde{r})$ where 
\begin{align}
	\tilde{r} &= r - \sinh(2r) \frac{\gammaheat}{\Gamma} \comma 
	\label{eqn:FiniteTemp:TrappedIon:effectiver}
\end{align}
and with an effective thermal occupation number $n_\mathrm{th}$ given by
\begin{align}
	n_\mathrm{th} &= \cosh(2r) \frac{\gammaheat}{\Gamma} \fullstop
	\label{eqn:FiniteTemp:TrappedIon:effectiventh}
\end{align}
Kienzler \emph{et al.} achieved a squeezed ground state with a fidelity larger than $88\,\%$ at $-12.6,\,\mathrm{dB}$ of squeezing \cite{Kienzler2015us}, which corresponds to $n_\mathrm{th} \approx 0.14$ or, equivalently, a motional heating rate $\gammaheat/\Gamma \approx \ValuefiniteTemperatureEvenOddFixedGammaCollgammacoll$. 
Note that this is a worst-case estimate of the fidelity since its reported value includes measurement errors. 
Consequently, the true value of $\gammaheat$ in the experiment is likely lower. 
However, even with this pessimistic values the even-odd effect can be observed in experiments, as we will show below in Sec.~\ref{sec:implementation:evenodd}.

%%%%%%%%%%%%%%%%%%%%%%%%%%%%%%%%%%%%%%%%%%
%%%%%%%%%%%%%%%%%%%%%%%%%%%%%%%%%%%%%%%%%%
\subsection{Solid-state spins in an optomechanical crystal}

In solid-state platforms, the spin ensemble in our scheme could be realized using defect centers in a semiconductor, \eg, NV-center defect spins in diamond.  
These spins can be implanted in a structure which in turn realizes an optomechanical crystal:  a patterned photonic crystal beam with a defect that localizes both a mechanical mode and an optical mode \cite{Eichenfield2009}.  
We note that high-Q diamond optomechanical crystals have been realized experimentally \cite{Loncar2016}, with a recent experiment even integrating such a system with NV-center defect spins \cite{Cady19}. 
The localized mechanical mode plays the role of the bosonic ``cavity'' in \cref{eq:realisticMasterEq2}.
The spins and mechanical motion exhibit an intrinsic coupling due to the strain dependence of spin-level transitions, and the coupling could be further enhanced by incorporating the high strain sensitivity of excited states through phonon-assisted Raman transitions \cite{Cady19,Golter:2016cd,Albrecht:2013vc}.

In this kind of setup, the optomechanical coupling between the localized mechanical and optical cavity mode provides a mechanism for the dissipative squeezing of the mechanical mode.  
If one is in the sideband-resolved regime (where the mechanical frequency is larger than the optical cavity decay rate), then this dissipative mechanical squeezing can be realized by driving the optical cavity by two control lasers that are resonant to the red and blue motional sidebands, respectively \cite{Kronwald2014}.  
We stress that these drives are classical, coherent state drives.    
Ignoring the non-linear coupling that is usually negligibly weak in most platforms, the optomechanical coupling is well approximated by
\begin{equation}
    \hat{H}_\textrm{OM} = G_\textrm{OM}^{(-)} \am \dg \hat{b} + G_\textrm{OM}^{(+)} \am \hat{b} + \hc \comma
    \label{eq:HamiltonianOM}
\end{equation}
where $G_\textrm{OM}^{(-)}$ ($G_\textrm{OM}^{(+)}$) is the red (blue) sideband optomechanical coupling strength, and $\hat{b}$ is the annihilation operator of the optical cavity mode. 
The squeezing strength is determined by the ratio of the red and blue sideband coupling, \ie, $\tanh(r) = |G_\textrm{OM}^{(+)}/G_\textrm{OM}^{(-)}|$, which can be tuned by varying the amplitude of the driving tones.  We note that this kind of dissipative squeezing of mechanical motion via optomechanics has been realized in several experiments \cite{Wollman2015,Teufel2015,Sillanpaa2015,Lei2016}.  
Our protocol thus provides a means of harnessing this capability to generate spin squeezing. 
Finally, in solid-state settings inhomogeneous broadening of the spin ensemble is almost always an issue; this is typically mitigated by using dynamical decoupling techniques. 
For spin-squeezing protocols based on OAT dynamics, a very simple decoupling sequence can be used, which repeatedly applies $\pi$ pulses about the $x$ axis \cite{bennett2013phonon}.
This very simple strategy fails in our case because it transforms the $\Sigmaop{}$ decay term in Eq.~\eqref{eq:realisticMasterEq2} into a $\Sigmaop{}^\dagger$ anti-damping term. 
However, as discussed in Sec.~\ref{sec:App:DynDec}, this unwanted excitation dynamics can be canceled by two additional $\pi$ pulses about the $z$ and $y$ axis, which makes our protocol compatible with dynamical decoupling.

%%%%%%%%%%%%%%%%%%%%%%%%%%%%%%%%%%%%%%%%%%
%%%%%%%%%%%%%%%%%%%%%%%%%%%%%%%%%%%%%%%%%%
\subsection{Superconducting microwave cavities}

Superconducting microwave cavities and circuit QED are another promising class of systems for implementing our ideas. 
Our basic building block of a bosonic mode coupled to a spin ensemble could be realized by coupling a single microwave cavity mode to either a set of superconducting qubits \cite{Macha2014,Shulga2017,Song2019,Brehm2021}, or to electronic spins in substrate (\eg, Bi donors implanted in Si \cite{Bienfait2017,Albanese2020}).
The second ingredient, a mechanism for the dissipative generation of microwave squeezing, could also be implemented in different ways.   
One approach is to inject squeezed microwave radiation directly into the cavity using the output of a Josephson parametric amplifier \cite{CastellanosBeltran2008,Zhou2014,Macklin2015}. 
This has already been achieved experimentally in Ref.~\onlinecite{Bienfait2017}, in a system where a cavity has been coupled to a spin ensemble.  
An alternate approach which has the advantage of not being limited by insertion losses (associated with transporting a squeezed state) is to mimic the same dissipative squeezing protocols used in optomechanics to squeeze a mechanical mode.  
This can be accomplished by coupling three microwave modes via a Josephson ring modulator \cite{Bergeal2010}, which generates a three-wave mixing term $(\hat{p} + \hat{p}^\dagger)(\hat{a} + \hat{a}^\dagger)(\hat{b} + \hat{b}^\dagger)$ between the modes $\hat{a}$, $\hat{b}$, and $\hat{p}$ \cite{Markovic2018}. 
By driving the pump mode $\hat{p}$ coherently at the sum and difference frequency of the $\hat{a}$ and $\hat{b}$ modes, $\omega_\pm$, one can engineer an interaction of the form of \cref{eq:HamiltonianOM}, where the prefactors $G_\mathrm{OM}^{(\pm)}$ depend on the strength of the drives at $\omega_\pm$, respectively. 
Adiabatic elimination of the strongly-damped $\hat{b}$ mode generates an effective squeezed bath for the $\hat{a}$ mode as shown in \cref{eq:masterEqCavitySpins}.
A recent experiment implementing this approach has demonstrated up to $-8\,\mathrm{dB}$ of intracavity squeezing of the $\hat{a}$ mode \cite{Dassonneville2021}.

\subsection{Experimental viability of the even-odd effect}
\label{sec:implementation:evenodd}

%%%%%%%%%%%%%
\begin{figure*}
	\centering
	\includegraphics[width=0.47\textwidth]{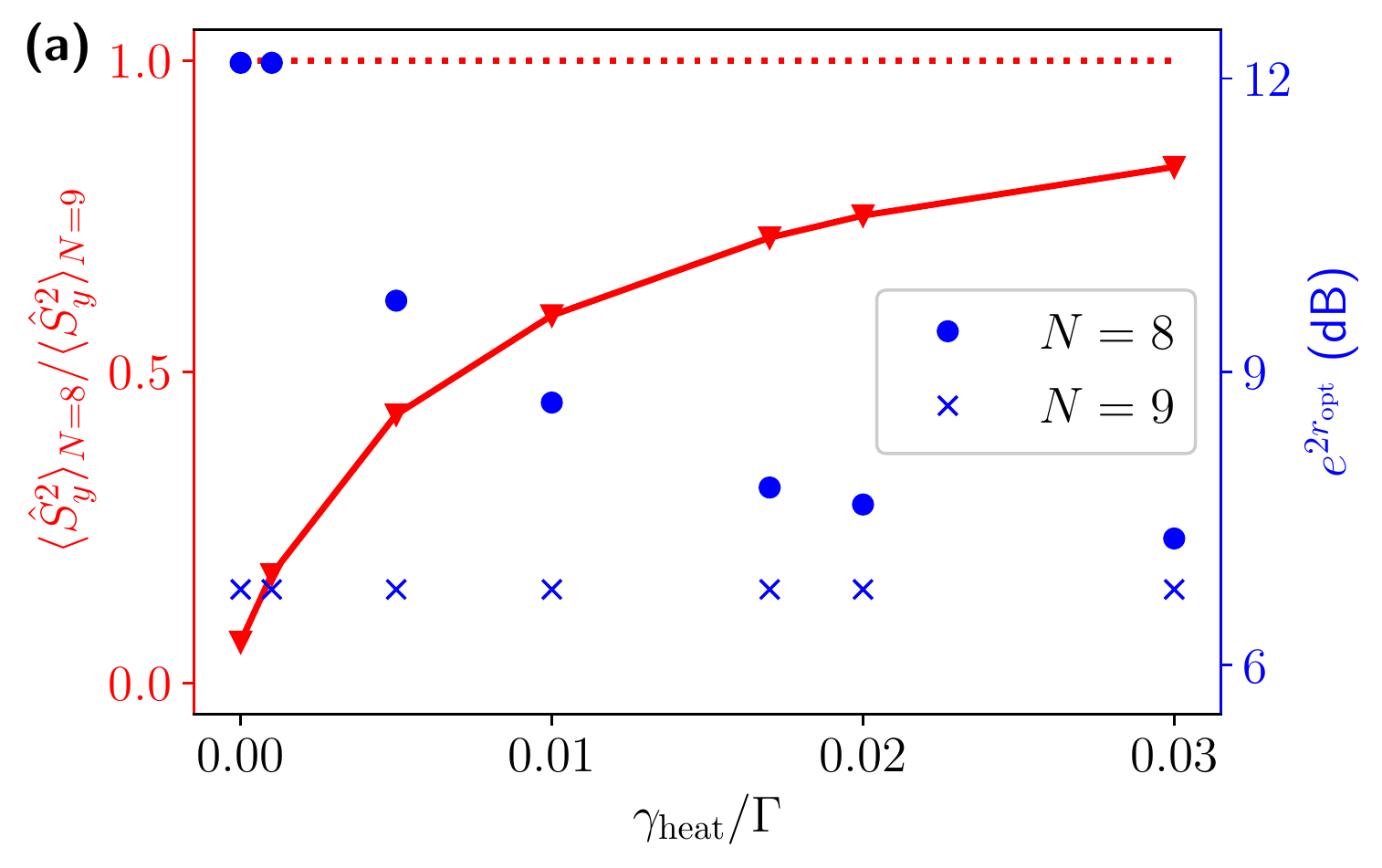}
	\includegraphics[width=0.48\textwidth]{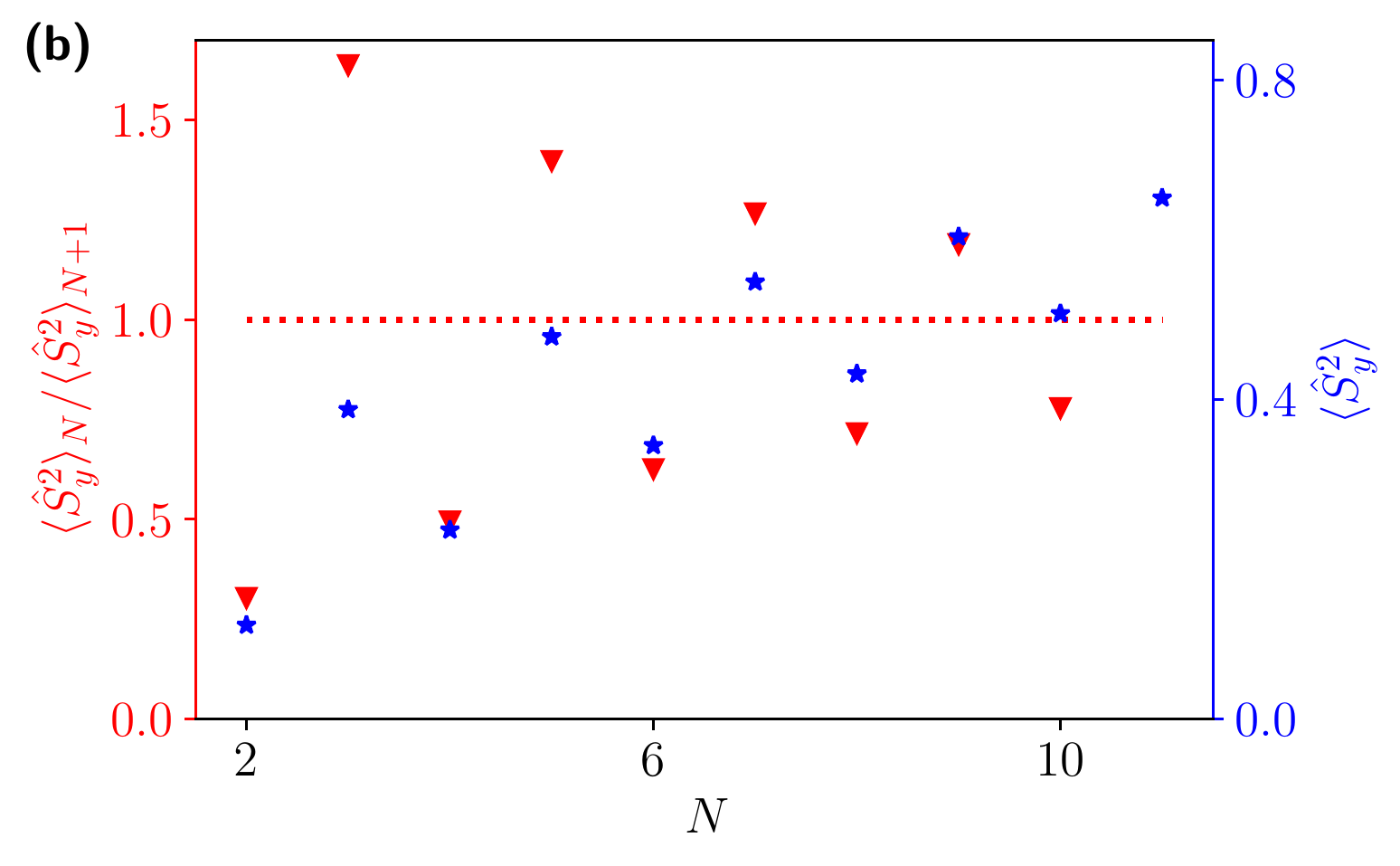}
	\caption{
		Impact of an impure squeezed reservoir on the even-odd effect. 
		For the trapped-ion setup introduced in Sec.~\ref{sec:implementation:trappedions}, motional heating at a rate $\gammaheat$ is the dominant source of imperfections, which can be modeled by Eq.~\eqref{eq:masterEqFiniteTempForSigma} using an effective squeezing parameter $\tilde{r}$ and an effective temperature $n_\mathrm{th}$ introduced in Eqs.~\eqref{eqn:FiniteTemp:TrappedIon:effectiver} and~\eqref{eqn:FiniteTemp:TrappedIon:effectiventh}.
	\textbf{(a)}
		Ratio between $\ave{\Sy^2}$ for $N=\ValuefiniteTemperatureEvenOddFixedNN$ vs.\ $\ValuefiniteTemperatureEvenOddFixedNNpo$ as a function of the motional heating rate $\gammaheat$ (red triangles). 
		The dotted red line is a guide to the eye to highlight a ratio of unity, \ie, no even-odd effect.
		For each value of $N$ and $\gammaheat$, we optimize the squeezing parameter over the range $0 \leq e^{2r} \leq 12 \,\mathrm{dB}$ to minimize the variance $\ave{\Sy^2}$ (blue data points).
	\textbf{(b)}
		Ratio of $\ave{\Sy^2}$ for $N$ vs.\ $N+1$ spins as a function of $N$ for an experimentally realistic parameter of $\gammaheat/\Gamma = \ValuefiniteTemperatureEvenOddFixedGammaCollgammacoll$ (red triangles).
		Again, the dotted red line is a guide to the eye to highlight a ratio of unity, \ie, no even-odd effect.
		For each $N$, we optimize the squeezing parameter over the same range $0 \leq e^{2r} \leq 12\,\mathrm{dB}$ as in (a) to minimize the variance $\ave{\Sy^2}$ (blue stars). 
	}
	\label{fig:EvenOdd:TrappedIon}
\end{figure*}
%%%%%%%%%%%%

Given the results of the Sec.~\ref{sec:Temperature}, one may worry that a finite effective temperature of the engineered reservoir will substantially decrease spin squeezing and mask the even-odd effect introduced in Sec.~\ref{sec:evenOdd}. 
Here, we show that this is not always the case and, in particular, that the even-odd effect can be observed in experimentally accessible parameter regimes. 

For concreteness, we focus on the trapped-ion platform introduced in Sec.~\ref{sec:implementation:trappedions}, which is most mature. 
With a first experimental observation of the even-odd effect in mind, we consider a modest number of spins, $N \lesssim 10$. 
In this platform, motional heating due to classical trap noise is the dominant source of imperfections of the engineered squeezed reservoir. 
As discussed in Sec.~\ref{sec:implementation:trappedions} and App.~\ref{sec:Appendix:FiniteTemperatureModel}, its impact can be modeled using Eq.~\eqref{eq:masterEqFiniteTempForSigma} with an effective squeezing parameter $\tilde{r}$ and an effective thermal occupation number $n_\mathrm{th}$ given by Eqs.~\eqref{eqn:FiniteTemp:TrappedIon:effectiver} and~\eqref{eqn:FiniteTemp:TrappedIon:effectiventh}, respectively. 

Compared to the bare squeezing parameter $r$ (which is related to the amplitude of the red and blue sideband drives), the effective squeezing parameter strength $\tilde{r}$ is reduced by the motional heating rate $\gammaheat$.
The effective thermal occupation number $n_\mathrm{th}$ grows with increasing $r$, \ie, a large squeezing parameter reduces the purity of the squeezed state. 
For a given number $N$ of ions, the variance $\langle \Sy^2 \rangle$ will thus take a minimum value at an optimal squeezing parameter $r_\mathrm{opt}$, whose value depends on $N$ and $\gammaheat/\Gamma$, as shown in Fig.~\ref{fig:EvenOdd:TrappedIon}(a).
Comparing these minimum variances for $N=\ValuefiniteTemperatureEvenOddFixedNN$ vs.\ $\ValuefiniteTemperatureEvenOddFixedNNpo$ ions, we find an even-odd difference of more than $20\,\%$ over a wide range of motional heating rates $\gammaheat$. 
The even-odd difference decreases with effective temperature $n_\mathrm{th}$ and will disappear if the impurity of the squeezed reservoir is sufficiently large. 
Note that even for the upper bound of the motional heating rate derived in Sec.~\ref{sec:implementation:trappedions}, $\gammaheat/\Gamma = \ValuefiniteTemperatureEvenOddFixedGammaCollgammacoll$, the even-odd difference is clearly visible.

An alternative way of probing the even-odd effect is shown in Fig.~\ref{fig:EvenOdd:TrappedIon}(b). 
There, the motional heating rate $\gammaheat$ is kept fixed but the number of ions in the trap is varied.
Again, we optimize $r$ for each data point individually to minimize the variance $\ave{\Sy^2}$. 
The ratio between these variances shows pronounced oscillations for $N \leq 10$ ions. 
Note that we restrict the optimal squeezing parameter $e^{2r_\mathrm{opt}}$ in both scenarios to be smaller than the experimentally achievable $-12.6\,\mathrm{dB}$.
Thus, our results suggest that the even-odd effect is realizable on state-of-the-art trapped ion platforms.

Instead of varying $r$ for each $N$ \emph{individually} to minimize the variances of even and odd $N$ independently, one could also use the same squeezing parameter $r$ for a pair of $N$ and $N+1$ spins and maximize the \emph{ratio} of their variances $\ave{\Sy^2(r)}_N/\ave{\Sy^2(r)}_{N+1}$. 
In this case, the even-odd effect would be even more pronounced than the data shown in Fig.~\ref{fig:EvenOdd:TrappedIon}.

%%%%%%%%%%%%%%%%%%%%%%%%%%%%%%%%%%%%%%%%%%%%%%%%%%%
%%%%%%%%%%%%%%%%%%%%%%%%%%%%%%%%%%%%%%%%%%%%%%%%%%%
%%%%%%%%%%%%%%%%%%%%%%%%%%%%%%%%%%%%%%%%%%%%%%%%%%%

\section{Connection to previous works}
\label{sec:PreviousWork}
Here we review previous works on dissipative spin squeezing \cite{AgarwalPuri1989,AgarwalPuri1990,AgarwalPuri1994,Kuzmich1997,dalla2013dissipative,borregaard2017one}, summarize their results and point out the differences to this work. 

Agarwal and Puri discussed the idealized spin-only quantum master equation~\eqref{eq:masterEqIdeal1} with an additional collective decay term $\mD{\hat{S}_-}{\rhoop}$ and a coherent drive $\Omega \hat{S}_+ + \Omega^* \hat{S}_-$ \cite{AgarwalPuri1989,AgarwalPuri1990}. 
They derived explicit expressions for the steady state and pointed out that the steady state of Eq.~\eqref{eq:masterEqIdeal1} is pure for even $N$ (where $\Sigmaop$ has a zero eigenvalue) and mixed for odd $N$ (where $\Sigmaop$ has only nonzero eigenvalues) \cite{AgarwalPuri1989}.
Moreover, they discussed the pairwise excitation structure of the even-$N$ steady state illustrated in Fig.~\ref{fig:EvenOdd1}(a) and showed numerical results for the odd-$N$ population distribution in the regime  $e^{2r} \gtrsim N$ \cite{AgarwalPuri1989}.
In a follow-up article \cite{AgarwalPuri1994}, they calculated the Wineland spin-squeezing parameter $\wineland$ of a state of the form $\ket{\psi} \propto \exp(\theta \hat{S}_z) \exp(-i \pi \hat{S}_y/2) \ket{j,m}$, which contains the even-$N$ steady state~\eqref{eq:darkState0} as a special case, but not the odd-$N$ steady state. 
They also discussed squeezing in the presence of a coherent drive \cite{AgarwalPuri1990}.

Unlike our work, Agarwal and Puri did not evaluate the spin-squeezing properties of the odd-$N$ steady state.
Consequently, they did not find the dramatic difference in spin squeezing between the undriven even-$N$ and odd-$N$ steady states for $e^{2r} \gtrsim N$, which is one of the central results of this work. 
Note that the spin-squeezing parameter $r$ required to see an even-odd effect will decrease if the number $N$ of spins is lowered, which makes the regime $e^{2r} \gtrsim N$ attainable on state-of-the-art experimental platforms, as discussed in Sec.~\ref{sec:Implementation}.
Therefore, another new and experimentally relevant result of our work is that the even-odd effect is not a mere mathematical complication, but a real and testable physical effect.

Agarwal and Puri suggested to generate spin squeezing by illuminating the spins with squeezed radiation, which is experimentally very challenging. 
In a later work, Kuzmich \emph{et al.}\ proposed a related alternate scheme to generate spin squeezing using V-type atoms illuminated by squeezed light \cite{Kuzmich1997}. 
They did not discuss dissipative spin squeezing and, consequently, did not comment on the even-odd effect at all.
In contrast to the proposals by Agarwal \emph{et al.}\ and Kuzmich \emph{et al.}, our work does not require any injection of nonclassical light, which lowers the experimental challenges significantly.

Dalla Torre \emph{et al.}\ proposed another method to implement dissipative spin-squeezing dynamics in a specific multilevel atomic system using Raman transitions \cite{dalla2013dissipative}.
They use a pure dark state of the type of Eq.~\eqref{eq:darkState1} to analyze the ideal situation when there is only collective loss of the form of Eq.~\eqref{eq:sigmaPp}.
In addition, they discuss the impact of dissipation due to single-atom Raman scattering into free space. 
They do neither comment on an even-odd effect nor mention that their pure-state analysis is strictly speaking only valid in the case of even $N$.

Our work proves that, while this mathematical treatment is indeed admissible if squeezing is not too strong, $e^{2r} \ll N$, substantial changes to the spin-squeezing physics will show up at $e^{2r} \gtrsim N$, which have not been discussed in the literature before. 
Regarding experimental versatility, Kuzmich \emph{et al.}\ and Dalla Torre \emph{et al.}\ discuss very specific implementations, which are not applicable to a generic ensemble of two-level systems. 
In contrast, we propose a more generic mechanism that has not been discussed in the literature before. 
Importantly, our approach is compatible with a wide range of systems including solid-state implementations (see Sec.~\ref{sec:Implementation}) and is perhaps the most flexible and experimentally viable implementation. 
This versatility allowed us to discover the prethermalization physics discussed in Sec.~\ref{sec:slowTimescale}, which is not present in the scheme of Ref.~\onlinecite{dalla2013dissipative} because of their very specific decay mechanism that couples collective and local dissipation.

Finally, Borregaard \emph{et al.}\ considered $\Lambda$-type atoms driven by multiple laser drives and identified a dissipative spin-squeezing scheme as its resonant limit \cite{borregaard2017one}.
They use this observation to interpret their results, but they do not discuss the dissipative spin-squeezing scheme in detail nor explore any of its consequences revealed in this work, like the even-odd physics and prethermalization.

Finally, we note that in contrast to our work,  the previous works on dissipative spin squeezing reviewed above did not discuss or analyze the consequences of an imperfect engineered reservoir (see Sec.~\ref{sec:Temperature}).

%%%%%%%%%%%%%%%%%%%%%%%%%%%%%%%%%%%%%%%%%%%%%%%%%%%
%%%%%%%%%%%%%%%%%%%%%%%%%%%%%%%%%%%%%%%%%%%%%%%%%%%
%%%%%%%%%%%%%%%%%%%%%%%%%%%%%%%%%%%%%%%%%%%%%%%%%%%

\section{Conclusions}
\label{sec:Conclusions}

In this work, we have revisited the reservoir-engineering approach to preparing and stabilizing spin-squeezed states.  
We analyzed in detail a particular implementation strategy that had not previously been studied, but that is compatible with a number of experimental platforms:  employ a hybrid-systems approach where one first uses bosonic reservoir-engineering techniques to stabilize a bosonic squeezed state, and then uses this state (via a standard Tavis-Cummings-type coupling) to dissipatively squeeze a spin ensemble.  
We also discussed how this approach compared favourably to the standard one-axis-twist method for spin squeezing in the presence of single-spin relaxation.

Our work also addressed fundamental aspects of dissipative spin squeezing, with a focus on two general but surprising phenomena.  
The first was an extreme, macroscopic sensitivity of the steady state to the parity of the number $N$ of spins in the ensemble.  
We analyzed both how this effect could be avoided (if the goal was to generate spin squeezing without any parity sensitivity), and how it might be harnessed for a powerful new sensing modality.  
The second general effect we studied was the emergence of a surprisingly long slow timescale and ``prethermalization" behavior when weak single-spin dephasing is added to our model.  
While the steady state in this regime exhibits at best limited squeezing, the intermediate time quasi-steady state can be highly squeezed.  
Moreover, the reduction of steady-state spin squeezing can be almost completely suppressed by deliberately introducing a small amount of single-spin relaxation.

Finally, we investigated the impact of an engineered reservoir stabilizing an impure steady state. 
We discovered a strong sensitivity of the Wineland parameter to impurity if the squeezing parameter $r$ is large. 

We hope our work will lay the groundwork for near-term experimental implementations of reservoir-engineered spin squeezing in a variety of systems.  
In future theoretical work, it will be interesting to explore extensions of the models analyzed here.  
For example, it is well known that collective Hamiltonian interactions that are not truly infinite range can still generate large amounts of spin squeezing \cite{Gorshkov2016,Perlin2020}.  
Is the same true with dissipative spin-spin interactions, and if so, are the requirements more or less forgiving than in the coherent case?  
It would also be interesting to study in more detail the effects of disorder, \eg, due to inhomogeneous broadening, both on spin squeezing and on the parity-sensing scheme proposed here.

\begin{acknowledgments}
This work was primary supported by the DARPA DRINQS program (agreement D18AC00014).
We also acknowledge partial support by the University of Chicago Materials Research Science and Engineering Center, which is funded by the National Science Foundation under Grant No. DMR-1420709. 
A.\ A.\ C.\ acknowledges support from the Simons Foundation through a Simons Investigator award. 
We thank A.-M.\ Rey and M.\ Kloc for useful discussions.
\end{acknowledgments}

%%%%%%%%%%%%%%%%%%%%%%%%%%%%%%%%%%%%%%%%%%%%%%%%%%%
%%%%%%%%%%%%%%%%%%%%%%%%%%%%%%%%%%%%%%%%%%%%%%%%%%%
%%%%%%%%%%%%%%%%%%%%%%%%%%%%%%%%%%%%%%%%%%%%%%%%%%%

\begin{appendix}

%%%%%%%%%%%%%%%%%%%%%%%%%%%%%%%%%%%%%%%%%%%%%%%%%%%
%%%%%%%%%%%%%%%%%%%%%%%%%%%%%%%%%%%%%%%%%%%%%%%%%%%
%%%%%%%%%%%%%%%%%%%%%%%%%%%%%%%%%%%%%%%%%%%%%%%%%%%

\section{Adiabatic elimination of a cavity coupled to a squeezed bath}
\label{sec:App:AdiabaticElimination}

In this Appendix, we outline the derivation of the effective quantum master equation~\eqref{eq:realisticMasterEq2} of the main text.
Our starting point is \cref{eq:masterEqCavitySpins} describing a collection of spins interacting with a squeezed bosonic mode. 
For the moment, we ignore the terms in \cref{eq:masterEqCavitySpins} describing local dissipation of the spins, 
\begin{align}
	\frac{\mathrm{d}}{\mathrm{d} t} \rhoop 
		&= -i g \comm{\Sp \hat{a} + \Sm \hat{a}^\dagger}{\rhoop} + \kappaint \mD{\hat{a}}{\rhoop} \nonumber \\
		&\phantom{=}\ + \kappasqz \mD{\cosh(r) \am + \sinh(r) \am \dg}{\rhoop} \fullstop
\end{align}
Assuming that the cavity evolves on a much shorter timescale than the spins, 
\begin{align}
	\kappaint + \kappasqz \gg g \sqrt{N} \comma
\end{align}
we adiabatically eliminate the cavity by a projection operator technique \cite{GardinerZoller2000} similar to the calculation outlined in Ref.~\onlinecite{AgarwalPuri1997}.
To this end, we split the quantum master equation into two superoperators,
\begin{align}
	\frac{\mathrm{d}}{\mathrm{d} t} \rhoop &= \mathcal{L}_\mathrm{cav} \rhoop + \mathcal{L}_\mathrm{int} \rhoop \comma 
	\label{eq:App:AdiabaticEliminationQME} \\
	\mathcal{L}_\mathrm{cav} \rhoop &= \kappaint \mD{\hat{a}}{\rhoop} + \kappasqz \mD{\cosh(r) \hat{a} + \sinh(r) \hat{a}^\dagger}{\rhoop} \comma \nonumber \\
	\mathcal{L}_\mathrm{int} \rhoop &= -i g \comm{\Sp \hat{a} + \Sm \hat{a}^\dagger}{\rhoop} \comma \nonumber 
\end{align}
where $\mathcal{L}_\mathrm{int} \rhoop$ is considered to be constant on the timescale defined by $\mathcal{L}_\mathrm{cav} \rhoop$. 
Using this approximation, we can formally solve \cref{eq:App:AdiabaticEliminationQME}, 
\begin{align}
	\rhoop(t) = e^{\mathcal{L}_\mathrm{cav} t} \rhoop(0) + e^{\mathcal{L}_\mathrm{cav} t} \int_0^t \mathrm{d} t' e^{- \mathcal{L}_\mathrm{cav} t'} \mathcal{L}_\mathrm{int} \rhoop(t') \fullstop
\end{align}
Performing a Born approximation, we decompose the state as $\rhoop(t) \approx \rhoop_\mathrm{sp}(t) \otimes \rhoop_\mathrm{cav}^\mathrm{ss}$, where $\rhoop_\mathrm{sp}(t)$ is the reduced density matrix of the spin system and $\rhoop_\mathrm{cav}^\mathrm{ss}$ is the steady state of $\mathcal{L}_\mathrm{cav}$.
The equation of motion of the reduced spin density matrix is
\begin{align}
	&\frac{\mathrm{d}}{\mathrm{d} t} \rhoop_\mathrm{sp}(t) \nonumber \\
	&= \int_0^t \mathrm{d} t' \Tr_\mathrm{cav} \left[ \mathcal{L}_\mathrm{int} e^{\mathcal{L}_\mathrm{cav} (t - t')} \mathcal{L}_\mathrm{int} \rhoop_\mathrm{sp}(t') \otimes \rhoop_\mathrm{cav}^\mathrm{ss} \right] \fullstop
	\label{eq:App:BornApproximationQME}
\end{align}
Inserting the explicit form of $\mathcal{L}_\mathrm{int}$, we find that the integral on the right-hand side of \cref{eq:App:BornApproximationQME} depends on the cavity correlation functions
\begin{align}
	&\Tr_\mathrm{cav} \left[ \hat{a}^{(\dagger)} e^{\mathcal{L}_\mathrm{cav} t} \hat{a}^{(\dagger)} \rhoop_\mathrm{cav}^\mathrm{ss} \right] \nonumber \\
	= &\Tr_\mathrm{cav} \left[ \hat{a}^{(\dagger)} e^{\mathcal{L}_\mathrm{cav} t} \rhoop_\mathrm{cav}^\mathrm{ss} \hat{a}^{(\dagger)} \right] \nonumber \\
	= &\frac{\kappasqz}{\kappasqz + \kappaint} \sinh(r) \cosh(r) e^{- (\kappasqz + \kappaint) t/2} \comma \displaybreak[3] \\
	&\Tr_\mathrm{cav} \left[ \hat{a} e^{\mathcal{L}_\mathrm{cav} t} \hat{a}^\dagger \rhoop_\mathrm{cav}^\mathrm{ss} \right] %\nonumber \\
	= \Tr_\mathrm{cav} \left[ \hat{a}^\dagger e^{\mathcal{L}_\mathrm{cav} t} \rhoop_\mathrm{cav}^\mathrm{ss} \hat{a} \right] \nonumber \\
	= & \frac{\kappasqz \cosh^2(r) + \kappaint}{\kappasqz + \kappaint} e^{- (\kappasqz + \kappaint) t/2} \comma \displaybreak[3] \\
	&\Tr_\mathrm{cav} \left[ \hat{a}^\dagger e^{\mathcal{L}_\mathrm{cav} t} \hat{a} \rhoop_\mathrm{cav}^\mathrm{ss} \right] %\nonumber \\
	= \Tr_\mathrm{cav} \left[ \hat{a} e^{\mathcal{L}_\mathrm{cav} t} \rhoop_\mathrm{cav}^\mathrm{ss} \hat{a}^\dagger \right] \nonumber \\
	= & \frac{\kappasqz}{\kappasqz + \kappaint} \sinh^2(r) e^{- (\kappasqz + \kappaint) t/2} \fullstop
\end{align}
These correlation functions decay fast compared to the timescale on which $\rhoop_\mathrm{sp}$ evolves, therefore, we can perform the Markov approximation $\rhoop_\mathrm{sp}(t') \approx \rhoop_\mathrm{sp}(t)$ and rewrite \cref{eq:App:BornApproximationQME} as follows:
\begin{align}
	\frac{\mathrm{d}}{\mathrm{d} t} \rhoop_\mathrm{sp} 
	&= \frac{4 g^2}{(\kappasqz + \kappaint)^2} \Big( \kappaint \mD{\Sm}{\rhoop_\mathrm{sp}} \nonumber \\
	&\phantom{=}\ + \kappasqz \mD{\cosh(r) \Sm - \sinh(r) \Sp}{\rhoop_\mathrm{sp}} \Big) \fullstop
\end{align}
Taking into account the remaining terms in \cref{eq:masterEqCavitySpins} describing single-spin dissipation, we recover \cref{eq:realisticMasterEq2} of the main text.

%%%%%%%%%%%%%%%%%%%%%%%%%%%%%%%%%%%%%%%%%%%%%%%%%%%
%%%%%%%%%%%%%%%%%%%%%%%%%%%%%%%%%%%%%%%%%%%%%%%%%%%
%%%%%%%%%%%%%%%%%%%%%%%%%%%%%%%%%%%%%%%%%%%%%%%%%%%

\section{Mean-field theory equations of motion}
\label{sec:App:MeanFieldTheory}

In this section, we provide the set of nonlinear equations of motion for symmetrized products of spin operators, for the effective spin-only model considered in the main text, namely \cref{eq:realisticMasterEq2}.
While such a system of equations is not closed, we neglect third-order cumulants (equivalently performing a $2^{\rm nd}$-order cumulant expansion) \cite{kubo1962generalized,zens2019critical}, which lets us approximate the third-order expectation values of various operators as
\begin{align}
\ave{\A_{i} \A_{j} \A_{k}}  \approx & \ave{\A_{i} \A_{j}} \ave{\A_{k}} + \ave{\A_{i} \A_{k}} \ave{\A_{j}} \nonumber \\
&+ \ave{\A_{j} \A_{k}} \ave{\A_{i}}  - 2 \ave{\A_{i}}\ave{\A_{j}}\ave{\A_{k}} \fullstop
\end{align}
We stress that the above approximation is applied to expectation values of symmetrized operators, defined according to the following convention:
\begin{align}
    \ave{\A_{i} \A_{j} \A_{k}}_{s}  = &  \frac{1}{6}\bigg(
    \ave{\A_{i} \A_{j} \A_{k}}
    +  \ave{\A_{i} \A_{k} \A_{j}} \\\nonumber
    & +  \ave{\A_{j} \A_{i} \A_{k}}
    +  \ave{\A_{j} \A_{k} \A_{i}} \\\nonumber
    & +  \ave{\A_{k} \A_{i} \A_{j}}
    +  \ave{\A_{k} \A_{j} \A_{i}} \bigg)
\end{align}
when $i \ne j \ne k$, and 
\begin{align}
    \ave{\A_{j}^{2} \A_{k}}_{s}  = &  \frac{1}{2}\bigg(
    \ave{\A_{j}^{2} \A_{k}} 
    + \ave{\A_{k} \A_{j}^{2} } \bigg)
\end{align}
when $i \ne j$. 

Given the initial state with spins completely polarized along the $-z$ direction (\ie, $\ave{\Sz} = -N/2$), the evolution is governed by the equations 
\begin{align} 
\partial_{t}\ave{\Sz} =& \left(-\frac{e^{-2 r} \Gamma}{2}-\frac{e^{2 r} \Gamma}{2}-\gammacoll -\gammarel \right) \ave{\Sz}\nonumber \\
& +\left(-\Gamma -\gammacoll \right) \left( \ave{\Sx^2} +  \ave{\Sy^2} \right)-\frac{\gammarel N}{2} \comma \displaybreak[3] \\
\partial_{t}\ave{\Sx^2} =& \left(e^{2 r} \Gamma +\gammacoll \right) \ave{\Sz}^{2} \nonumber \\
&+\left(\left(2 \Gamma +2 \gammacoll \right) \ave{\Sx^2}-\frac{\Gamma}{2}-\frac{\gammacoll}{2}\right) \ave{\Sz} \nonumber \\
 &+\left(-e^{2 r} \Gamma -\gammacoll -2 \gamma_{\phi} -\gammarel \right) \ave{\Sx^2}\nonumber \\
 &+\left(e^{2 r} \Gamma +\gammacoll \right) \ave{\CSzSz}+\frac{N \left(\gamma_{\phi} +\frac{\gammarel}{2}\right)}{2} \comma \displaybreak[3] \\
\partial_{t}\ave{\Sy^2} =& \left(e^{-2 r} \Gamma + \gammacoll \right) \ave{\Sz}^{2}  \nonumber \\ 
&+\left(\left(2 \Gamma +2 \gammacoll \right) \ave{\Sy^2}-\frac{\Gamma}{2}-\frac{\gammacoll}{2}\right) \ave{\Sz} \nonumber \\
&+\left(-e^{-2 r} \Gamma -\gammacoll -2 \gamma_{\phi} -\gammarel \right) \ave{\Sy^2} \nonumber \\
&+\left(e^{-2 r} \Gamma +\gammacoll \right) \ave{\CSzSz}+\frac{N \left(\gamma_{\phi} +\frac{\gammarel}{2}\right)}{2} \comma \displaybreak[3] \\
\partial_{t}\ave{\CSzSz} =& \left(\gammacoll +\Gamma +\gammarel \right) \ave{\Sz}+\left(e^{2 r} \Gamma +\gammacoll \right) \ave{\Sx^2} \nonumber \\
&+ \left(-e^{-2 r} \Gamma -e^{2 r} \Gamma -2 \gammacoll -2 \gammarel \right) \ave{\CSzSz}\nonumber \\
&+\left(e^{-2 r} \Gamma +\gammacoll \right) \ave{\Sy^2} +\frac{\gammarel N}{2},
\end{align}
where $\ave{\CSzSz}= \ave{\Sz^{2}} - \ave{\Sz} \ave{\Sz}$. 
We stress that if we assume that \cref{eq:realisticMasterEq2} is a result of coupling the spin system to a cavity interacting with an engineered squeezed reservoir with photon loss $\kappaint$, then we have
\begin{align}
    \Gamma&= \frac{4g^{2}}{\left(\kappasqz + \kappaint\right)^{2}}   \kappasqz \comma
    \label{eq:kappasqzApp}
\end{align}
and
\begin{align}
    \gammacoll&= \frac{4g^{2}}{\left(\kappasqz + \kappaint \right)^{2}}   \kappaint \comma
    \label{eq:gammacollApp}
\end{align}
as discussed in the main text and shown in detail in \cref{sec:App:AdiabaticElimination}.

%%%%%%%%%%%%%%%%%%%%%%%%%%%%%%%%%%%%%%%%%%%%%%%%%%%
%%%%%%%%%%%%%%%%%%%%%%%%%%%%%%%%%%%%%%%%%%%%%%%%%%%
%%%%%%%%%%%%%%%%%%%%%%%%%%%%%%%%%%%%%%%%%%%%%%%%%%%

\section{Cooperativity scaling of the $\wineland$ parameter}
\label{sec:App:coopScalingDerivation}

In this Appendix, we provide a derivation of the cooperativity scaling of the Wineland parameter $\wineland$.
We concentrate our analysis on the weak dephasing limit, and start with the case where $\gammaphi=0$ and where only the local decay $\gammarel$ as well as the collective cavity-induced decay $\gammacoll$ are present. 
A scenario where local spin dephasing is dominant can lead to substantially altered behavior of the system, and is the subject of \cref{sec:slowTimescale,sec:App:LiouvillianPerturbationTheorySlowTimescale}.

\subsection{Analytical derivation}

We begin by linearizing the mean-field-theory equations of motion shown in \cref{sec:App:MeanFieldTheory} by focusing on the limit where $\ave{\Sz}$ stays fixed at $-N/2$. 
This approximation closely reflects the true system dynamics when the spin number $N$ is large and when the single cooperativities $\etaphi$ or $\etarel$ are not much larger than unity, resulting in effective spin squeezing that is far from the Heisenberg limit. 
Hence, taking $\ave{\Sz} = -N/2$ (\ie, spins keeping their polarization throughout the evolution and in the steady state), the Wineland parameter takes a simple form, 
\begin{align}
    \wineland =&  \frac{4}{N}\ave{\Sy^{2}}_{\rm ss} \comma
    \label{eq:winelandSimple1App}
\end{align}
which we can write using the results in \cref{sec:App:MeanFieldTheory} as 
\begin{align} 
    \xi_{R}^{2} &= \frac{\left(N +1\right) \gammacoll + \left(N e^{-2 r}+1\right) \Gamma + \gammarel}{\left( N + 1\right) \gammacoll +\left( N +  e^{-2 r}\right) \Gamma + \gammarel} \fullstop
\end{align}
Note that $\wineland$ gets smaller as $r$ increases and hence, in what follows, we will take the limit $r \rightarrow \infty $. 
It is worth pointing out, however, that choosing a finite $r$ which satisfies $\exp(-2r) \ll  1/\sqrt{\Crel}$ is sufficient to reproduce the scaling of $\wineland$ derived below. 
In the large-$r$ limit, we find
\begin{align} 
    \xi_{R}^{2} &= \frac{(N+1) \gammacoll +\Gamma +\gammarel}{(N+1) \gammacoll + \Gamma N +\gammarel} \fullstop
\end{align}
Next, we use \cref{eq:kappasqz,eq:gammacoll} of the main text to rewrite the above expression as
\begin{align} 
    \xi_{R}^{2} &= \frac{\frac{N+1}{N} \kappaint + \frac{\left( \kappasqz + \kappaint \right)^{2}}{4G^{2}}  \gammarel   + \frac{\kappasqz}{N}}{\frac{N+1}{N} \kappaint + \frac{\left( \kappasqz + \kappaint \right)^{2}}{4G^{2}}  \gammarel   + \kappasqz} \fullstop
    \label{eq:winelandSimple3App}
\end{align}
We consider a limit where $N\rightarrow \infty$, while $G=\sqrt{N}g$ stays fixed.
In such a case, the last term of the numerator can be dropped.
Here it is crucial to point out that in an experimental setting, one will typically not have much control over $\kappaint$ and $\gammarel$, while $\kappasqz$ can be tuned at will through appropriate reservoir engineering (see \cref{sec:Implementation}). 
Hence, it is important to understand what value of $\kappasqz$ should be chosen to maximize the amount squeezing that this protocol can achieve. 
At first glance, one might think that choosing $\kappasqz$ as large as possible (\ie, $\kappasqz \rightarrow \infty$) is ideal as that maximizes the amount of bosonic squeezing that the spin-coupled cavity experiences. 
From \cref{eq:kappasqz}, however, we see that such a choice will actually limit the value of $\Gamma$, which directly impacts the strength of squeezed-vacuum reservoir that the spins see [see \cref{eq:realisticMasterEq2} in the main text], resulting in the squeezing performance being strongly limited by the value of $\gammarel$. Hence, as we shall see shortly, the right thing to do is to still choose $\kappasqz \gg \kappaint, \gammarel$, but yet not too large, so that the $\Gamma$-controlled process is dominant over the local spin decay $\gammarel$. To see this explicitly, we minimize \cref{eq:winelandSimple3App} with respect to $\kappasqz$.
Assuming $N \gg 1$, this leads to
\begin{align} 
    \xi_{R}^{2} &\approx 2 \left(\sqrt{\frac{4G^{2}}{\kappaint \gammarel} + 1} + 1 \right)^{-1} \nonumber \\
    &=  \frac{2}{\sqrt{\Crel}} + \bigoh{\frac{1}{\Crel}} \comma
    \label{eq:scaling1App}
\end{align}
where, in the second line, we used \cref{eq:collCoop} of the main text to express the result in terms of the collective cooperativity $\Crel$ and then expanded in the limit of large $\Crel$. 
Our above expression shows the $\Crel^{-1/2}$ cooperativity scaling for the dissipative protocol, which outperforms the $\Crel^{-1/3}$ behavior of the OAT method \cite{lewis2018robust}, in the case where spin decay is the dominant local noise process. 
The optimal value of $\kappasqz$ that results in \cref{eq:scaling1App} reads
\begin{align}
    \kappasqz^{\text{opt}} =&   \left( \kappaint^{2} +  \frac{4G^{2} \kappaint}{\gammarel  } \right)^{\frac{1}{2}} \nonumber \\
    & \approx 2 G \sqrt{\frac{\kappaint}{\gammarel}} = \kappaint \sqrt{\Crel} \comma 
    \label{eq:optkappasqzApp}
\end{align}
which confirms the need for $\kappasqz \gg \kappaint$ (given large $\Crel$), while also showing that it should not be infinitely large. 
Finally, we stress that $\wineland$ is ultimately limited by $~\kappaint/(\kappaint + \kappasqz)$, hence it is crucial that an appropriate $\kappasqz$ can be realized in an experimental setting.

While the above result has been calculated in the limit where $\gammaphi=0$, a similar expression is valid when some local dephasing is present (\ie, $\gammaphi \ne 0$). 
In such a case, one can simply assume $\gammarel \rightarrow \gammarel + 2 \gammaphi$ in \cref{eq:winelandSimple3App}. 
As discussed in more detail in \cref{sec:slowTimescale} of the main text, however, this is only true when $\gammaphi$ is not too large, namely when $\gammaphi \lesssim N \gammarel $. 
Otherwise, a local dephasing process can have a significant impact on the evolution and therefore dramatically limit the steady-state performance of the protocol.

%%%%%%%%%%%%%%%%%%%%%
%%%%%%%%%%%%%%%%%%%%%

\subsection{Mean-field theory simulations}
\label{sec:App:coopScalingSimsMFT}

In this section, we present mean-field-theory simulations of the dissipative protocol obtained using the full (nonlinear) equations shown in \cref{sec:App:MeanFieldTheory}. 
We consider the case where local spin decay dominates over local spin dephasing, and in particular work in the limit of $\gammaphi=0$. 

\begin{figure}[t]
	\centering
	\includegraphics[width=0.45\textwidth]{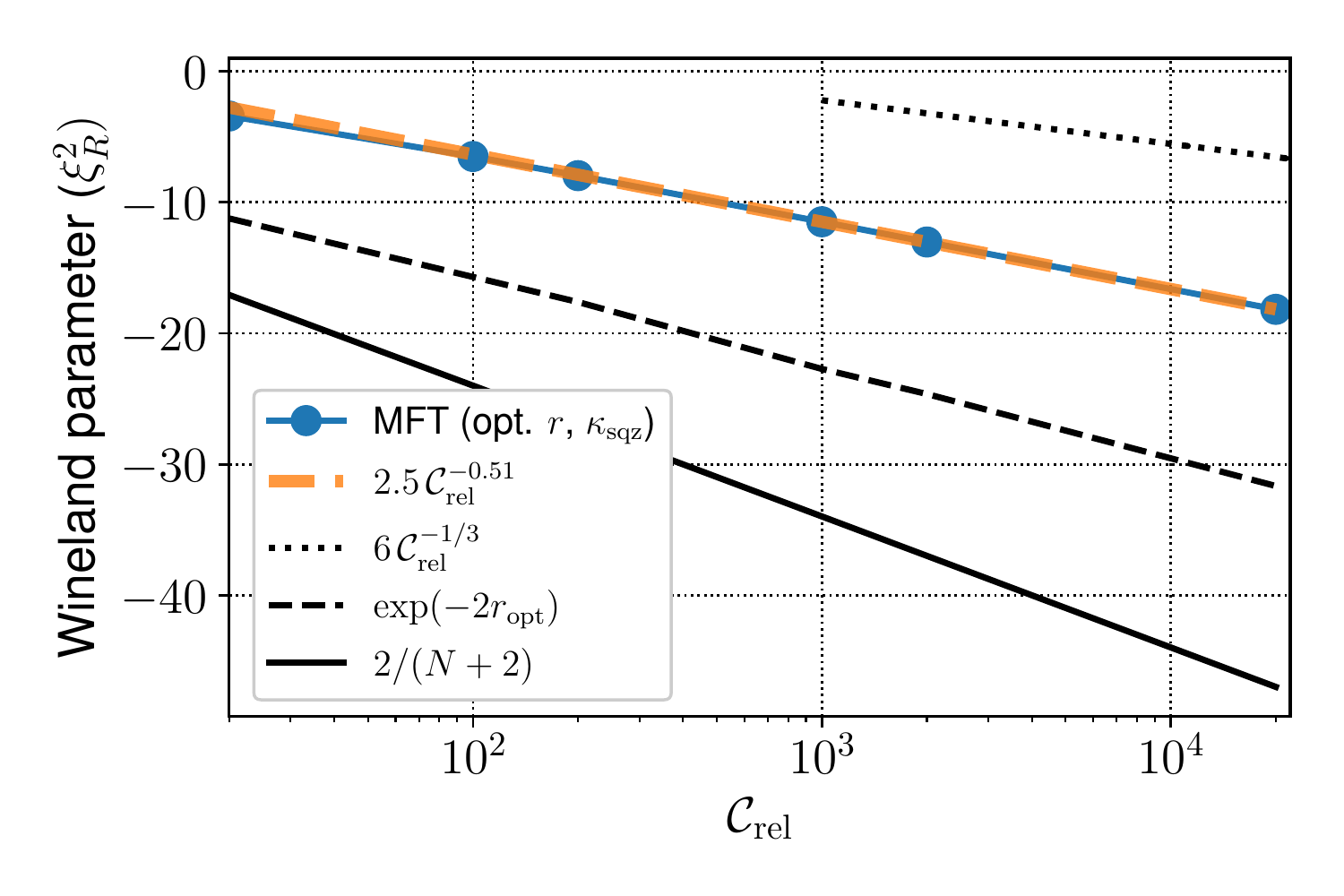} 
    \caption{
        Scaling of the Wineland parameter $\wineland$ as a function of collective cooperativity $\Crel$.   
        The blue curve corresponds to $\wineland$ calculated by evolving the full (nonlinear) mean-field equations of motion for the dissipative system (see \cref{sec:App:MeanFieldTheory}). 
        Here $\kappaint=500g$, $\gammarel=0.04g$, giving $\etarel=0.2$. The number $N$ of spins is changed in order to vary $\Crel$.
        At each blue point, both $r$ and $\kappasqz$ are optimized in order to minimize $\wineland$. The orange dashed curve shows the corresponding fit (calculated over the three data points with the largest $\Crel$). 
        The black dashed line
        describes the optimized squeezing of the engineered bosonic reservoir. 
        The solid black line shows an ideal Heisenberg scaling $2/(N+2)$.
        Finally, the black dotted curve shows the OAT scaling as calculated in \cite{lewis2018robust}. 
	}
	\label{fig:scalingMFT}
\end{figure}

The plot in \cref{fig:scalingMFT}, shows the scaling of the Wineland parameter as a function of the collective cooperativity $\Crel$. 
The parameters are $\kappaint=500g$, $\gammarel=0.04g$, giving $\etarel=0.2$, while the number $N$ of spins is varied in order to modify $\Crel$.
At each blue point, both $r$ and $\kappasqz$ are optimized in order to minimize $\wineland$. 
The orange curve shows the corresponding fit, which is calculated using the three data points with largest $\Crel$. 
We see good agreement with the cooperativity scaling discussed in the main text and derived in detail in the section above (where we have linearized the equations of motion). 
For comparison, the black dashed line describes the optimized squeezing of the engineered bosonic reservoir. 
The black solid line shows an ideal Heisenberg scaling $2/(N+2)$. 
We also plot the dotted black curve, which corresponds to the squeezing one would get from the OAT protocol in the limit where $\gammarel$ dominates over $\gammaphi$ (in the large $\Crel$ limit)  -- see Ref.~\onlinecite{lewis2018robust}. 
The simulations confirm that the dissipative protocol can indeed outperform the OAT approach.   

%%%%%%%%%%%%%%%%%%%%%%%%%%%%%%%%%%%%%%%%%%%%%%%%%%%
%%%%%%%%%%%%%%%%%%%%%%%%%%%%%%%%%%%%%%%%%%%%%%%%%%%

%%%%%%%%%%%%%%%%%%%%%%%%%%%%%%%%%%%%%%%%%%%%%%%%%%%
%%%%%%%%%%%%%%%%%%%%%%%%%%%%%%%%%%%%%%%%%%%%%%%%%%%
%%%%%%%%%%%%%%%%%%%%%%%%%%%%%%%%%%%%%%%%%%%%%%%%%%%

\section{Even-odd effect}
\label{sec:App:EvenOddEffect}

In this Appendix, we briefly review previous results on the dissipative steady state of \cref{eq:masterEqIdeal1} in the main text and we derive \cref{eq:SteadyStateOddN} of the main text.
We then comment on variance detection measurements required to use the even-odd effect as a sensor, and we discuss the impact of local dissipation.

\subsection{Properties of the steady state}

Agarwal and Puri derived that the steady state of \cref{eq:masterEqIdeal1} is 
\begin{align}
	\hat{\rho}_\mathrm{ss}^{(j)} \propto \Sigmaop^{-1} ( \Sigmaop^\dagger )^{-1} = (\Sigmaop^\dagger \Sigmaop )^{-1} \comma
	\label{eq:App:SteadyStateAP}
\end{align}
if the Hermitian operator $\Sigmaop^\dagger \Sigmaop$ is invertible \cite{AgarwalPuri1990}. 
Using the non-unitary transformation $e^{\theta \hat{S}_z}$, where $\theta(r) = \ln \sqrt{\tanh(r)}$, one can express the jump operator $\Sigmaop(r)$ as
\begin{align}
	\Sigmaop(r) = - 2 i \sqrt{\sinh(r) \cosh(r)} e^{\theta(r) \hat{S}_z} \hat{S}_y e^{- \theta(r) \hat{S}_z} \fullstop
\end{align}
Therefore, the eigenstates of $\Sigmaop(r)$ are 
\begin{align}
	\ket{j,m(r)} \propto e^{\theta(r) \hat{S}_z} \ket{j,m}_y \comma
\end{align}
where $\ket{j,m}_y$ are the eigenstates of $\Sy$ corresponding to an eigenvalue $m \in \{-j, \dots, j\}$ \cite{AgarwalPuri1990}.

If $N$ is odd, both $\Sigmaop(r)$ and $\Sigmaop^\dagger(r)$ have only nonzero eigenvalues and $\Sigmaop(r)^\dagger \Sigmaop(r)$ is invertible. 
Defining the eigenstates of $\Sigmaop(r)^\dagger \Sigmaop(r)$, 
\begin{align}
	\Sigmaop(r)^\dagger \Sigmaop(r) \ket{\psi_k} = \lambda_k \ket{\psi_k} \comma
\end{align}
with positive eigenvalues $0 < \lambda_0 \leq \dots \leq \lambda_{2j}$,
one can evaluate \cref{eq:App:SteadyStateAP} and obtains 
\begin{align}
	\hat{\rho}_\mathrm{ss}^{(j)} = \frac{1}{\sum_{k=0}^{2j} \frac{1}{\lambda_k}} \sum_{k=0}^{2j} \frac{1}{\lambda_k} \ket{\psi_k}\bra{\psi_k} \comma
	\label{eq:MixedSteadyState}
\end{align}
which is expression~\eqref{eq:SteadyStateOddN} of the main text. This is the generic form for the steady state of a master equation with a single jump operator that has no zero eigenvalues \cite{Schirmer2010}.

If $N$ is even, $\Sigmaop(r)$ has a zero eigenvalue in each subspace of angular momentum $j$ and the associated eigenstates are the dark states
\begin{align}
	\psidk{j;r} \propto e^{\theta(r) \hat{S}_z} \ket{j,0}_y \comma
\end{align}
given by \cref{eq:darkState1} of the main text.
These dark states are zero eigenstates of $\Sigmaop(r)^\dagger \Sigmaop(r)$ too, therefore, $\Sigmaop(r)^\dagger \Sigmaop(r)$ is not invertible.
Informally speaking, if one inverted $\Sigmaop(r)^\dagger \Sigmaop(r)$ in the presence of a zero eigenvalue, the term in \cref{eq:MixedSteadyState} associated with the zero eigenvalue would diverge and only the dark state would contribute to $\hat{\rho}_\mathrm{ss}^{(j)}$ after normalization, $\hat{\rho}_\mathrm{ss}^{(j)}(r) = \psidk{j;r} \bra{\psi_\mathrm{dk}[j;r]}$.

%%%%%%%%%%%%%%%%%%%%%%%%%%%%%%%%%%%%%%%%%%%%%%%%%%%
\begin{figure}[t]
	\centering
    \includegraphics[width=0.48\textwidth]{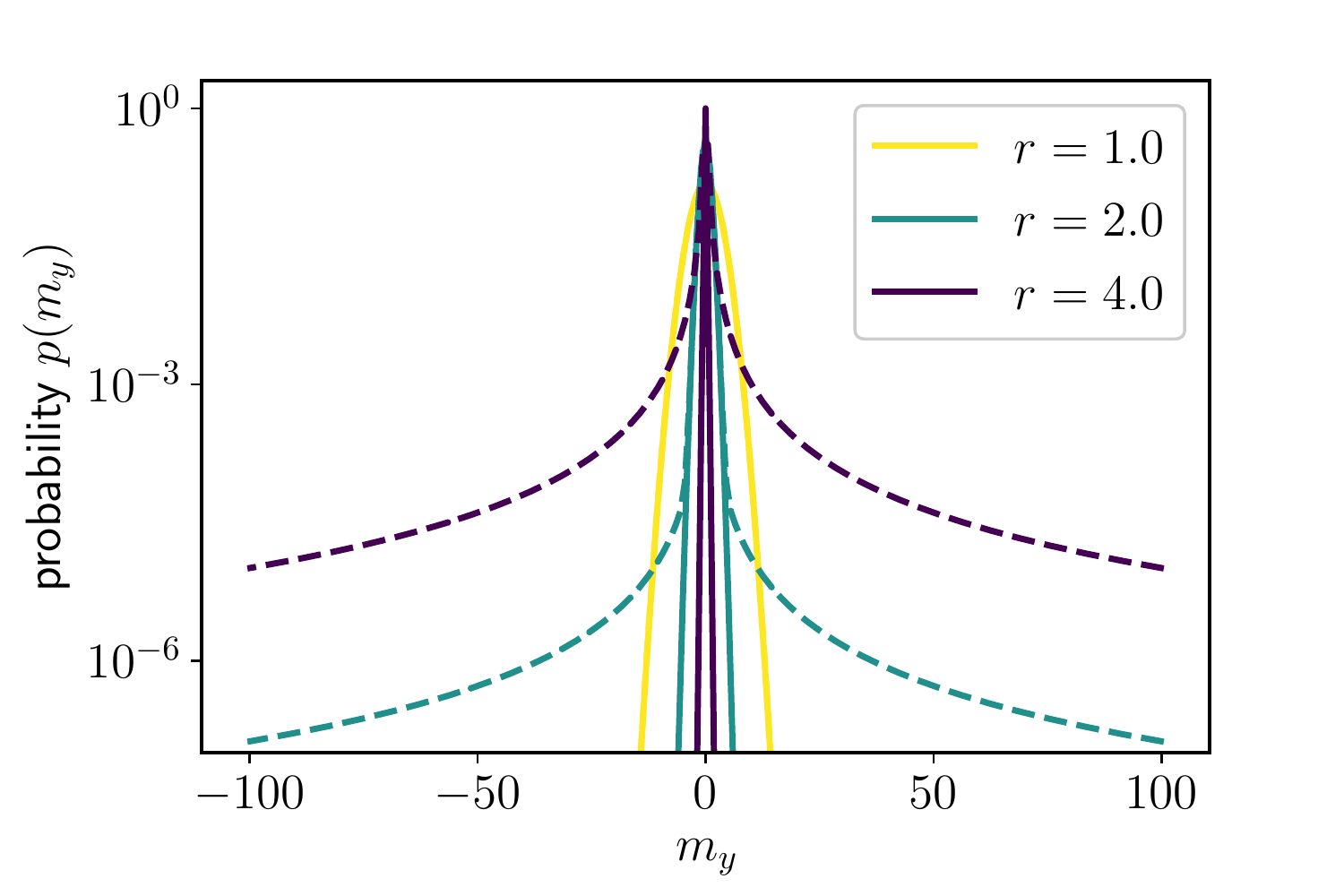}
	\caption{
		$\hat{S}_y$ probability distribution of the steady state of \cref{eq:masterEqIdeal1} for $N=\ValueevenOddSyProbabilityDistributionNeven$ (solid lines) vs.\ $N=\ValueevenOddSyProbabilityDistributionNodd$ (dashed lines). 
		The probability distributions for even and odd $N$ differ if the relation $e^{2r} \gtrsim N$ holds.
		In the limit $e^{2r} \gg N$, the odd-$N$ distribution develops a fat tail of large fluctuations.
	}
	\label{fig:EvenOdd6}
\end{figure}
%%%%%%%%%%%%%%%%%%%%%%%%%%%%%%%%%%%%%%%%%%%%%%%%%%%

\subsection{Using the even-odd effect for sensing}

As described in \cref{subsec:EvenOddSensing} of the main text, the sensitivity of the steady state on the parity of the number $N$ of spins can be used for sensing. 
Experimentally, sudden changes in the parity of $N$ can be induced by various mechanisms. 
Trapped atoms can be physically lost from the trap by collisions with background gas, internal collisions, and photon-assisted processes \cite{Grimm2000}.
If the spin-$1/2$ degree of freedom is a subspace of an atomic multi-level structure, undesired internal transitions can occur, which take the atom out of the spin-$1/2$ subspace and effectively remove it from the collective dynamics even though it may still be trapped \cite{dalla2013dissipative}. 
Moreover, one could devise a system where the coupling strength of a single spin to the cavity and, thus, to the collective spin depends on an external parameter. 
A change of this single-spin coupling strength modifies the number of collective spins, which is collectively amplified and yields a large change of the steady state.

Note that such effective atom loss events do not change the collective expectation value $\ave{\Sy} = 0$ of the distribution.
However, the statistics of the fluctuations $\ave{\Sy^2}$ depends on the parity, as shown in Fig.~\ref{fig:EvenOdd6}.
The parity of $N$ can thus be inferred by imposing a threshold condition on the variance $\ave{\Sy^2}$ measured using spin-noise spectroscopy \cite{Appel2009,Takano2009,SchleierSmith2010,Bohnet2016,Beguin2018,Guarrera2021,Luecke2014}.

%%%%%%%%%%%%%%%%%%%%%%%%%%%%%%%%%%%%%%%%%%%%%%%%%%%
\begin{figure}[t]
	\centering
    \includegraphics[width=0.48\textwidth]{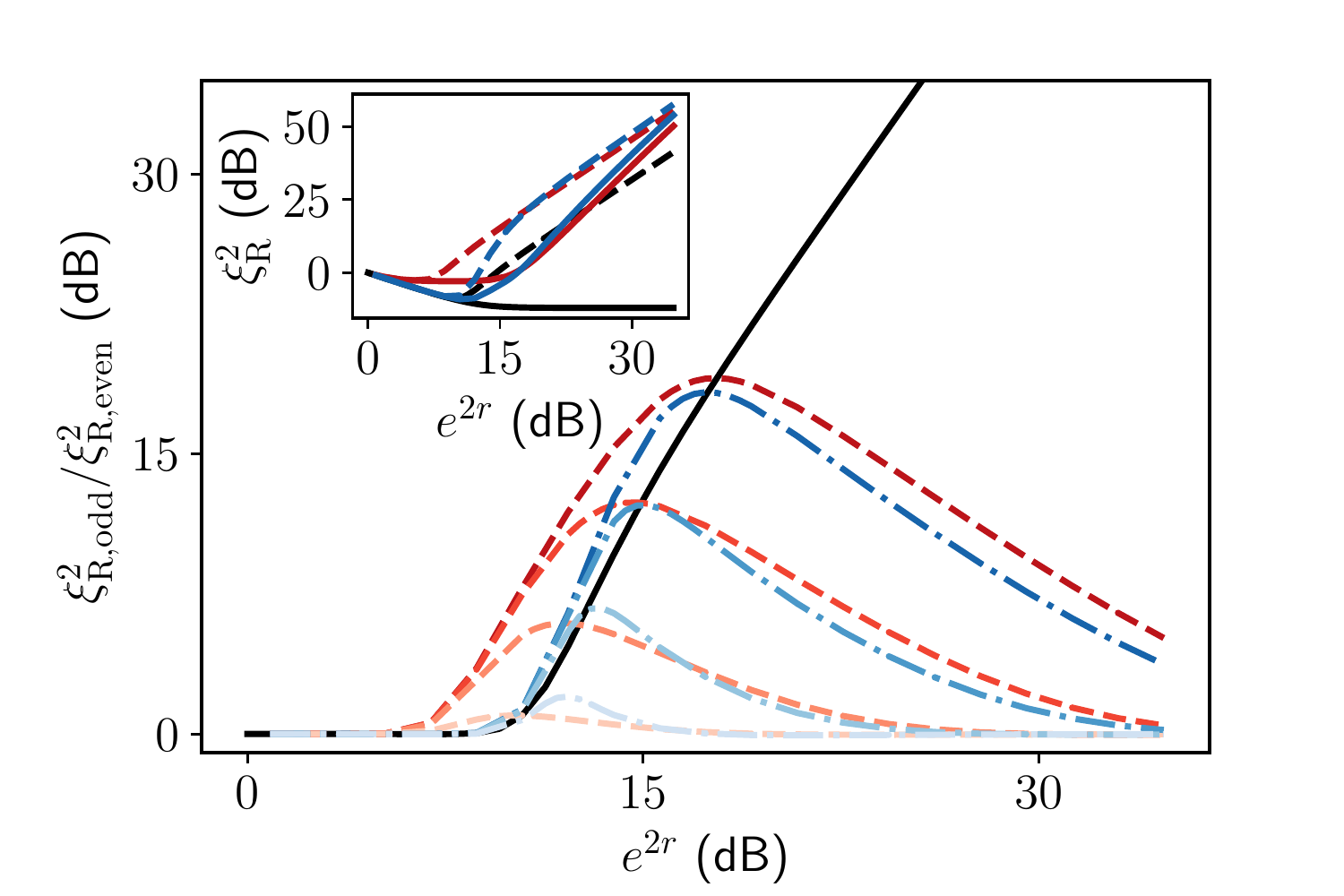}
	\caption{
		Ratio between the steady-state Wineland parameters for even and odd $N$ without local dissipation (solid black line), with local dephasing (dashed lines), and with local relaxation (dash-dotted lines), calculated using the quantum master equation \cref{eq:realisticMasterEq2} of the main text. 
		The parameters are from top to bottom: 
		$\gammaphi/\Gamma = (\ValueevenOddLocalDissipationgammaphiInseta, \ValueevenOddLocalDissipationgammaphiInsetb, \ValueevenOddLocalDissipationgammaphiInsetc, \ValueevenOddLocalDissipationgammaphiInsetd)$ 
		and $\gammarel/\Gamma = 0$ for the red curves 
		and $\gammarel/\Gamma = (\ValueevenOddLocalDissipationgammarelInseta, \ValueevenOddLocalDissipationgammarelInsetb, \ValueevenOddLocalDissipationgammarelInsetc, \ValueevenOddLocalDissipationgammarelInsetd)$ 
		and $\gammaphi/\Gamma = 0$ for the blue curves.
		\textbf{Inset}:
		Corresponding plot of the Wineland parameters for $N=\ValueevenOddLocalDissipationNeven$ (solid lines) vs.\ $N=\ValueevenOddLocalDissipationNodd$ (dashed lines) without local dissipation (black lines), with local dephasing (red lines, $\gammaphi/\Gamma=\ValueevenOddLocalDissipationgammaphi$, $\gammarel/\Gamma=0$), and with local relaxation (blue lines, $\gammaphi/\Gamma=0$, $\gammarel/\Gamma = \ValueevenOddLocalDissipationgammarel$). 
	}
	\label{fig:EvenOdd5}
\end{figure}
%%%%%%%%%%%%%%%%%%%%%%%%%%%%%%%%%%%%%%%%%%%%%%%%%%%

\subsection{Impact of local dissipation}

So far, our analysis of even-odd effects in the steady state has focused on the idealized case without any single-spin dissipation:  $\gammarel=\gammaphi=\gammacoll=0$.  We found that the Wineland parameters for even and odd $N$ differ strongly in the regime $e^{2r} \gtrsim N$, as shown in Fig.~\ref{fig:EvenOdd2} of the main text.
Figure~\ref{fig:EvenOdd5} shows that if local dissipation is taken into account, spin squeezing is reduced but the ratio between the Wineland parameters for even and odd $N$ remains large. 
Moreover, for a fixed value of the squeezing parameter $r$, the ratio of the Wineland parameters in the presence of local dephasing can even be larger than the corresponding ratio obtained for $\gammaphi=0$. 
For fixed local dissipation rates, the ratio is largest around the onset of the even-odd effect. 
At this optimum squeezing parameter $r_\mathrm{max}$, effective single-spin cooperativities much larger than unity, $\Gamma/\gammaphi \gg 1$ or $\Gamma/\gammarel \gg 1$, are required to observe a ratio of the Wineland parameters greater than two.

%%%%%%%%%%%%%%%%%%%%%%%%%%%%%%%%%%%%%%%%%%%%%%%%%%%
%%%%%%%%%%%%%%%%%%%%%%%%%%%%%%%%%%%%%%%%%%%%%%%%%%%
%%%%%%%%%%%%%%%%%%%%%%%%%%%%%%%%%%%%%%%%%%%%%%%%%%%

\section{Liouvillian perturbation theory of the slow timescale}
\label{sec:App:LiouvillianPerturbationTheorySlowTimescale}

In this Appendix, we use Liouvillian perturbation theory \cite{Li2014} to analyze the emergence of the long relaxation timescale in the presence of local dephasing, which has been discussed in Sec.~\ref{sec:slowTimescale} of the main text. 
We also provide a simple physical argument to understand this effect.

\subsection{Hilbert space of $N$ spin-$1/2$ systems and permutational invariance}
\label{sec:App:LiouvillianPerturbationTheorySlowTimescale:StructureHS}

Addition of angular momenta of $N$ spin-$1/2$ systems gives rise to $\lfloor N/2 \rfloor + 1$ subspaces of total angular momentum $j$, where $j$ takes values between $j_\mathrm{max}=N/2$ and $j_\mathrm{min}=0$ ($1/2$) if $N$ is even (odd) \cite{BrinkSatchler}. 
For $N > 2$, all but the maximum-angular-momentum subspace are degenerate since there are multiple ways to combine $N$ spin-$1/2$ systems to a total angular momentum $j < N/2$ \cite{Dicke1954} (for an illustration, see, \eg, Ref.~\onlinecite{Damanet2016}). 
If local dissipative processes act \emph{identically} on each spin-$1/2$ system, the equations of motion are invariant under permutation of the spins \cite{ChaseGeremia2008}. 
Consequently, if the system is initialized in a permutationally invariant state, \eg, any state in the subspace $j = j_\mathrm{max}$, the collective and local dissipative processes will preserve the permutational symmetry. 
Exploiting this symmetry, one can derive an effective quantum master equation which requires significantly less degrees of freedom to describe the system \cite{ChaseGeremia2008} and gives rise to efficient numerical simulation of large spin ensembles \cite{shammah2018open}.

\subsection{Analysis of the slow timescale}
\label{sec:App:LiouvillianPerturbationTheorySlowTimescale:Analysis}

Our starting point, the quantum master equation~\eqref{eq:realisticMasterEq2} of the main text, belongs to the class of permutationally invariant systems described above.
In the following, we focus on the case $\gammacoll = \gammarel = 0$.
Introducing the dimensionless time $\tau = \Gamma t$, the equation can be rewritten in the form $\mathrm{d}\hat{\rho}/\mathrm{d} \tau = \mathcal{L}_0 \hat{\rho} + \varepsilon \mathcal{L}_1 \hat{\rho}$, where we introduced the dimensionless superoperators 
\begin{align}
    \mathcal{L}_0 &= \mathcal{D}\left[ \Sigmaop(r) \right] \comma \\
    \mathcal{L}_1 &= \sum_{k=1}^N \mathcal{D} \left[ \frac{\sz^{(k)}}{2} \right] \comma
\end{align}
and the dimensionless perturbation strength $\varepsilon = 2 \gammaphi/\Gamma$.
In the absence of local dephasing, $\varepsilon = 0$, the superoperator $\mathcal{L}_0$ has $\lfloor N/2 \rfloor + 1$ different steady states $\hat{\rho}_0^{(j)}$, each of them living in a different subspace of collective angular momentum $j$.
Weak local dephasing, $\gammaphi \ll \Gamma$, enables incoherent transitions between adjacent angular-momentum subspaces \cite{ChaseGeremia2008}, 
which can be visualized as trajectory in a triangular $(j,m)$ state space \cite{Zhang2018}.
This perturbation lifts the degeneracy of the steady states and opens a new dissipative gap that determines the relaxation timescale towards the new, unique steady state.

\begin{figure}[t]
    \centering
    \includegraphics[width=0.5\textwidth]{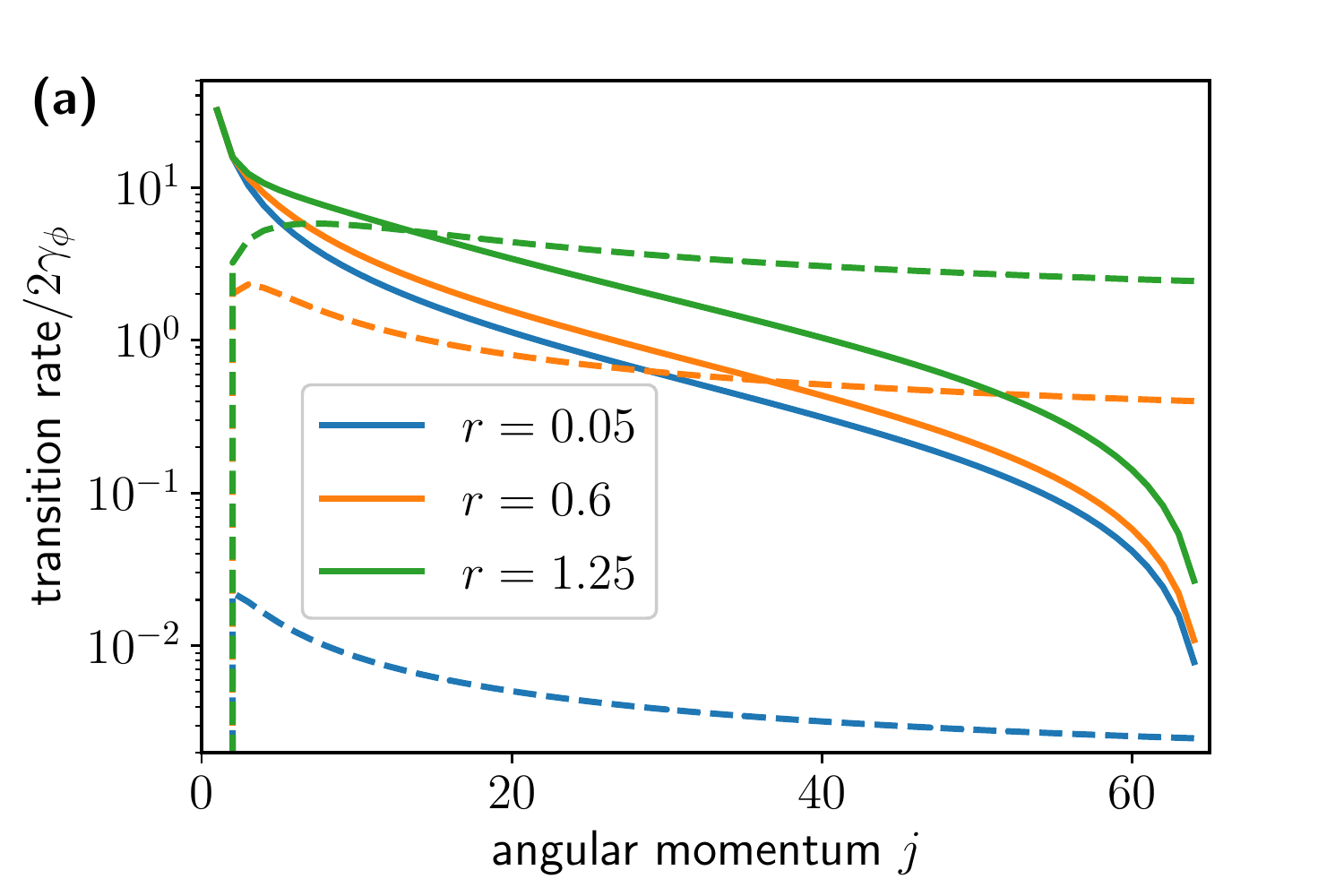}
    \includegraphics[width=0.5\textwidth]{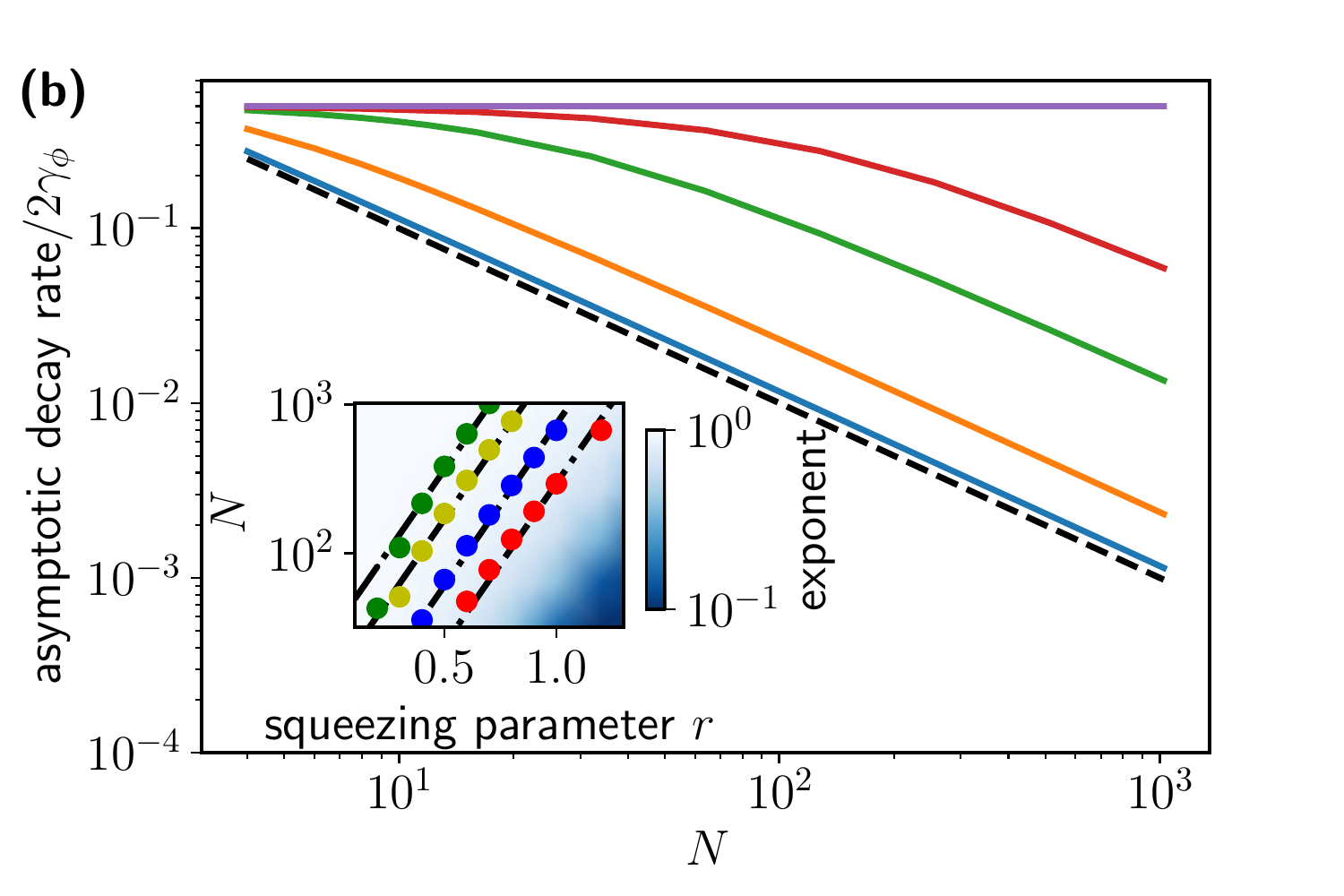}
    \caption{
    \textbf{(a)} Transition rates $j-1 \to j$ (solid) and $j \to j-1$ (dashed) between different angular momentum subspaces due to local dephasing for $N=\ValueslowRateTransitionRatesN$ and $\Gamma=1$. 
   	\textbf{(b)} Asymptotic decay rate in the presence of local dephasing for $r=\ValueslowRateDecayRatera$, $\ValueslowRateDecayRaterb$, $\ValueslowRateDecayRaterc$, $\ValueslowRateDecayRaterd$, and $\ValueslowRateDecayRatere$ (bottom to top). 
   		The dashed black line indicates $1/N$ scaling obtained for $r = 0$. 
	\textbf{Inset} Scaling exponent $a$ in $1/N^a$ as a function of $N$ and $r$.
		The data points indicate the positions where $a$ becomes smaller than $\ValueslowRateDecayRatecd$, $\ValueslowRateDecayRatecc$, $\ValueslowRateDecayRatecb$, and $\ValueslowRateDecayRateca$ (left to right). 
		The dash-dotted black lines are a guide to the eye and indicate $e^{5 r}$ scaling. 
    }
    \label{fig:slowTimeScale2}
\end{figure}

The first-order corrections to the vanishing eigenvalues of $\mathcal{L}_0$ are given by the eigenvalues of the tridiagonal matrix 
\begin{align}
	M_{j,j'} = \Tr[ \hat{\mathds{1}}^{(j')} \mathcal{L}_1 \hat{\rho}_0^{(j)}] 
\end{align}
containing the transition rates $j \to j'$ between collective angular momentum subspaces.
Here, $\hat{\mathds{1}}^{(j)}$ is the identity operator in the angular-momentum subspace $j$. 
For even $N$, the transition rates are shown in Fig.~\ref{fig:slowTimeScale2}(a). 
They depend on the structure of the dark state~\eqref{eq:darkState1} given in the main text,
\begin{align}
	\Gamma_{j \to j-1} &= \sum_{m=-j+1}^{j-1} \Gamma_{j;m,m}^{(5)} \abs{c_m^{(j)}}^2 \comma 
	\label{eq:App:LPT:transitionRate1} \\
	\Gamma_{j \to j+1} &= \sum_{m=-j}^j \Gamma_{j;m,m}^{(6)} \abs{c_m^{(j)}}^2 \comma
	\label{eq:App:LPT:transitionRate2}
\end{align}
where $c_m^{(j)}$ are the expansion coefficients of the dark state and $\Gamma_{j;m,m}^{(5,6)}$ are the transition rates derived in Ref.~\onlinecite{ChaseGeremia2008} using the notation introduced in Ref.~\onlinecite{shammah2018open}. 
Note that, in our case, $\mathcal{L}_1$ is dimensionless such that the transition rates~\eqref{eq:App:LPT:transitionRate1} and~\eqref{eq:App:LPT:transitionRate2} are dimensionless, too. 
The asymptotic decay rate, \ie, the absolute value of the smallest gap in the spectrum of $M$, is shown in Fig.~\ref{fig:slowTimeScale2}(b).

%%%%%%%%%%%%%%%%%%%%%%%%%%%%%%%%%%%%%%%%%%%%%%%%%%%
%%%%%%%%%%%%%%%%%%%%%%%%%%%%%%%%%%%%%%%%%%%%%%%%%%%

For $r = 0$, the unperturbed steady states are the ground states of each angular-momentum subspace, $\hat{\rho}_0^{(j)} = \ket{j,-j}\bra{j,-j}$. 
Therefore, transitions are only possible towards subspaces of larger angular momentum, $\Gamma_{j \to j-1} = 0$, and the relaxation dynamics is dominated by the bottleneck of the smallest nonzero transition rate, $\Gamma_{N/2-1,-N/2+1 \to N/2,-N/2+1} = 1/N$.

For $r \neq 0$, transition rates $\Gamma_{j \to j-1}$ are nonzero and dominate over the rate $\Gamma_{N/2-1 \to N/2}$ if the condition $r > 1/\sqrt{N}$ holds. 
As a consequence, an initial state in the maximum-angular-momentum subspace $j=N/2$ will undergo a directed hopping process towards lower angular momentum subspaces until it reaches a subspace $j_0$ where ``downward'' and ``upward'' rates are balanced, $\Gamma_{j_0 \to j_0-1} \approx \Gamma_{j_0-1 \to j_0}$. 
Note that the downward rates $\Gamma_{j \to j-1}$ are almost constant as a function of $j$ whereas the upward rates $\Gamma_{j \to j+1}$ depend strongly on $j$, as shown in \cref{fig:slowTimeScale2}(a). 
The asymptotic decay rate towards the steady state is proportional to $1/N$ if the transition rates in the vicinity of the equilibrium point $j_0$ scale proportional to $1/N$. 
Inspection of the rates $\Gamma_{j;m,m}^{(5,6)}$ listed in Ref.~\onlinecite{shammah2018open} shows that this is the case if $j \gg N/2$ and $m \ll 0$. 
For a given squeezing parameter $r$, these conditions can be fulfilled if $N$ is sufficiently large, 
\begin{align}
	N \gg e^{a r} \comma
\end{align}
as shown in Fig.~\ref{fig:slowTimeScale2}(b). 
Numerically, we find an exponent $a \approx 5$, see inset of Fig.~\ref{fig:slowTimeScale2}(b).

In the limit $r \to \infty$, the asymptotic decay rate converges to the constant value $1/2$.

\subsection{Physical argument for the slow timescale}
\label{sec:App:LiouvillianPerturbationTheorySlowTimescale:Argument}

The existence of a bottleneck relaxation rate causing a $1/N$ scaling of the asymptotic decay rate for local dephasing can be understood by an intuitive argument. 
To explain it, we focus on the transition rate $\Gamma_{N/2-1,-N/2+1 \to N/2,-N/2+1}$, which is the bottleneck determining the asymptotic decay rate in the limit $r < 1/\sqrt{N}$.
The states that are involved in this transition can be parametrized as  \cite{Bacon2006,Vetter2016}
\begin{align}
	\overline{\ket{p}} = \frac{1}{\sqrt{N}} \sum_{j=1}^N e^{2 \pi i j p/N} \ket{j} \comma
	\label{eq:ExplicitFormulaDegenerateStates}
\end{align} 
where $p \in \{0, \dots, N-1\}$.
Here, $\ket{j}$ denotes the $N$-particle state where the $j$th spin is in the excited state and all others are in the ground state. 
The $p=0$ state has total angular momentum $j=N/2$, \ie, we can identify it with the state
\begin{align}
    \overline{\ket{0}} &\equiv \ket{N/2,-N/2+1} 
\end{align}
in the maximum-angular-momentum subspace.
In contrast, the $N-1$ states with $p>0$ have total angular momentum $j=N/2-1$.
Therefore, the index $p>0$ allows us to label the $N-1$ degenerate states in the $j=N/2-1$ subspace,
\begin{align}
	\overline{\ket{p}} &\equiv \ket{N/2-1,-N/2+1,p} \text{ for } p>0 \fullstop
\end{align}
Local dephasing of spin $n$ changes one sign in the superposition~\eqref{eq:ExplicitFormulaDegenerateStates},
\begin{align}
	\frac{1}{2} \sz^{(n)} \overline{\ket{p}} = - \frac{1}{2} \overline{\ket{p}} + \frac{1}{\sqrt{N}} e^{2 \pi i n p/N} \ket{n} \comma
\end{align}
and thus creates an overlap between the orthogonal states $\overline{\ket{0}}$ and $\overline{\ket{p>0}}$ that is proportional to $1/N$, 
\begin{align}
	V_{0,p}^{(n)} = \overline{\bra{0}} \frac{1}{2} \sz^{(n)} \overline{\ket{p>0}} = \frac{1}{N} e^{2 \pi i n p/N} \fullstop
	\label{eq:App:OverlapV0p}
\end{align}
For identical dephasing processes on all $N$ spins and for a collective initial state, \ie, a uniform statistical mixture of all $N-1$ states $\ket{N/2-1,-N/2+1,p}$, the total upward transition rate between the two collective angular momentum subspaces is 
\begin{align}
	&\Gamma_{N/2-1,-N/2+1 \to N/2,-N/2+1} \nonumber \\
	&= \sum_{p=1}^{N-1} \frac{1}{N-1} \sum_{n=1}^N \abs{V_{0,p}^{(n)}}^2 
	= \frac{1}{N} \comma
	\label{eq:SlowRateScaling}
\end{align}
which is the bottleneck of the relaxation process and features the $1/N$ scaling with system size.
Note that the corresponding downward rate is of the order of unity because we have to sum over all $N-1$ possible target states $\overline{\ket{p>0}}$, too,
\begin{align}
	&\Gamma_{N/2,-N/2+1 \to N/2-1,-N/2+1} \nonumber \\
	&= \sum_{p=1}^{N-1} \sum_{n=1}^N \abs{V_{0,p}^{(n)}}^2 
	= \frac{N-1}{N} \fullstop
\end{align}
Also note that local relaxation does \emph{not} lead to a similar emergence of the slow timescale because the overlap corresponding to \cref{eq:App:OverlapV0p} will only be proportional to $ 1/\sqrt{N}$ and is thus canceled by the summation performed in \cref{eq:SlowRateScaling}.

%%%%%%%%%%%%%%%%%%%%%%%%%%%%%%%%%%%%%%%%%%%%%%%%%%%
%%%%%%%%%%%%%%%%%%%%%%%%%%%%%%%%%%%%%%%%%%%%%%%%%%%
%%%%%%%%%%%%%%%%%%%%%%%%%%%%%%%%%%%%%%%%%%%%%%%%%%%

\section{Optimal parameters in master equations}
\label{sec:App:optParamsME}

%%%%%%%%%%%%%%%%%%%%%%%%%%%%%%%%%%%%%%%%%%%%%%%%%%%
\begin{figure*}[th]
	\centering
    \hspace{-0.2cm}
    \includegraphics[width=0.33\textwidth]{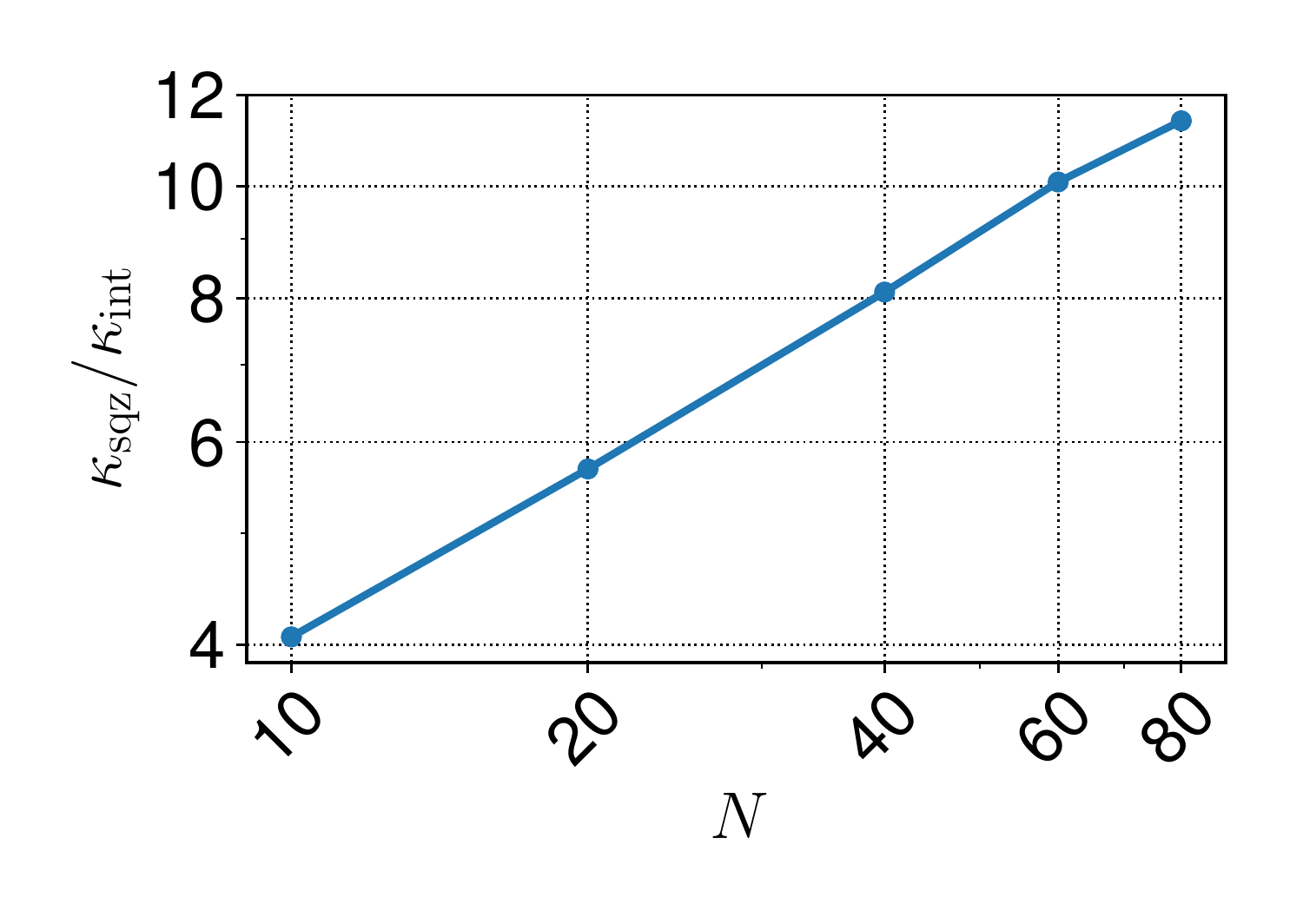}
    \hspace{-0.3cm}
    \includegraphics[width=0.33\textwidth]{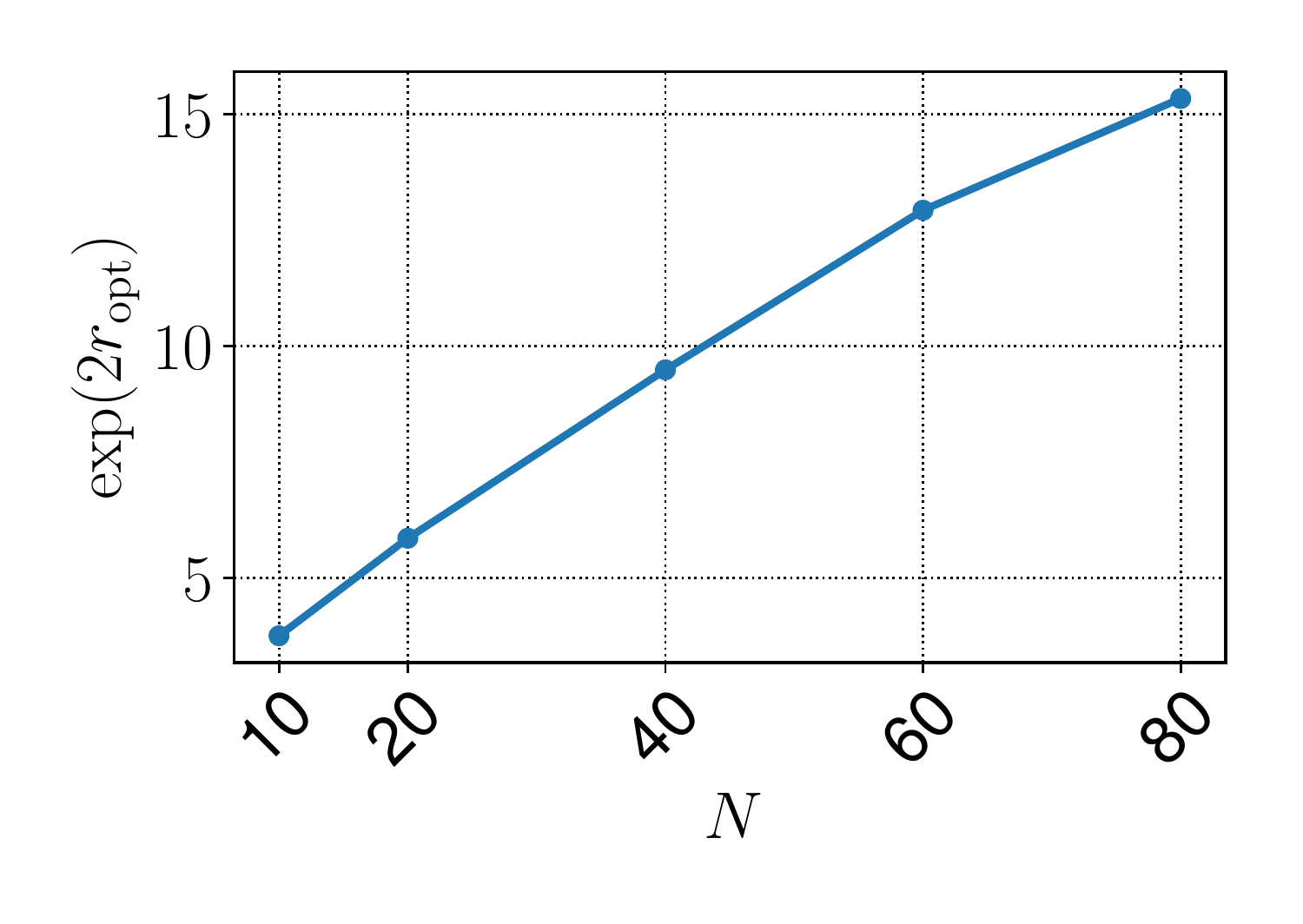}
    \hspace{-0.3cm}
    \includegraphics[width=0.33\textwidth]{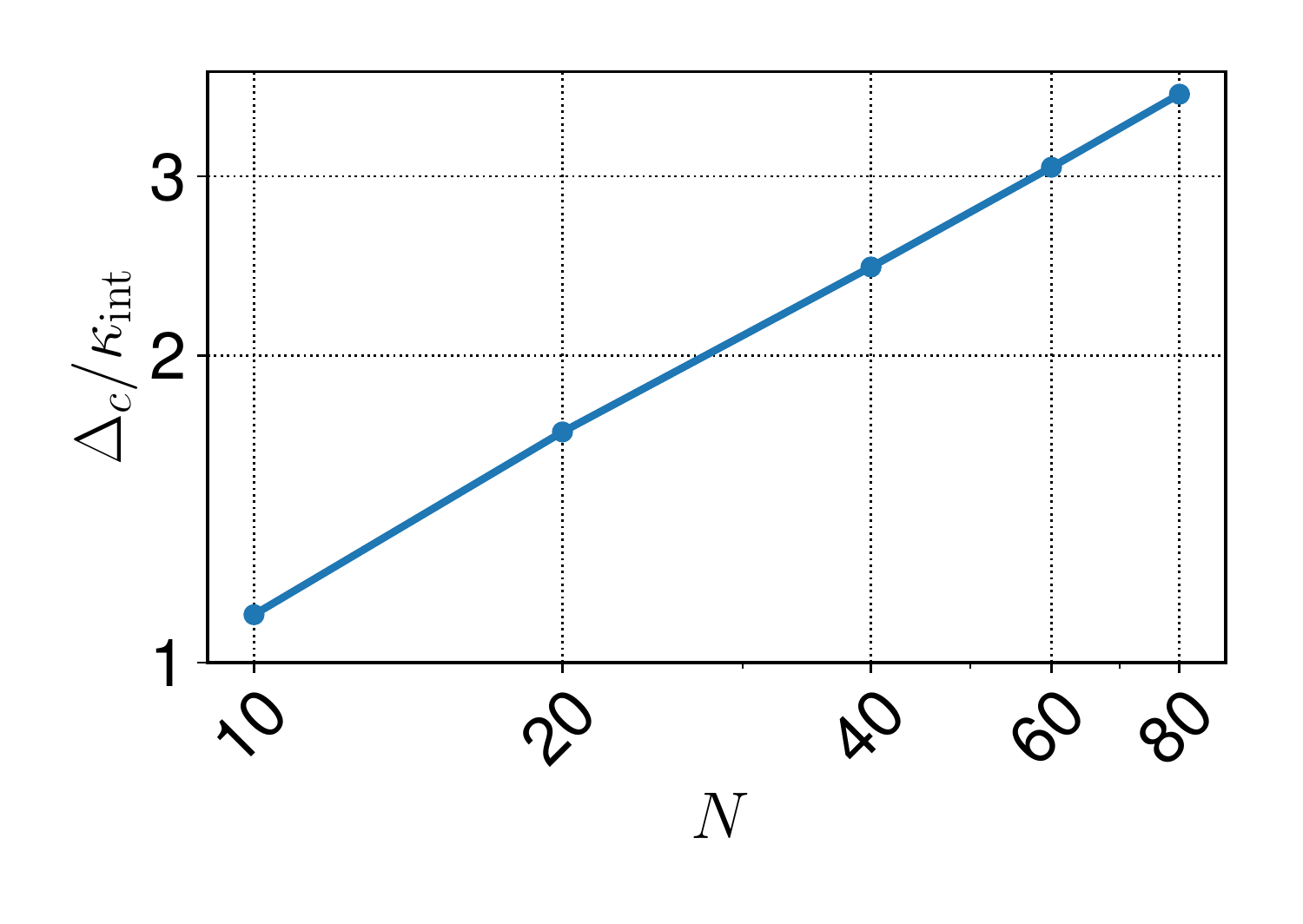}
    \caption{
        Values of optimal parameters obtained from master equation simulations of both the dissipative as well as the OAT protocol, which were used to generate \cref{fig:meOATvsDissip} of the main text. 
        Left and center panels show the optimal values of $\kappaint$ and $r$ [plot shows $\exp(2r)$] that resulted from the dissipative protocol simulations. 
        The right panel shows the optimal spin-cavity detuning that was used in for the OAT protocol. See \cref{fig:meOATvsDissip} for details about the rest of the parameters that were used. 
    } 
	\label{fig:optimalParamsME}
\end{figure*}
%%%%%%%%%%%%%%%%%%%%%%%%%%%%%%%%%%%%%%%%%%%%%%%%%%%

In this section, we show how the optimal protocol parameters vary as a function of increasing system size $N$ in simulations from \cref{fig:meOATvsDissip} of the main text. 
In the case of the dissipative protocols, the optimization included varying both $r$ as well as $\kappasqz$, whereas in the case of OAT, the spin-cavity detuning $\Delta_{c}$ (see \cref{sec:App:oat}) is was varied. 
The results are shown in \cref{fig:optimalParamsME}.

%%%%%%%%%%%%%%%%%%%%%%%%%%%%%%%%%%%%%%%%%%%%%%%%%%%
%%%%%%%%%%%%%%%%%%%%%%%%%%%%%%%%%%%%%%%%%%%%%%%%%%%
%%%%%%%%%%%%%%%%%%%%%%%%%%%%%%%%%%%%%%%%%%%%%%%%%%%

\section{Effective One-Axis-Twist quantum master equation}
\label{sec:App:oat}

In this Appendix, we present the effective model that we consider when discussing the OAT protocol both in the main text and in \cref{sec:App:coopScalingSimsMFT}. 
In particular, following \cite{lewis2018robust,bennett2013phonon}, we envision an ensemble of spins dispersively coupled to a bosonic cavity. 
After adiabatically eliminating the cavity, the spin-only quantum master equation can be approximated by \cite{lewis2018robust}
\begin{align}
    \dot\rhoop =& -i \comm{\chi  \left(\vec{\hat{S}}^{2} - \Sz^{2} \right)}{\rhoop}   
    + \frac{ \kappaint  g^{2}}{\Delta_{c}^{2} +  \left(\frac{\kappaint}{2}\right)^{2}}  \mD{\Sm}{\rhoop} \nonumber  \\
    &+ \gammarel \sum_{k} \mD{\smm^{(k)}}{\rhoop} \comma 
    \label{eq:effMasterEquationOAT}
\end{align}
with
\begin{align}
    \chi &= \frac{g^{2} \Delta_{c}} {\Delta_{c}^{2} 
    + \left(\frac{\kappaint}{2}\right)^{2}}
    \label{eq:chiOAT} \comma 
\end{align}
and with $\Delta_{c}$ representing the cavity-spin detuning, $g$ the cavity-spin coupling strength, $\kappaint$ the decay rate of the cavity, and $\gammarel$ the local spin decay.
We point out that we assume in the simulations that $\Delta_{c}$ is a tunable parameter, over which we optimize in order to maximize the amount of spin squeezing that the protocol can achieve.

%%%%%%%%%%%%%%%%%%%%%%%%%%%%%%%%%%%%%%%%%%%%%%%%%%%
%%%%%%%%%%%%%%%%%%%%%%%%%%%%%%%%%%%%%%%%%%%%%%%%%%%
%%%%%%%%%%%%%%%%%%%%%%%%%%%%%%%%%%%%%%%%%%%%%%%%%%%

\section{Scaling of the Wineland parameter $\wineland$ in the limit $\etarel \rightarrow \infty $}
\label{sec:App:directSqueezedDriving}

When analyzing the performance of the dissipative spin-squeezing protocol in the main text, as one means of implementation, we envisioned engineering the required dissipator by coupling a spin ensemble to a lossy cavity that in turn interacts with an appropriately engineered squeezed bath. 
Furthermore, in our cooperativity scaling analysis (see \cref{sec:scaling,sec:App:coopScalingDerivation}) we investigated the limit of weak single spin cooperativity $\etarel \leq 1 $.
It is also interesting to consider a different asymptotic regime, where the internal cavity loss $\kappaint$ is negligible, giving an extremely large $\etarel$.  We focus on the specific case where $\kappaint=0$, and the only undesired dynamics is due to single-spin relaxation at a rate $\gammarel$.  Such a situation could be realized without any cavity, by directly irradiating an ensemble of two-level atoms with squeezed light.  While this situation was analyzed in 
Refs.~\cite{AgarwalPuri1990,Kuzmich1997} the impacts of single spin relaxation were not studied.

The master equation in our chosen limit is thus
\begin{align}
    \dot\rho =&   \Gamma \mD{\Sigmaop[r]}{\rhoop} + \gammarel \sum_k \mD{\smm^{(k)}}{\rhoop} \fullstop
    \label{eq:directMasterEq1}
\end{align}
The key dimensionless parameter that describes the competition of the desired collective dissipative dynamics and the unwanted relaxation is
\begin{align}
    \etatilde &= \frac{\Gamma}{\gammarel} \fullstop
\end{align}
Once again concentrating our attention on the large-$N$ limit and fixing $\ave{\Sz} = -N/2$, we can approximate the Wineland parameter using the mean-field equations of \cref{sec:App:MeanFieldTheory} as
\begin{align} 
    \xi_{R}^{2} \approx & \frac{\etatilde + 1}{ N \etatilde + 1} \fullstop
    \label{eq:winelandKappaintZero}
\end{align}
In the above expression we have already taken the limit $r \rightarrow \infty$, which minimizes $\wineland$. 
As one would expect, achievable squeezing increases as $\etatilde$ gets larger, but more importantly we have that $\wineland \propto 1/N$. We can also define a quantity analogous to a collective cooperativity in this simplified system, 
\begin{align}
    \Ctilde &= N \etatilde \comma 
\end{align}
which then lets us write 
\begin{align} 
    \xi_{R}^{2} \propto &  \frac{1}{\Ctilde} \comma 
    \label{eq:winelandKappaintZero2}
\end{align}
assuming $\etatilde \leq 1$ and $\Ctilde \gg 1$.

%%%%%%%%%%%%%%%%%%%%%%%%%%%%%%%%%%%%%%%%%%%%%%%%%%%
%%%%%%%%%%%%%%%%%%%%%%%%%%%%%%%%%%%%%%%%%%%%%%%%%%%
%%%%%%%%%%%%%%%%%%%%%%%%%%%%%%%%%%%%%%%%%%%%%%%%%%%

\section{Modeling experimental imperfections as an effective-temperature squeezed reservoir}
\label{sec:Appendix:FiniteTemperatureModel}
In this Appendix, we show that the generalized model~\eqref{eq:masterEqFiniteTempForSigma}, describing an engineered reservoir that stabilizes an impure steady state, can capture the impact of collective excitation and relaxation if the squeezing parameter $r$ and the effective thermal occupation number $n_\mathrm{th}$ are adjusted properly. 
To proof this, we start with the fully general quantum master equation
\begin{align}
	\dot{\rhoop} &= \Gamma \mD{\Sigmaop}{\rhoop} + \Gamma' \mD{\Sigmaop^\dagger}{\rhoop} \nonumber \\
	&+ \gamma \mD{\Sm}{\rhoop} + \gamma' \mD{\Sp}{\rhoop} \comma
	\label{eq:App:FiniteTemp:QMEStart}
\end{align}
which can model, \eg, interaction with an impure squeezing reservoir and a finite-temperature collective-decay reservoir if the ratios $\Gamma/\Gamma'$ and $\gamma/\gamma'$ are chosen properly. 
Equation~\eqref{eq:App:FiniteTemp:QMEStart} is equivalent to a quantum master equation of the form given in Eq.~\eqref{eq:masterEqFiniteTempForSigma}, 
\begin{align}
	\dot{\rhoop} &= \tilde{\Gamma} \mD{\tilde{\Sigma}}{\rhoop} + \tilde{\Gamma}' \mD{\tilde{\Sigma}^\dagger}{\rhoop} \comma \\
	\tilde{\Sigma} &= \cosh(\tilde{r}) \hat{S}_- - \sinh(\tilde{r}) \hat{S}_+ \comma
\end{align}
where we defined a new squeezing parameter $\tilde{r}$ and new decay rates $\tilde{\Gamma}$, $\tilde{\Gamma}'$ as follows:
\begin{align}
	\frac{1}{\tanh (2 \tilde{r})} 
		&= \frac{1}{\tanh(2r)} + \frac{\gamma + \gamma'}{\Gamma + \Gamma'} \frac{1}{\sinh(2r)} \comma 
		\label{eq:App:FiniteTemp:rtilde}\\
	\tilde{\Gamma} 
		&= \frac{(\Gamma + \Gamma') \cosh(2r) + \gamma + \gamma'}{2 \cosh(2 \tilde{r})} \nonumber \\
		&+ \frac{\Gamma - \Gamma' + \gamma - \gamma'}{2} \comma 
		\label{eq:App:FiniteTemp:Gammatilde} \\
	\tilde{\Gamma}' &= \tilde{\Gamma} - \Gamma + \Gamma' - \gamma + \gamma' \fullstop
		\label{eq:App:FiniteTemp:Gammaptilde} 
\end{align}
Condition~\eqref{eq:App:FiniteTemp:rtilde} can be satisfied for arbitrary nonnegative rates $\Gamma$, $\Gamma'$, $\gamma$, and $\gamma'$. 
Collective excitation or decay, $\gamma' \neq 0$ or $\gamma \neq 0$, respectively, will decrease the squeezing parameter, \ie, we always have $\tilde{r} \leq r$. 
Sufficient (but not necessary) conditions to obtain nonnegative decay rates $\tilde{\Gamma}$ and $\tilde{\Gamma}'$ are
\begin{align}
	\Gamma + \frac{\gamma - \gamma'}{2} \geq 0 \comma \\
	\Gamma' + \frac{\gamma' - \gamma}{2} \geq 0 \fullstop
\end{align}

Note that these conditions are satisfied if $\gamma = \gamma'$, which is the case for the trapped-ion implementation dicussed in Sec.~\ref{sec:Implementation}. 
There, we have $\Gamma' = 0$ and $\gamma = \gamma' = \gammaheat \ll \Gamma$ (see Sec.~\ref{sec:Implementation}).
Expanding the general results~\eqref{eq:App:FiniteTemp:rtilde} to~\eqref{eq:App:FiniteTemp:Gammaptilde} in the small parameter $\gammaheat/\Gamma$, we find
\begin{align}
	\tilde{r} &\approx r - \sinh(2r) \frac{\gammaheat}{\Gamma}  \comma \\
	\tilde{\Gamma} &\approx \Gamma \left( 1 + \cosh(2r) \frac{\gammaheat}{\Gamma} \right) \comma \\
	\tilde{\Gamma}' &\approx \Gamma \cosh(2r) \frac{\gammaheat}{\Gamma} \fullstop
\end{align}
This corresponds to an impure squeezed reservoir with reduced squeezing parameter $\tilde{r} < r$, effective thermal occupation number $n_\mathrm{th} = \cosh(2r) \gammaheat/\Gamma$, and unchanged decay rate $\Gamma$, as shown in Eqs.~\eqref{eqn:FiniteTemp:TrappedIon:effectiver} and~\eqref{eqn:FiniteTemp:TrappedIon:effectiventh}. 

Similarly, for a perfect squeezing reservoir and small zero-temperature collective decay, \ie, $\Gamma' = \gamma' = 0$ and $\gamma = \gammacoll \ll \Gamma$, one finds 
\begin{align}
	\tilde{r} &\approx r - \frac{1}{2} \sinh(2r) \frac{\gammacoll}{\Gamma} \comma \\
	\tilde{\Gamma} &\approx \Gamma \left( 1 + \frac{\gammacoll}{\Gamma} \right) \left( 1 + n_\mathrm{th} \right) \comma \\
	\tilde{\Gamma}' &\approx \Gamma \left( 1 + \frac{\gammacoll}{\Gamma} \right) n_\mathrm{th} \comma \\
	n_\mathrm{th} &\approx \frac{\sinh^2(r)}{1 + \frac{\gammacoll}{\Gamma}} \frac{\gammacoll}{\Gamma} \comma
\end{align}
\ie, the presence of collective decay can be understood in terms of an impure squeezing reservoir with decreased squeezing parameter $\tilde{r} < r$, effective temperature $n_\mathrm{th}$, and an enhanced decay rate $\Gamma + \gammacoll$.

\section{Optimal squeezing parameter for an impure squeezed reservoir}
\label{sec:Appendix:r_opt}

%%%%%%%%%%%%%
\begin{figure}
	\includegraphics[width=0.48\textwidth]{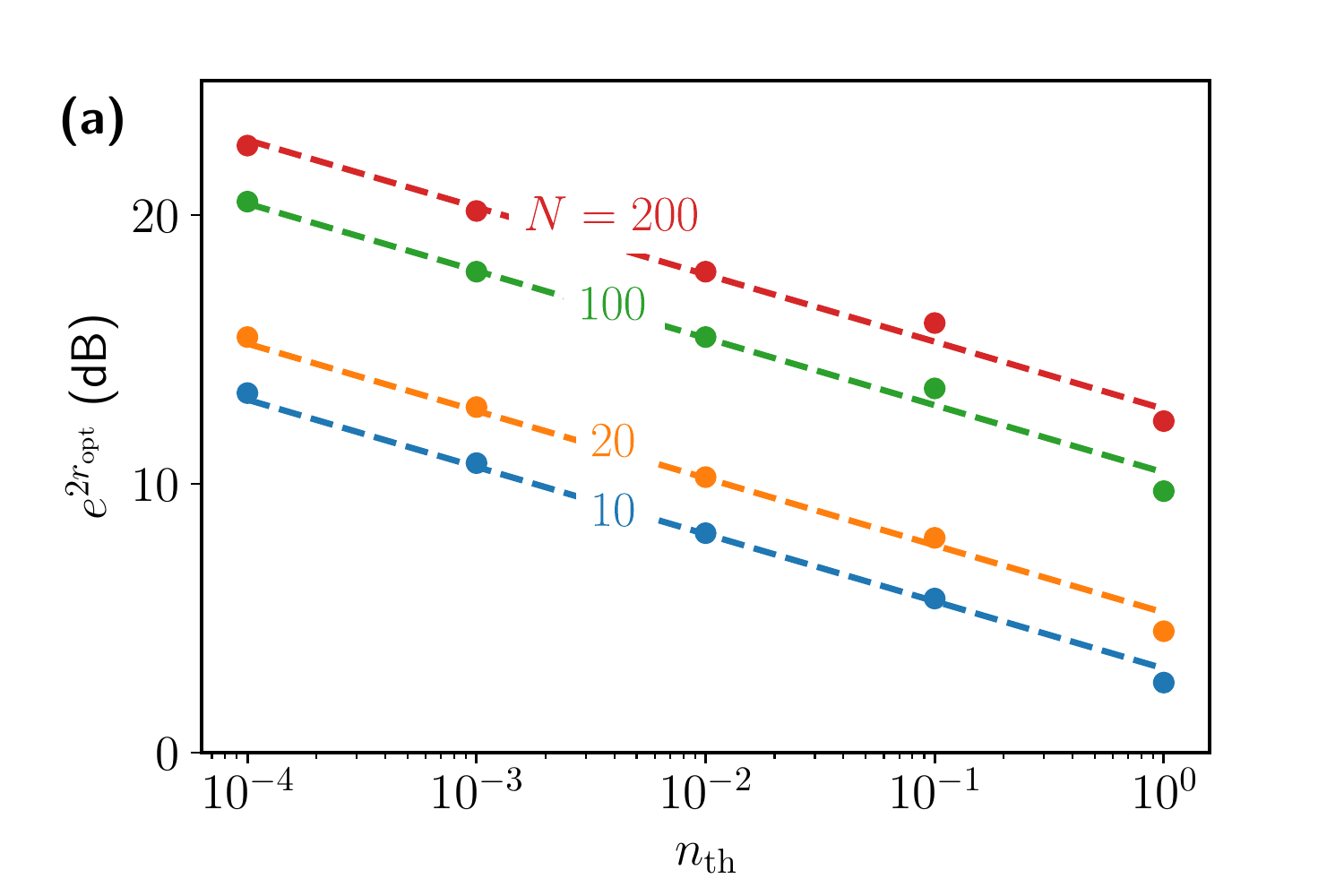}
	\includegraphics[width=0.48\textwidth]{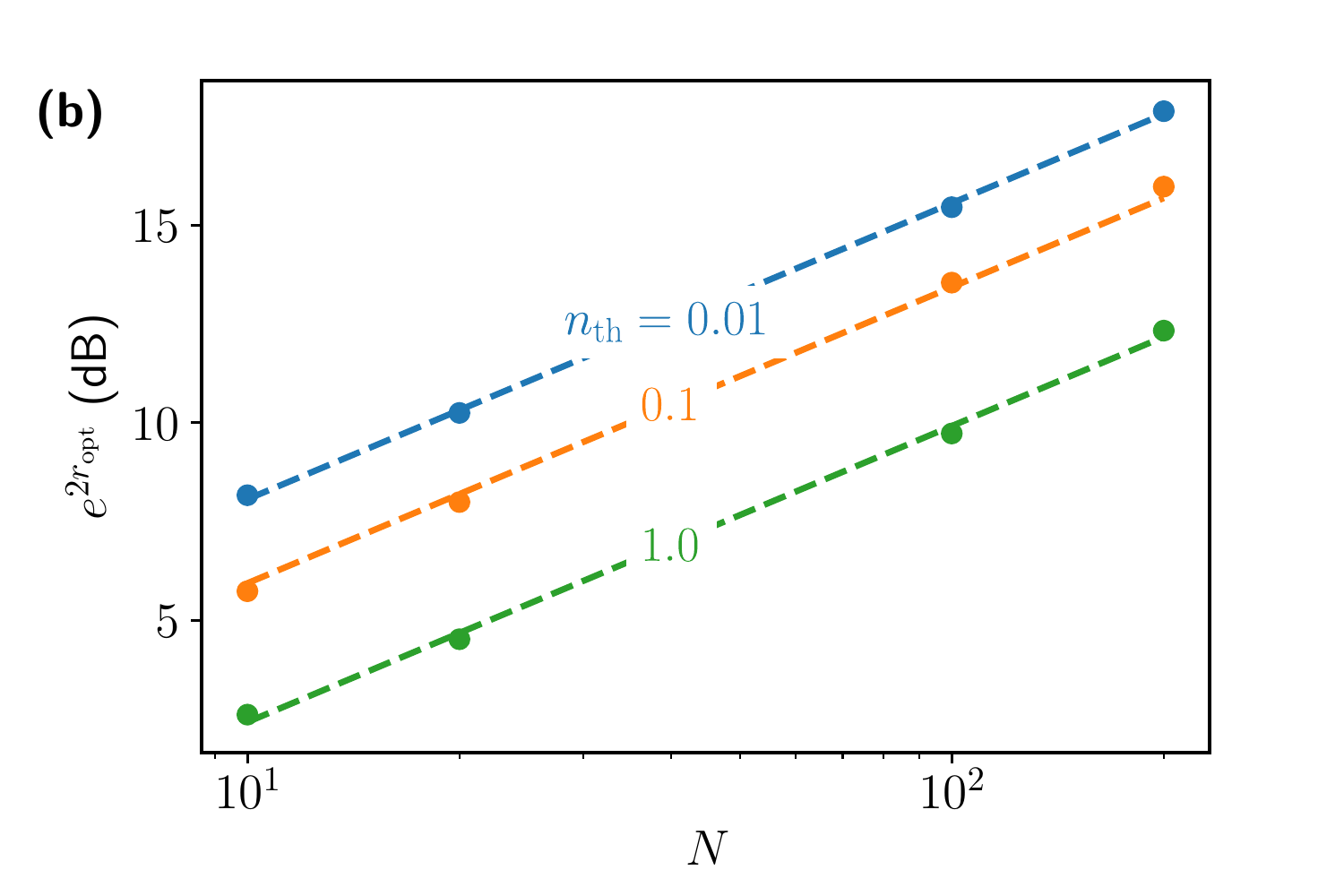}
	\caption{
		Scaling of the optimal squeezing parameter $r_\mathrm{opt}$ introduced in Sec.~\ref{sec:Temperature} and of the steady-state purity if the engineered reservoir stabilizes an impure squeezed state [as modeled by Eq.~\eqref{eq:masterEqFiniteTempForSigma}]. 
		For each value of $n_\mathrm{th}$ and $N$, $r_\mathrm{opt}$ is determined by numerically minimizing $\wineland$. 
		\textbf{(a)} 
			Scaling of $e^{2 r_\mathrm{opt}}$ with $n_\mathrm{th}$. 
			The dashed lines are fits to a power law $n_\mathrm{th}^{\ValuefiniteTemperatureROptVsNthexp}$. 
		\textbf{(b)}
			Scaling of $e^{2 r_\mathrm{opt}}$ with $N$. 
			The dashed lines are fits to a power law $N^{\ValuefiniteTemperatureROptVsNexp}$. 
	}
	\label{fig:App:finiteTemp:Squeezing}
\end{figure}
%%%%%%%%%%%%%

In this appendix, we summarize the numerical results for the optimal squeezing parameter $r_\mathrm{opt}$ introduced in Sec.~\ref{sec:Temperature}.
Numerical optimizations of the Wineland parameter (see Fig.~\ref{fig:App:finiteTemp:Squeezing}) indicate that the optimal squeezing parameter $r_\mathrm{opt}$ is approximately consistent with a power-law scaling
\begin{align}
	e^{2 r_\mathrm{opt}} \propto N^a \times n_\mathrm{th}^b \comma
\end{align}
where $a \approx \ValuefiniteTemperatureROptVsNexp$ and $b \approx \ValuefiniteTemperatureROptVsNthexp$.

In the limits $r \gg r_\mathrm{opt}$ and $N \gg 1$, these results can qualitatively be understood using the heuristic picture introduced in Sec.~\ref{sec:Temperature}.
An approximate steady state of Eq.~\eqref{eq:masterEqFiniteTempForSigma} is obtained by truncating the true steady state to its two most important pure-state contributions,
\begin{align}
	\rhoop_\mathrm{ss} &\approx \frac{n_\mathrm{th} + 1}{2 n_\mathrm{th} + 1} \ket{\psi_\mathrm{dk}}\bra{\psi_\mathrm{dk}} \nonumber \\
	&+ \frac{n_\mathrm{th}}{2 n_\mathrm{th} + 1} \frac{\Sigmaop^\dagger \ket{\psi_\mathrm{dk}} \bra{\psi_\mathrm{dk}} \Sigmaop}{-2 \ave{\Sz}_\mathrm{dk}} \fullstop
\end{align}
In the large-$N$ and large-$r$ limit, the mean-spin length $\ave{\Sz}$ and the variance $\ave{\Sy^2}$ scale as
\begin{align}
	\ave{\Sz}_\mathrm{ss} &\to - \frac{n_\mathrm{th} + 1}{2 n_\mathrm{th} + 1} \frac{N^2}{4} e^{-2r} \comma \\
	\ave{\Sy^2}_\mathrm{ss} &\to + \frac{n_\mathrm{th} + 1}{2 n_\mathrm{th} + 1} \left( \frac{N^2}{8} e^{-4r} + 1 \right) \comma
\end{align}
where the constant term in the brackets is due to the finite $\ave{\Sy^2}$ variance of the ``first excited'' state $\propto \Sigmaop^\dagger \psidk{r}$.
Thus, the Wineland parameter will diverge exponentially if the constant becomes relevant, \ie, if we have roughly
\begin{align}
	e^{2r} \approx N \sqrt{\frac{n_\mathrm{th}+1}{n_\mathrm{th}}} \fullstop
	\label{eq:App:finiteTemp:ScalingExp2r}
\end{align}
Note that this result is only valid in the limits $N \gg 1$ and $r \gg r_\mathrm{opt}$, \ie, it should not be expected to reproduce the numerical scaling observed in Fig.~\ref{fig:App:finiteTemp:Squeezing} quantitatively.

%%%%%%%%%%%%%%%%%%%%%%%%%%%%%%%%%%%%%%%%%%%%%%%%%%%
%%%%%%%%%%%%%%%%%%%%%%%%%%%%%%%%%%%%%%%%%%%%%%%%%%%
%%%%%%%%%%%%%%%%%%%%%%%%%%%%%%%%%%%%%%%%%%%%%%%%%%%

\section{Impure engineered reservoir: scaling of $\xi_{R}^{2}$ with mean-field theory}
\label{sec:App:ImpureResMFT}

%%%%%%%%%%%%%
\begin{figure}
	\includegraphics[width=0.5\textwidth]{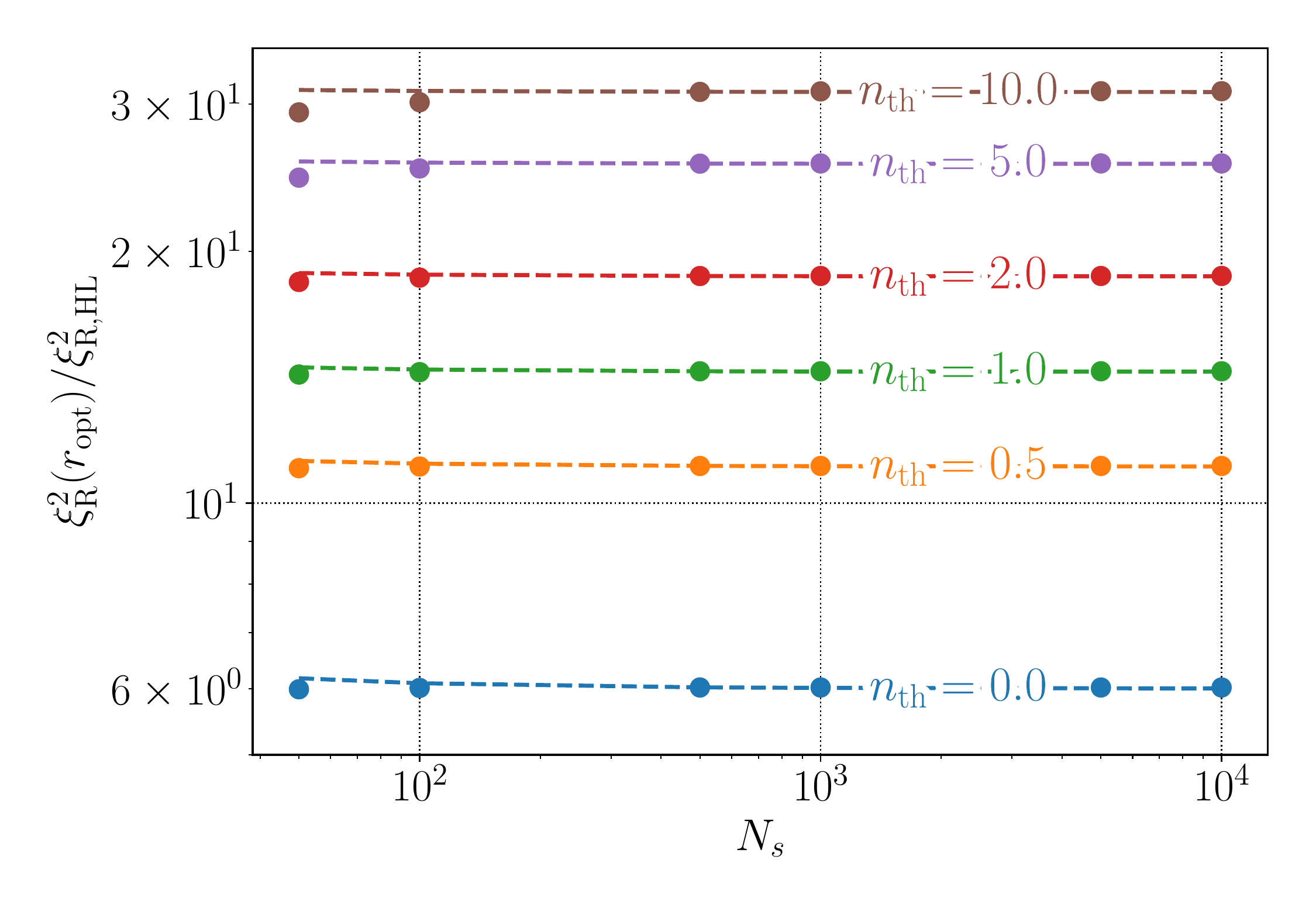}
	\caption{
            Ratio between the minimum steady-state Wineland parameter obtained by solving the mean-field equations of motion for the system governed by \cref{eq:App:masterEqFiniteTempForSigma}, and the ideal Heisenberg-limited value $\xi_\mathrm{R,HL}^2 = 2/(N+2)$. 
            Each dot corresponds a solution with optimized squeezing strength $r$, while the dashed curves are fits to a horizontal line (taken over the largest four values of $N$ for each $n_{\rm th}$). 
	}
	\label{fig:App:finiteTempMFT}
\end{figure}
%%%%%%%%%%%%%

In this Appendix we provide a brief companion discussion to \cref{sec:Temperature} of the main text, and explore a thermal squeezed light reservoir model that has imperfect purity, using mean-field theory. 
In particular we consider the spin-only master equation (\ref{eq:masterEqFiniteTempForSigma}) of the main text, which we reproduce here for completeness:
\begin{align}
    \dot{\rhoop} = \Gamma (n_\mathrm{th} + 1) \mD{\Sigmaop}{\rhoop} + \Gamma n_\mathrm{th} \mD{\Sigmaop^\dagger}{\rhoop}.
    \label{eq:App:masterEqFiniteTempForSigma}
\end{align}
Following \cref{sec:App:MeanFieldTheory}, we can once again write the corresponding mean-field equations setting third cumulants to zero, which can be readily solved numerically. 
Of particular interest is the scaling of the Wineland parameter $\wineland$, with the spin number $N$. 
The results are presented in \cref{fig:App:finiteTempMFT}, where we show the ratio between the minimum steady-state Wineland parameter obtained by solving the mean-field equations of motion, and the ideal Heisenberg-limited value $\xi_\mathrm{R,HL}^2 = 2/(N+2)$ (which full theory predicts at $n_{\rm th}=0$). Each dot corresponds a solution with optimized squeezing strength $r$, while the dashed curves are fits to a horizontal line taken over the largest four values of $N$ for each $n_{\rm th}$. From the plot it is clear that the mean-field theory predicts $\sim 1/N$ scaling at large $N$ which is weakly skewed at smallest values of $N$ that we consider. These results are consistent with our full master equation simulations presented in \cref{sec:Temperature} of the main text.
Similarly to a simple bosonic theory prediction, we also find that the minimum steady state value of Wineland parameter scales linearly with growing $n_{\rm th}$. In particular, to a good approximation it satisfies the phenomenological equation 
\begin{align}
    \xi_{R}^{2} & \approx 2.8 (1 + 2 n_{\rm th}) \exp (-2 r_{\rm opt} ) ,
\end{align}
with $r_{\rm opt}$ separately optimized for each $n_{\rm th}$.
Finally, we point out that, as can be seen from the blue curve of \cref{fig:App:finiteTempMFT}, the mean-field theory solution does not correctly predict the prefactor of the $1/N$ in the case of $n_{\rm th} =0$ (i.e., where the corresponding ratio $\xi_{R}^{2}(r_{\rm opt}) /  \xi_\mathrm{R,HL}^2 $ should be equal to 1). This is a somewhat expected behavior, as in parameter regimes where the $\xi^{2}_{R}$ may be either at, or near its Heisenberg limited value, the state of the spin ensemble is not Gaussian, which may substantially lower the ability of mean-field theory to quantitatively describe its behavior.

\end{appendix}

%%%%%%%%%%%%%%%%%%%%%%%%%%%%%%%%%%%%%%%%%%%%%%%%%%%
%%%%%%%%%%%%%%%%%%%%%%%%%%%%%%%%%%%%%%%%%%%%%%%%%%%
%%%%%%%%%%%%%%%%%%%%%%%%%%%%%%%%%%%%%%%%%%%%%%%%%%%

\bibliography{library}

%%%%%%%%%%%%%%%%%%%%%%%%%%%%%%%%%%%%%%%%%%%%%%%%%%%
%%%%%%%%%%%%%%%%%%%%%%%%%%%%%%%%%%%%%%%%%%%%%%%%%%%
%%%%%%%%%%%%%%%%%%%%%%%%%%%%%%%%%%%%%%%%%%%%%%%%%%%

\end{document}